\documentclass[twocolumn,superscriptaddress,aps,prb,floatfix]{revtex4-2}
\usepackage[T1]{fontenc}
\usepackage{amsfonts}
\usepackage{comment}
\usepackage{amsmath,amssymb,epsfig,color}
\usepackage{float}
\usepackage{amsmath}
\usepackage{amssymb}
\usepackage{tabularx}
\usepackage{url}
\usepackage{graphicx}
\usepackage{dcolumn}
\usepackage{bbold}
\usepackage{physics}
\usepackage{multirow}
\usepackage{pifont}
\usepackage[normalem]{ulem}
\usepackage{standalone}

\usepackage[unicode,bookmarks=false]{hyperref}

\DeclareMathOperator{\sgn}{sgn}
\begin{document}
\title{Extrinsic anomalous Hall effect in altermagnets}
\author{A. Osin}
\affiliation{Racah Institute of Physics, Hebrew University of Jerusalem, Jerusalem 91904, Israel}
\author{A. Levchenko}
\affiliation{Department of Physics, University of Wisconsin--Madison, Madison, Wisconsin 53706, USA}
\author{M. Khodas}
\affiliation{Racah Institute of Physics, Hebrew University of Jerusalem, Jerusalem 91904, Israel}
\date{\today }

\begin{abstract}
We find the extrinsic anomalous Hall conductivity (AHC) to be comparable to the intrinsic one in roughly half of the altermagnetic spin Laue groups in the limit of large exchange splitting.
In materials with a finite Dzyaloshinskii-Moriya type interaction, the extrinsic contribution is essential even in the clean limit. 
In other altermagnets it is mostly negligible.
This peculiar behavior is linked to the nonanalytic dependence of the intrinsic AHC on spin-orbit coupling.
Both originate from the lifting of the spin degeneracy along the nodal planes as the weak spin-orbit coupling breaks the nonrelativistic spin symmetry.
\end{abstract}

\maketitle
\section{Introduction}

Traditionally, magnetic materials have been classified into two major categories: ferromagnets and antiferromagnets.
Ferromagnets possess a finite magnetization ($\mathbf{M}$), which has led to their widespread technological applications.
In antiferromagnets, the N\'eel vector ($\mathbf{N}$) is finite, while the magnetic moments of the two sublattices compensate each other, resulting in zero net magnetization.

Magnetically compensated altermagnets represent a new research frontier \cite{Hayami2018,Voleti2020,Yuan2020,Ma2021,Mazin2021,Smejkal2022,Smejkal2022a,Mazin2022,Fernandes2024,Bai2024}.
These unconventional magnets are distinguished from conventional antiferromagnets by their specific symmetries.
The antiunitary symmetry combines time reversal ($\mathcal{T}$) with a unitary operation $\mathcal{O}$ that exchanges the sublattices while preserving the individual magnetic moments.
In altermagnets, $\mathcal{O}$ is a rotation--proper or improper, symmorphic or nonsymmorphic--that is neither inversion ($\mathcal{P}$) nor a translation $T_{\mathbf{t}}$ by a vector $\mathbf{t}$.

%Altermagnetism is a novel type of magnetic order that shares some properties with ferromagnets and some with antiferomagnets \cite{Hayami2019,Voleti2020,Yuan2020,Junwei2021,Mazin2021,Smejkal2022,Smejkal2022a,Mazin2022,Fernandes2024,Bai2024}.
%Similar to antiferromagnets, an altermagnet has zero nonrelativistic net magnetization ($\mathbf{M}$).
%At the same time, altermagnets exhibit a finite anomalous Hall effect normally observed in ferromagnets.
%This implies that the magnetic symmetry group does not include elements combining time reversal $\mathcal{T}$ with inversion $\mathcal{P}$ or with spatial translation $T_\mathbf{t}$ by some vector $\mathbf{t}$.

Breaking of the $\mathcal{T}T_{\mathbf{t}}$ and $\mathcal{T} \mathcal{P}$ symmetries lifts the spin degeneracy at a generic momentum, $\mathbf{k}$ even in nonrelativistic limit of zero spin-orbit coupling (SOC). 
The resulting nonrelativistic spin splitting, $\Delta_{\mathrm{A}}$ is typically of the order of the electronic bandwidth.
At zero SOC, the spin degeneracies along planes determined by the operation $\mathcal{O}$ and the index two site symmetry subgroup, $S$ of a crystallographic point group, $P = S + \mathcal{O}S$.
The resulting $d$-, $g$- and $i$-wave  spin texture in momentum space is invariant under $S$ and flips the sign under $\mathcal{O}S$. 
Such spin splitting has been recently observed for MnTe and CrSb $g$-wave candidates \cite{Ding2024, Krempasky2024}.
These textures alternatively can be envisioned as arising from the Zeeman coupling of electrons and  multipolar moments of real space magnetization density, respectively \cite{Hayami2019,Voleti2020}. 

The above nonrelativistic spin degeneracies are protected by the symmetries forming the spin-group that by definition leaves a given arrangement of magnetic moments invariant \cite{Brinkman1966,Litvin1974,Lifshitz2005}.
The spin groups of collinear magnets have a continuous spin-only  subgroup $\infty {}^{'} 2$.
This group contains the SO(2) spin rotations around $\mathbf{N}$ as well as the two-fold rotations ${'} 2$ around an arbitrary $\mathbf{N}_\perp \perp \mathbf{N}$ combined with $\mathcal{T}$. 
The spin-only subgroup makes the Bloch Hamiltonian even in momentum $\mathbf{k}$, including crystals that nominally lack inversion center \cite{Smejkal2022}. 
For this reason,  the crystallographic point groups supporting altermagnetism contain inversion are referred to as the spin Laue groups (SLG).

The anti-unitary ${'} 2$ symmetry forbids the anomalous Hall effect in collinear and coplanar magnets \cite{Liu2025}.
As SOC breaks ${'} 2$ a finite anomalous Hall conductivity (AHC) is observed \cite{Feng2022, Reichlova2024}.
In the weak SOC limit, the nonanalytic Berry curvature is localized at the nodal planes of spin degeneracies protected by SLG \cite{Smejkal2020}.
Therefore, even though the relativistic SOC is essential for anomalous Hall effect, it is the nonrelativistic spin-symmetry that makes AHC non-analytic \cite{Attias2024} and/or linear in SOC \cite{Roig2024}.
Similar argument imply the enhanced transport nonlinearities specific to altermagnets \cite{Fang2024}.

The magnetization is odd under $\mathcal{T}\mathcal{O}$ symmetry.
Hence, SLG enforces $\mathbf{M}=0$ \cite{McClarty2024,Schiff2025}.
At zero SOC the magnetization may be induced by strain in the form of ferrimagnetism \cite{Borovik2013,Ma2021,Khodas2025}.
In contrast, at finite SOC the magnetization is finite if it is allowed by the magnetic point group, dependent in turn on $\mathbf{N}$ orientation \cite{Birss1964}.
The strain, therefore is an efficient way to manipulate altermagnets \cite{Ma2021,Aoyama2024,Chakraborty2024a,Belashchenko2025,Takahashi2025,Karetta2025}.

In relativistic problem, AHC and $\mathbf{M}$ are closely related.
The AHC is an antisymmetric part, of the conductivity tensor $\hat{\sigma}$. 
When contracted with the Levi-Civita tensor, it uniquely defines an axial vector which transforms as magnetization $\mathbf{M}$ under all unitary and nonunitary operations.
Therefore, the AHC and $\mathbf{M}$ coexist or both vanish per given magnetic point group.

Despite the close relationship between AHC and $\mathbf{M}$,
the former varies only weakly across the candidates, while the latter varies substantially \cite{Kluczyk2024}. 
According to Ref.~\cite{Roig2025} the magnetization is large if it is induced by the N\'eel order $\mathbf{N}$ already to the linear order in SOC.
The moments canting is shown to result from the bilinear coupling of the Dzyaloshinskii–Moriya type that is odd under exchange of $\mathbf{M}$ and $\mathbf{N}$. 
In cases where such coupling is inconsistent with the magnetic point group the magnetization appears at least in the second order in SOC and, therefore, is expected to be much weaker. 

\begin{table*}[t]
\centering
\caption{
The SOC at the $\Gamma$-point $\boldsymbol{\lambda}_{\mathbf{k}=0}$ AHC is finite (zero) in the class A (B) altermagnets (see Fig.~\ref{fig:classes} for specific representatives of the two classes). 
The table is in one-to-one correspondence with Tab. I of Ref.~\cite{Roig2025}.
Class A materials have the magnetization linear in SOC via the free energy contribution $\propto \boldsymbol{\lambda}_{\mathbf{k}=0}\cdot (\mathbf{N} \times \mathbf{M})$ of the Dzyaloshinskii–Moriya type. 
Concomitantly, the extrinsic AHC is similar to the intrinsic AHC. 
In class B materials the induced magnetization appears at second or higher order in SOC.
At the same time, the extrinsic AHC is negligible.}
\label{tab:AB}
\begin{tabular}{|c||c c c c||c c c c||c c|}
\hline
 &    $d$-wave & & & & $g$-wave & & & & $i$-wave & \\
\hline
         P      &  $C_{2h}$    &  \multicolumn{1}{|c|}{$D_{2h}$}  & \multicolumn{1}{c|}{$C_{4h}$} & $D_{4h}$  & $D_{4h}$ & \multicolumn{1}{|c|}{$D_{3d}$}  & \multicolumn{1}{c|}{$C_{6h}$} & $D_{6h}$ & $D_{6h}$ & \multicolumn{1}{|c|}{$O_h$} \\
    %\hline
    $\Gamma_N$    &  $B_g$       &   \multicolumn{1}{|c|}{$B_{1g}$}  & \multicolumn{1}{c|}{$B_g$} & $B_{2g}$ & $A_{2g}$ & \multicolumn{1}{|c|}{$A_{2g}$} & \multicolumn{1}{c|}{$B_g$} & $B_{1g}$ & $A_{2g}$ & \multicolumn{1}{|c|}{$A_{2g}$} \\
\hline
    SLG         &  ${}^22/{}^2m$  &  \multicolumn{1}{|c|}{${}^2m{}^2m{}^1m$} & \multicolumn{1}{c|}{${}^24/{}^1m$} & ${}^24/{}^1m{}^2m_y{}^1m_d$ &  ${}^14/{}^1m{}^2m{}^2m$ & \multicolumn{1}{|c|}{${}^1\bar{3}{}^2m$} & \multicolumn{1}{c|}{${}^26/{}^2m$} &  ${}^26/{}^2m{}^2m_y{}^1m_x$ & ${}^26/{}^1m^2m^2m$  & \multicolumn{1}{|c|}{${}^1m{}^1\bar{3}^2{m}$} \\
    \hline
    $\boldsymbol{\lambda}_{\mathbf{k}=0}$ & $\hat{x},\hat{y}$ & \multicolumn{1}{|c|}{$\hat{z}$} & \multicolumn{1}{c|}{0} & 0 & $\hat{z}$ & \multicolumn{1}{|c|}{$\hat{z}$} & \multicolumn{1}{c|}{0} & $0$ & $\hat{z}$ & \multicolumn{1}{|c|}{0}
\\ \hline
\end{tabular}
\end{table*}

In the existing literature the AHC in altermagnets is treated as intrinsic \cite{Smejkal2020,Roig2024,Attias2024,Sato2024,Roig2025}.
Namely, as determined by the geometrical properties of the Bloch wave-functions.
This intrinsic AHC, $\hat{\sigma}_{\mathrm{in}}$ is insensitive to disorder and is quantized in two dimensions in topological insulators.

In contrast, in gapless magnets such as ferromagnetic metals, the disorder produces a nonzero contribution, $\hat{\sigma}_{\mathrm{ex}}$ to the AHC known as extrinsic \cite{Crepieux2001,Nagaosa2010}.
In a metal, a momentum relaxing scattering is necessary in order to make the dc conductivity well defined.
The resulting extrinsic contribution to AHC $\hat{\sigma}_{\mathrm{ex}}$ depends on the strength as well as on the physical origin of the disorder scattering.
In cases where the scattering is spatially asymmetric with respect to the spin orientation the skew scattering contribution, $\hat{\sigma}_{\mathrm{ex}}^s$ scales as the inverse of the skew scattering rate
and, therefore, dominates the AHC in the clean limit.

In magnets with vanishing $\hat{\sigma}_{\mathrm{ex}}^s$ the AHC is finite in the clean limit.
Importantly, the clean limit of AHC is distinct from the AHC of a clean system. 
We show that even when the skew scattering is suppressed, $\hat{\sigma}_{\mathrm{ex}}$ may be of the same order of magnitude as $\hat{\sigma}_{\mathrm{in}}$.
The limiting extrinsic AHC depends on how the clean limit is approached.
There are two distinct ways of to reach it.
The disorder scattering rate may tend to zero either because of the vanishing impurity concentration, $n_{\mathrm{imp}}$ or because the scattering potential $U$ is taken to zero at fixed $n_{\mathrm{imp}}$.
In this work we adopt the latter point of view fixing $n_{\mathrm{imp}}$ and letting $U$ tend to zero.

Here we study the altermagnets that possess the inversion center at one of the lattice sites.
Introducing the Pauli matrices vector $\boldsymbol{\sigma}$ to describe spin, the
SOC Hamiltonian $\propto \boldsymbol{\lambda}_{\mathbf{k}}\cdot \boldsymbol{\sigma} $.
The inversion center makes the Hamiltonian including SOC even in momentum, $\boldsymbol{\lambda}_{\mathbf{k}} = \boldsymbol{\lambda}_{-\mathbf{k}}$ \cite{Roig2024}.
Hence, the current operator is strictly odd.

Furthermore, we focus on the short range disorder potential.
As the current operator is odd in $\mathbf{k}$, such disorder does not renormalize the current vertex.
Besides, it produces no skew scattering.
In result, the extrinsic AHC attains a finite value in the clean limit, referred to as extrinsic AHC for shortness.
Our conclusions hold qualitatively as long as the Fermi surface encloses $\mathbf{k}=0$ ($\Gamma$-point).

\begin{figure}[ht]
    \includegraphics[width=0.483\textwidth]{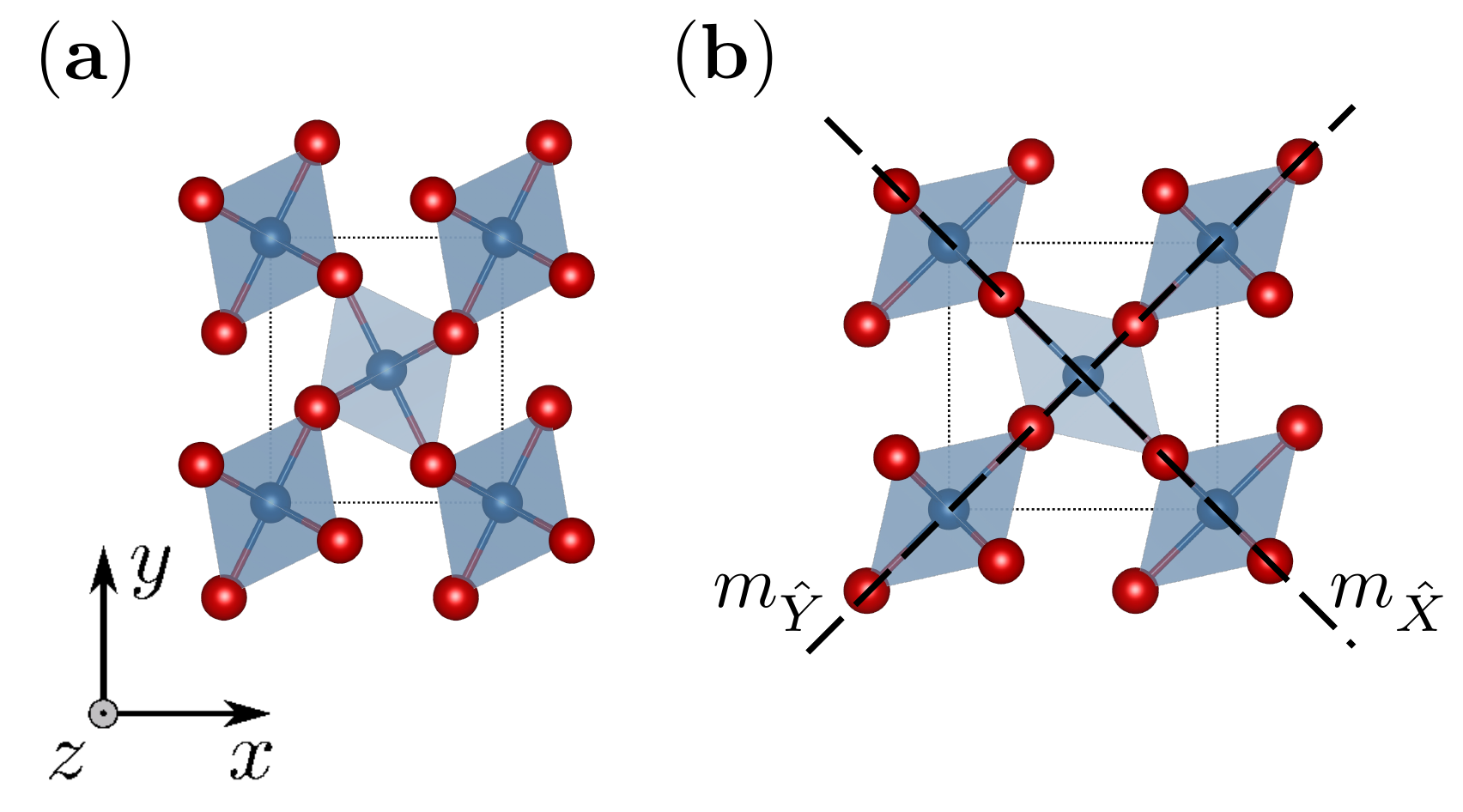}
    \caption{(a) The representative ${}^1m_z {}^2m_x {}^2m_y$ of class A. (b) The representative ${}^24/{}^1m{}^2m_y{}^1m_d$ of class B. 
    The horizontal mirror symmetry, $m_{\hat{z}}$ is common to the two systems. 
    The symmorphic mirror symmetry operations, $m_{\hat{X}}$ and $m_{\hat{Y}}$, where $\hat{X} =( \hat{x} + \hat{y} )/\sqrt{2}$,  $\hat{Y} =( \hat{x} - \hat{y} )/\sqrt{2}$ are present in the class B representative only. 
    These symmetries ensure the SOC vanishes along the symmorphic mirror plane intersections.
    In particular, while in class B the spectrum at $\Gamma$-point remains spin degenerate at finite SOC, in the class A this degeneracy is lifted.}
    \label{fig:classes}
\end{figure}

Altermagnets fall into two distinct categories, denoted as class A and class B, depending on whether the SOC lifts the spin degeneracy at the $\Gamma$-point (class A) or not (class B), see Tab.~\ref{tab:AB}.
We find that in class A altermagnets, the extrinsic AHC is comparable in magnitude to the intrinsic contribution and hence is essential.
In contrast, in class B altermagnets, the extrinsic AHC is negligible.
The distinction between the two classes is illustrated in Fig.~\ref{fig:classes} for two representative $d$-wave altermagnetic classes.

%%%%%%%%%

Our classification directly relates to the strength of the induced magnetization for the generic orientation of the N\'eel vector.
In materials of class A the magnetization is linear in SOC and is relatively large, while in materials of class B it appears at higher orders in SOC, and is typically weak \cite{Roig2025}.
Within the minimal models of Ref.~\cite{Roig2024} the magnetization results from the one-loop contribution to Landau free energy, $\Delta F = \boldsymbol{\lambda}_w\cdot (\mathbf{N} \times \mathbf{M})$, where $\boldsymbol{\lambda}_w$ is $\boldsymbol{\lambda}_{\mathbf{k}}$ properly averaged over the Brillouin Zone \cite{Roig2025}.
As $\boldsymbol{\lambda}_w \sim \boldsymbol{\lambda}_{\mathbf{k}=0}\equiv \boldsymbol{\lambda}_{0}$ our point is that when the induced magnetization is substantial the extrinsic AHC is comparable to the intrinsic one.
In contrast, for the materials with weak magnetization the extrinsic contribution is similarly negligible. 

The parallelism between the relative magnitude of $\mathbf{M}$ and extrinsic AHC is manifest in the one-to-one correspondence between the Tab.~\ref{tab:AB} and last column of Tab. I of Ref.~\cite{Roig2025}.
For altermagnets with magnetization linear in SOC, $\boldsymbol{\lambda}_0 \neq 0$, and extrinsic contribution is large.
In all other cases $\boldsymbol{\lambda}_0 = 0$ and the induced magnetization as well as the extrinsic AHC are suppressed.

%Contrary to a naive expectation, the large extrinsic contribution makes AHC of class A altermagnets universal in the following way.
%It is worth noticing that the conditions for a significant extrinsic contribution is essentially the same as the condition for a large magnetization identified recently in the Ref.~\cite{Roig2025}.
%In other words, class A altermagnets have a relatively large magnetization. 
%The actual magnitude of magnetization varies across the candidates.
%This in itself makes AHC dependent on the microscopic parameters that often are not known. 
%In such cases, the extrinsic contribution is a good fraction of the total AHC.
%It therefore reduces the non-universality introduced by the magnetization.

The paper is organized as follows. 
In Sec.~\ref{sec:Model} we formulate the microscopic models for the two representative candidates. 
The Sec.~\ref{sec:AHE} summarizes of the Kubo-St\v{r}eda formalism.
In Sec.~\ref{sec:Intrinsic} we recap the intrinsic AHE in the two-band limit, and consider the effect of the finite magnetization in the limit of weak SOC.
The Sec.~\ref{sec:Extrinsic} contains our main results on the extrinsic AHC.
We conclude in Sec.~\ref{sec:Conclusions}.

%%%%%%%%%%%%%%%%%%%%%%%%%%
\section{Model formulation}
\label{sec:Model}

%The model includes the band-structure Hamiltonian, $H_b$ and the disorder Hamiltonian $H_{\mathrm{dis}}$.
%As summarized in Introduction and detailed below altermagnets split into two classes deliniated in Tab.~\ref{tab:AB}.
%The class A altermagnets possess a relatively large magnetization and at the same time exhibit a large extrinsic AHC. 
%The rest of the altermagnets comprise a class B and show very small magnetization and negligible extrinsic AHC.
%
Although our discussion is general we consider the two specific representative of each class.
We chose them to be as similar structurally as possible to highlight the important differences of the two classes.
%To be specific, we consider the two representatives one for each class. 
We take the FeSb$_2$ as the representative of class A and rutile structures as the candidate of a class B.
The FeSb$_2$ is a $d$-wave altermagnet with ${}^2m_x{}^2m_y{}^1m_z$ SLG predicted to be metallic and magnetic upon doping \cite{Mazin2021}, see Fig.~\ref{fig:classes}a.
The epitaxial thin films of Mn$_5$Si$_3$ are experimentally realized candidates with the same SLG \cite{Reichlova2024}. 
These systems feature a large AHC and have a vanishingly small net magnetic moment.
%Our conclusions are based on symmetry, and hold qualitatively regardless of a particular realization.

The rutile structure has a  ${}^24/{}^1m{}^2m_y{}^1m_d$ SLG and falls into class B category, see Fig.~\ref{fig:classes}b.
Although initially RuO$_2$ has been proposed as the altermagnetic candidate it has been shown to be nonmagnetic in its bulk form \cite{Wenzel2025}.
The recently discovered oxyselenide altermagnets \cite{Ma2021,Cui2023,KV2Se2O,RbV2Te2O} have the same SLG as rutile structure.
%In what follows we analyze the microscopic models of the rutile structure.

The two representative systems are $d$-wave altermagnets that are structurally similar, see Fig.~\ref{fig:classes}.
Moreover, in both systems the spin degeneracy in the nodal planes $k_x=0$, $k_x=\pi$, $k_y=0$ and $k_y=\pi$ is enforced by the corresponding spin symmetry.
These similarities enable us to introduce the microscopic model for both at the same time while highlighting the key differences.

The minimal models of altermagnetism have been developed in a number of works, \cite{Roig2024,Attias2024,Brekke2023,Parshukov2024,Rao2024,Antonenko2025}.
The two-band models ignoring the sublattice degree of freedom can be insufficient.
These models fail to reproduce the in-gap states localized at the impurity sites \cite{Gondolf2025}.
Similarly, the proper description of the superconducting correlations  requires the full four-band model as the starting point \cite{Maeda2025,Fukaya2025a,Khodas2025,Chakraborty2025a}. 
Consequently, we perform a systematic reduction of the minimal four-band models to an effective two-band model in the limit of large exchange splitting, $2 |\mathbf{N}|$.
Such a careful reduction is necessary to ensure the consistency of the effective two-band model.
The four-band Hamiltonian, $\hat{H}_4$ operates in the spin and sublattice spaces parameterized by the two sets of Pauli matrices, $\boldsymbol{\sigma}$ and $\boldsymbol{\tau}$, respectively. 
The corresponding unit matrices $\sigma_0$ and $\tau_0$ are omitted. 
Specifically, 
\begin{align}\label{model1}
    \hat{H}_4 = E_{0}(\mathbf{k}) + t_{x,\mathbf{k}} \tau_x + t_{z,\mathbf{k}} \tau_z + \tau_y \boldsymbol{\lambda}_{\mathbf{k}} \cdot \boldsymbol{\sigma} + \tau_z \mathbf{N} \cdot \boldsymbol{\sigma}\, . 
\end{align} 

We consider the following hierarchy of energy scales defined by Eq.~\eqref{model1}.
The exchange splitting, $2|\mathbf{N}|$ is the dominant energy scale.
The next in the row is the bandwidth, $E_0 \ll |\mathbf{N}|$ set by the first term, 
$E_{0}(\mathbf{k}) = -2E_0 (\cos k_x + \cos k_y + \cos k_z) + 6 E_0$.
Here we assume this form of $E_{0}(\mathbf{k})$ to hold without loss of generality for both candidates.  
The spin independent part of Eq.~\eqref{model1} that is odd under $\mathcal{O}$ defines the altermagnetic spin splitting, 
$t_{z,\mathbf{k}} \sim \Delta_{\mathrm{A}} \ll E_0$.
It is reasonable to estimate $t_{x,\mathbf{k}}  \sim t_{z,\mathbf{k}}$.
The other energy scale is the Fermi energy at zero spin splitting, $E_F \lesssim E_0$.

We set
\begin{align}\label{eq:tx_tz}
    t_{x,\mathbf{k}} &= t_{x} \cos \frac{k_x}{2} \cos \frac{k_y}{2} \cos \frac{k_z}{2}\, ,
    \notag \\
    t_{z,\mathbf{k}} & = t_A \sin k_x \sin k_y 
\end{align}
for both candidates.
%What sets the two candidates apart is SOC at $\mathbf{k}=0$, $\boldsymbol{\lambda}_0$.
In class A $\boldsymbol{\lambda}_0 \neq 0$ and in class B $\boldsymbol{\lambda}_0 = 0$.
For our class A candidate we take \cite{Roig2024,Attias2024}, 
\begin{align}\label{eq:SOC_A}
    \lambda_{x,\mathbf{k}} & = \lambda_x \sin \frac{k_x}{2} \cos \frac{k_y}{2} \sin \frac{k_z}{2}\, ,
    \notag \\
    \lambda_{y,\mathbf{k}} & = \lambda_y \cos \frac{k_x}{2} \sin \frac{k_y}{2} \sin \frac{k_z}{2} \, ,
    \notag \\
    \lambda_{z,\mathbf{k}} & = \lambda_z \cos \frac{k_x}{2} \cos \frac{k_y}{2} \cos \frac{k_z}{2}\, ,
\end{align}
where for simplicity we set $\lambda_{x} = \lambda_y = \lambda$.
In the model Eq.~\eqref{eq:SOC_A} $\boldsymbol{\lambda}_0 = \lambda_z \hat{z}$.
For the class B candidate we have \cite{Roig2024},
\begin{align}\label{eq:SOC_B}
    \lambda_{x,\mathbf{k}} & = \lambda \sin \frac{k_x}{2} \cos \frac{k_y}{2} \sin \frac{k_z}{2},
    \notag \\
    \lambda_{y,\mathbf{k}} & = -\lambda \cos \frac{k_x}{2} \sin \frac{k_y}{2} \sin \frac{k_z}{2},
    \notag \\
    \lambda_{z,\mathbf{k}} & = \lambda_z \cos \frac{k_x}{2} \cos \frac{k_y}{2} \cos \frac{k_z}{2}(\cos k_x - \cos k_y),
\end{align}
and $\boldsymbol{\lambda}_0 = 0$.

To gain a qualitative insight into AHC we consider the Hamiltonian \eqref{model1} in $\mathbf{k} \cdot \mathbf{p}$-approximation.
In this simplifying approach we retain the leading terms of the series expansion of Eq.~\eqref{model1} in $\mathbf{k}$ around the $\mathbf{k}=0$. 
In this approximation the spin- and lattice-independent part of the Hamiltonian $E_{0}(\mathbf{k}) \approx E_0 k^2$.
Furthermore, the Fermi surface at zero spin splitting is spherical, with the Fermi momentum, $k_F = \sqrt{E_F/E_0}$, and Fermi velocity $v_F = 2 E_0 k_F$.
Even though these assumptions grossly oversimplify the realistic Fermi surfaces, they qualitatively capture our main conclusions.

In $\mathbf{k} \cdot \mathbf{p}$-approximation in both Eqs.~\eqref{eq:SOC_A} and \eqref{eq:SOC_B} the $\lambda_{x,\mathbf{k}}$ and $\lambda_{y,\mathbf{k}}$ SOC components defines the energy scale, $\bar{\lambda} = \lambda k_F^2$.
The energy scale associated with the out of plane SOC induced spin splitting is $\lambda_z$ in class A and $\lambda_z k_F^2$ in class B.
Similarly, the altermagnetic splitting originating from $t_{z,\mathbf{k}}$ is characterized by the energy scale $\Delta_{\mathrm{A}} = t_A k_F^2$.
It is worth pointing out that all the energy scales related to SOC originate from the same microscopic atomic spin-orbit interaction.
Hence, we fix the ratio $\bar{\lambda}/\lambda_z$  to explore the the weak and strong SOC limits.

Finally, we specify the disorder potential $\hat{H}_{\mathrm{dis}}(\mathbf{r})$ as Gaussian and having short range correlation function, 
\begin{align}\label{dis_potential}
    \langle \hat{H}_{\mathrm{dis}}(\mathbf{r}) \hat{H}_{\mathrm{dis}}(\mathbf{r}')\rangle  = 
    n_{\mathrm{imp}} U^2 \delta(\mathbf{r}-\mathbf{r}')\, .
    \end{align}
The disorder potential, Eq.~\eqref{dis_potential} is spin independent.
Furthermore, in line with Ref.~\cite{Luttinger1958} we assume the scattering potential to have a finite range exceeding the unit cell and yet smaller than the inter-electron distance. 
As the tight-binding orbitals at the two sublattices can be assumed orthonormal our disorder potential causes mostly intra-sublattice  scattering.
%The approximation \eqref{dis_potential} is consistent with the $\mathbf{k}\cdot \mathbf{p}$ approximation provided inter-electron separation exceeds the potential range. 
%Here we think of the tight-binding limit of atomic orbitals strongly localized at the magnetic sites. 
%The disorder model \eqref{dis_potential} is not the most general.
In real situation the disorder may have a non-negligible matrix elements for intersublattice transitions.
This would be the case for the atomically localized interstitial defects \cite{Gondolf2025}.
As our goal is to highlight the universal features of the AHC we adopt the simplest model \eqref{dis_potential}.

%It scatters electrons within each sublattice independently and in exactly the same way. 
%Electron scattering rate off the disorder potential \eqref{dis_potential},
%$\Gamma_0 = 2 \pi \nu_F n_{\mathrm{imp}}W^2$, where $\nu_F$ is the density of states at the Fermi energy, and $n_{\mathrm{imp}}$ is the concentration of impurities. 
 
\subsection{Effective two-band low-energy Hamiltonian}
\label{sec:2-band}
We next construct the effective Hamiltonian in the limit of the large exchange splitting.
In this case we adopt the scheme proposed in Ref.~\cite{Attias2024}.
The last term of Eq.~\eqref{model1} taken alone defines the two doubly degenerate flat bands with energies $\pm |\mathbf{N}|$. 
The Bloch-periodic wave functions for the pair of bands with energies $+|\mathbf{N}|$ form the subspace,
$W_p =\mathrm{span} \{ u_a \bar{\psi}_+,u_b \bar{\psi}_-\}$ where the spinors are defined by the properties, $\hat{N}\cdot\boldsymbol{\sigma} \bar{\psi}_\pm = \pm \bar{\psi}_\pm$.
The orbital functions $u_{a(b)}$ describe states localized at the sublattices $a$ and $b$, respectively.
Similarly, the other two degenerate flat subbands at the energy $-|\mathbf{N}|$ define the subspace 
$W_n =\mathrm{span} \{ u_a \bar{\psi}_-,u_b \bar{\psi}_+\}$.
Hereinafter we refer to the $W_p$ and $W_n$ as positive and negative energy bands, respectively.

%%%%%%%%%
\begin{figure}[ht]
    \includegraphics[width=0.48\textwidth]{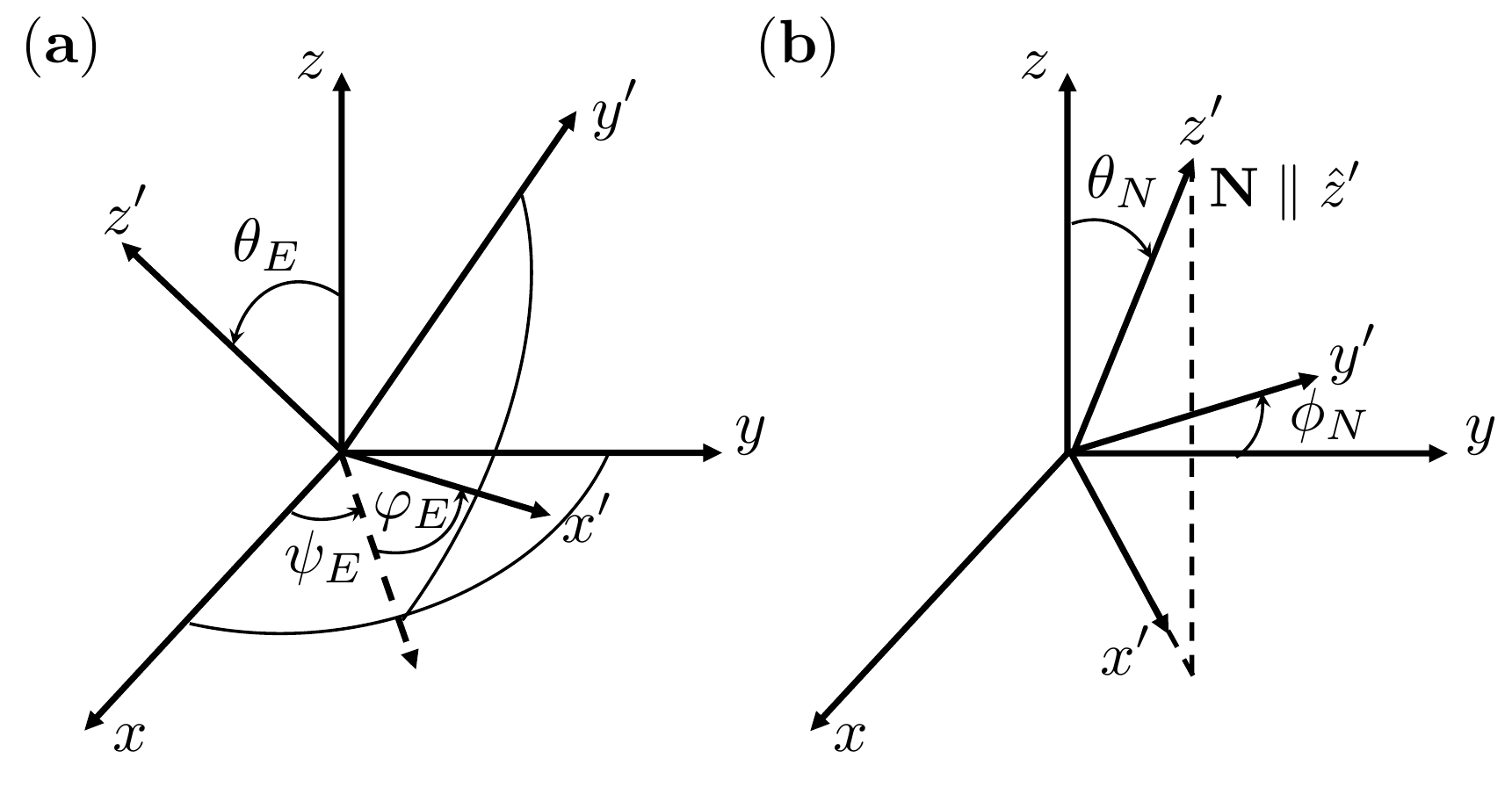}
    \caption{(a) The parametrization of the frame rotation by a three Euler angles, $\psi_E,\theta_E,\varphi_E$. The full rotation is the rotation around $\hat{z}$ by $\varphi_E$ followed by rotation around $\hat{x}$ by $\theta_E$ and by $\psi_E$ around $\hat{z}$.  (b) A rotation that aligns the $z'$ axis with the N\'{e}el vector $\mathbf{N}$ defined by $\theta_E = \theta_N$, $\psi_E = \phi_N + \pi/2$, and  we choose $\varphi_E = -\pi/2$, where $\theta_N$ and $\phi_N$ are polar and azimuthal angles of $\mathbf{N}$.}
    \label{fig:Euler}
\end{figure}

First, we rotate the spin quantization reference frame so that the $z$-axis of the rotated frame is aligned with $\mathbf{N}$, see Fig.~\ref{fig:Euler}.
Such a rotation, $R$ can be conveniently parametrized by Euler angles, $R = R(\psi_E,\theta_E,\varphi_E)$, see Fig.~\ref{fig:Euler}a.
The Pauli matrices rotate as components of the vector, such that $\sigma_i$ is transformed into  $\sum_{j=1}^3R_{ij}(\psi_E,\theta_E,\varphi_E)\sigma_j$.
Let us denote the polar and azimuthal angles of $\mathbf{N}$ as $\theta_N$ and $\phi_N$, respectively.
The Euler angles specifying $R$ are $\theta_E = \theta_N$, $\psi_E = \phi_N + \pi/2$, while $\varphi_E$ can be arbitrary, and we fix it to $\varphi_E = -\pi/2$ as shown in Fig.~\ref{fig:Euler}b. 
The matrix defining the transformation of the spin Pauli matrices then reads,
\begin{align}\label{eq:Lyub2}
    R=%(\theta_N, \!- \frac{\pi}{2},\phi_N\!+\!\frac{\pi}{2}) \!=\!
    \begin{pmatrix}
          \cos \theta_N \cos \phi_N &  - \sin \phi_N  & \cos \phi_N \sin \theta_N \\
        \cos \theta_N \sin \phi_N & \cos \phi_N  
        & \sin \phi_N \sin \theta_N \\
        - \sin \theta_N & 0 & \cos \theta_N       
    \end{pmatrix}\, .
\end{align}

Equation \eqref{eq:Lyub2} allows us to write the Hamiltonian, \eqref{model1} in the spin rotated basis.
In this basis formed by the two states of $W_u$ and two states of $W_d$ the Hamiltonian takes the block diagonal form, 
\begin{align}\label{eq:structure}
    \hat{H}_4 = \begin{pmatrix}
        \check{H}_p & \check{V} \\ \check{V}^\dagger & \check{H}_n
    \end{pmatrix}\, ,
\end{align}
where the 2 by 2 matrices are denoted by the 'check'-symbol, $\check{H}$, and 4 by 4 matrices are denoted by the 'hat'-symbol $\hat{H}$.
The Hamiltonians $\check{H}_{p(n)}$ are sufficient in the $1/|\mathbf{N}| =0$ limit.
In the spin rotated basis they take the form, 
\begin{align}\label{eq:check_H}
    \check{H}_{p(n)} = \pm |\mathbf{N}| + \check{h}_{p(n)}
\end{align}
where the two Hamiltonians describing the band splitting at each of the two quasi-degenerate  manifolds are conveniently parametrized by the pseudo-spin Pauli matrices $\boldsymbol{\rho}$,
\begin{align}\label{Hud}
    \check{h}_{p(n)}(\mathbf{k}) = E_0(\mathbf{k}) + \mathbf{h}^{p(n)}_{\mathbf{k}}\cdot \boldsymbol{\rho}\, .
\end{align}
%where 
\begin{subequations}\label{h_parameter}
%    \begin{align}\label{h0_ud}
%    h^{u(d)}_{0,\mathbf{k}} = & E_0(\mathbf{k}) \pm N \, .
%\end{align}
\begin{align}\label{hV_ud}
    \left[\mathbf{h}^{p(n)}_{\mathbf{k}}\right]_x \!\!= &  \pm(\lambda_{x,\mathbf{k}} \sin \phi_N -\lambda_{y,\mathbf{k}}\cos \phi_N), \notag \\
    \left[\mathbf{h}^{p(n)}_{\mathbf{k}}\right]_y \!\!= &  \left(\lambda_{x,\mathbf{k}}  \cos \phi_N 
   \! +\! \lambda_{y,\mathbf{k}} \sin \phi_N\right)\cos \theta_N \!-\! \lambda_{z,\mathbf{k}}\sin \theta_N ,
\notag \\
    \left[\mathbf{h}^{p(n)}_{\mathbf{k}}\right]_z \!\!= &  t_{z,\mathbf{k}}\, .
\end{align}
\end{subequations}

The off diagonal blocks describing the mixing of the positive and negative energy states read  
\begin{subequations}\label{Vud}
    \begin{align}\label{check_V}
    \check{V} = v_\mathbf{k} \rho_x + \check{V}_{\mathbf{M}}\, .  
\end{align}
where the first part 
\begin{align}\label{vk}
v_\mathbf{k} & =  t_{x,\mathbf{k}} 
\notag \\
\! - \! i & [ (\lambda_{x,\mathbf{k}} \cos \phi_N  \!+\!\lambda_{y,\mathbf{k}}  \sin \phi_N )\sin \theta_N \!+\! \lambda_{z,\mathbf{k}} \cos \theta_N]
\end{align}
results from the band structure including SOC,
and the second part 
\begin{align}
    \check{V}_{\mathbf{M}} = & M_x (\cos \theta_N \cos \phi_N   +   i \sin \phi_N \rho_z + i \cos \phi_N \sin \theta_N \rho_y)
   \notag \\
    + & M_y (\cos \theta_N \sin \phi_N    -i \cos \phi_N \rho_z + i \sin \phi_N \sin \theta_N \rho_y)
   \notag \\
  + &  M_z (- \sin \theta_N  + i \cos \theta_N \rho_y )
\end{align}
results from a finite magnetization, $\mathbf{M}$.
\end{subequations}
%For both candidates, the AHC is possible only if the N\'{e}el vector has a finite component in the basal $xy$-plane.
%In both cases the non-symmorphic mirror in a plane parallel to $z$-axis is the operation of the magnetic point group that forbids the magnetization along $\hat{z}$.
In summary, the band Hamiltonian is fixed by Eqs.~\eqref{eq:structure}, \eqref{Hud}, \eqref{h_parameter} and \eqref{Vud}.
\subsubsection{The choice of the N\'{e}el vector orientation}
For specific orientations of the N\'eel vector the AHC and magnetization are not allowed by the magnetic point group symmetry.
%The magnetic moments transform as axial vectors under MPG.
In both classes considered here the magnetization is forbidden for $\mathbf{N} \parallel \hat{z}$.
In this case $\mathbf{N}$ as any other axial vector flips under the mirror in any plane that is parallel to $\mathbf{N}$.
The $m_x$ and $m_y$ mirror operations are accompanied by a fraction of a lattice translation exchanging the two sublattices.
This restores the original orientation of N\'{e}el vector.
Hence, these non-symmorphic mirror operations are part of the magnetic point group for $\mathbf{N} \parallel \hat{z}$. 

The magnetization $\mathbf{M}$ is the same at the two sublattices.
Therefore, $\mathbf{M} \parallel \hat{z}$ is mapped to $-\mathbf{M}$ under the $m_x$ and $m_y$ operations.
This excludes $\mathbf{M} \parallel\hat{z}$.
Furthermore, $\mathbf{M}\parallel \hat{x} (\hat{y})$ is ruled out by $m_y$ and $m_x$, respectively, or alternatively by the symmorphic $m_z$.
In conclusion, the magnetization and anomalous Hall effect are not allowed for $\mathbf{N} \parallel \hat{z}$.
The arguments apply equally to both classes ${}^1m_z {}^2m_x {}^2m_y$ and ${}^24/{}^1m{}^2m_y{}^1m_d$ as well as to $g$-wave ${}^26/{}^2m{}^2m_y{}^1m_x$ candidates such as CrSb.
%The conclusion holds also for CrSb representative of $g$-wave ${}^26/{}^2m{}^2m_y{}^1m_x$ class.
In these examples the finite AHC conductivity results from the strain in the form of the elasto-Hall conductivity \cite{Takahashi2025}.

If on the other hand the N\'eel vector is in-plane, e.g. $\mathbf{N}\parallel \hat{y}$, the $m_x$, $\mathcal{T}m_y$ and $\mathcal{T}m_z$ symmetries allow for a finite $\mathbf{M}\parallel \hat{x}$-direction resulting in the canting of magnetic moments within the $xy$-plane \cite{Mazin2021,Attias2024}.  
This, again holds for the two classes considered.

In the case of ${}^24/{}^1m{}^2m_y{}^1m_d$ the diagonal mirror operations $m_d$ introduce the additional constrain.
The allowed $M_y$ for $\mathbf{N} = |\mathbf{N}| \hat{x}$ is the same as the allowed $M_x$ for $\mathbf{N} = |\mathbf{N}| \hat{y}$.
In fact, the bilinear coupling between $\mathbf{N}$ and $\mathbf{M}$ in the free energy $\propto (M_x N_y + M_y N_x)$ \cite{Roig2025}.
In contrast to the Dzyaloshinskii–Moriya coupling it is symmetric in the sublattice exchange.
For this reason, this coupling does not appear to the linear order in SOC \cite{Roig2025}.
One can also see it directly as the diagonal symmetry planes contains the atoms at the two sublattices, see Fig.~\ref{fig:classes}b.

%For the same reason as before, the magnetization is forbidden for $\mathbf{N} \parallel \hat{z}$ for ${}^24/{}^1m{}^2m_y{}^1m_d$ candidates belonging to the class B.
%In this case the $\mathbf{M} = M_y \hat{y}$ is allowed for $\mathbf{N} = |\mathbf{N}| \hat{x}$, and $\mathbf{M} = M_x \hat{x}$ is allowed for $\mathbf{N} = |\mathbf{N}| \hat{y}$, and moreover $M_x = M_y$.

Based on the above arguments below we set $\mathbf{N}= |\mathbf{N}| \hat{y}$ without loss of generality.
The only component of the AHC tensor allowed by symmetry is $\hat{\sigma}_{yz}$, and we omit the coordinate indices of AHC tensor when it does not cause confusion.

%The AHC is finite only for those configurations that allow for the finite magnetization.
%Let us set, $\mathbf{N}= |\mathbf{N}| \hat{y}$, and $\mathbf{M}=M \hat{x}$ for definitness.
%We now specify the model formulated for general orientation of the these vectors above for this specific orientation. 
%
%\subsubsection{The models for the N\'eel vector, \( \mathbf{N}= N \hat{y} \)}
The specific Hamiltonian for the N\'eel vector \( \mathbf{N}= |\mathbf{N}| \hat{y} \) is obtained by setting $\theta_N = \pi/2$ and $\phi_N=\pi/2$ in Eqs.~\eqref{hV_ud} and \eqref{vk}.
The Eq.~\eqref{hV_ud} takes the form  
\begin{align}\label{hV_ud_A}
    \mathbf{h}^{p(n)}_{\mathbf{k}} = &  (\pm \lambda_{x,\mathbf{k}} , - \lambda_{z,\mathbf{k}} ,t_{z,\mathbf{k}} )\, ,
    \end{align}
and the Eq.~\eqref{vk} simplifies to
\begin{align}\label{check_V1}
    \check{V} = (t_{x,\mathbf{k}}  - i   \lambda_{y,\mathbf{k}}) \rho_x + i M \rho_z\, .  
\end{align}

%\begin{align}\label{vk_A}
%v_\mathbf{k} & =  t_{x,\mathbf{k}}  - i   \lambda_{y,\mathbf{k}} \, .
%\end{align}
%The Hamiltonian for the  N\'eel vector \( \mathbf{N}= N \hat{x} \) is obtained by setting $\theta_N = \pi/2$ and $\phi_N=0$ in Eqs.~\eqref{hV_ud} and \eqref{vk} in exactly the same way.

\section{Kubo-St\v{r}eda formulation of AHC}
\label{sec:AHE}
%\subsection{Symmetry considerations}

In the Kubo-St\v{r}eda formulation, we employ here the Hall conductivity tensor splits into two contributions: $\hat{\sigma} = \hat{\sigma}^I + \hat{\sigma}^{II}$ having distinct physical meaning \cite{Streda1982}. 
The second term $\hat{\sigma}^{II}$ is proportional to the derivative of the total density with respect to the applied magnetic field. 
It has a topological interpretation and is insensitive to disorder. 
In contrast, $\hat{\sigma}^I$ describes the response of electrons at the Fermi surface.
When the bulk is gapped the only non-zero contribution is $\hat{\sigma}^{II}$.
Since the latter is also independent on the disorder the extrinsic contribution vanishes, 
and $\hat{\sigma}  = \hat{\sigma}^{II} = \hat{\sigma}_{in}$.

In metals both contributions are generally finite. 
Since our goal is the extrinsic contribution, it is sufficient to focus on $\hat{\sigma}^{I}$.
The extrinsic contribution is the difference of $\hat{\sigma}^{I}$ in the clean limit $U \rightarrow 0$ and $\hat{\sigma}^{I}$ in clean system, $U = 0$ at fixed $n_{\mathrm{imp}}$
\begin{align}\label{sigma_ex}
    \hat{\sigma}_{ex} = \lim_{U \rightarrow 0}\hat{\sigma}^{I}_{U} -\hat{\sigma}^{I}_{U=0} \, , \quad n_{\mathrm{imp}} = \mathrm{const}\, ,
\end{align}
where in what follows we omit the subscript $U$.
The Eq.~\eqref{sigma_ex} is finite if there is no skew scattering contribution as has been noticed early on in Ref.~\cite{Luttinger1958}.

Often one employs a different yet equivalent definition of the extrinsic contribution \cite{Burkov2014}.
Considering the ac Hall current at a finite frequency, the extrinsic contribution is the difference of the ballistic AHC at frequencies exceeding the disorder scattering rate and the AHC in the opposite, diffusive limit.
We prefer the equivalent definition \eqref{sigma_ex} defined strictly as the dc response.
This formulation additionally specifies our choice of weak scattering amenable to a simple Born approximation.
We note that if the clean limit is understood as vanishing concentration of possibly strong scatterers the results may be quantitatively different as in the case of Weyl semimetals \cite{Hou2015,Messica2023}.

We compute the $\hat{\sigma}^{I}$ which is given by the disorder averaging of the correlation function,
\begin{align}\label{eq:sigma^I}
    \sigma^{I}_{\alpha \beta} = \frac{e^2}{4 \pi} \Tr \left[ \hat{v}_{\alpha} \hat{G}^R \hat{v}_{\beta} \hat{G}^A - \hat{v}_{\beta} \hat{G}^R \hat{v}_{\alpha} \hat{G}^A \right]\, ,
\end{align}
where the $\Tr$ is taken over spin/sublattice as well as spatial degrees of freedom.
The velocity operator is diagonal in the Bloch momentum, with the diagonal matrix elements, $\hat{v}_{\alpha} = \partial \hat{H}_4/\partial \mathbf{k}$.
The retarded and advanced Green functions entering Eq.~\eqref{eq:sigma^I} are defined in the standard way, 
\begin{align}
    \hat{G}^{R(A)}= \left[\epsilon - \left(\hat{H}_4 + \hat{H}_{\mathrm{dis}}\right) + E_F \pm i 0 \right]^{-1}
\end{align}
and are evaluated at $\epsilon=0$.

Note that in Eq.~\eqref{eq:sigma^I} we deliberately kept the antisymmetric part of the conductivity tensor.
The other symmetric part describes the anisotropic magnetoresistance also known as planar Hall effect.
In specifying $\hat{\sigma}^I$ to the antisymmetric part we focus on AHC and relegate the anisotropic magnetoresistance to separate studies.
Physically, the two phenomenona give rise to a distinct angular dependence on the in-plane N\'eel vector orientation \cite{Gonzalez2024}.

Before addressing the extrinsic AHC we briefly review the intrinsic contribution in our model.
The bands form the two weakly spin split doublets of bands $2 |\mathbf{N}|$ apart in energy.
For definiteness we assume that the Fermi level crosses the positive energy bands, while the negative energy bands are fully occupied.

%%%%%%%%%%%%%%%%%%%%
%It might nevertheless be meaningful to discuss the anomalous Hall conductivity at zero magnetization.
%This is the case when the magnetization even if allowed by symmetry is zero or parametrically small. 
%Diagrammatic contributions to $\hat{\sigma}^I$ have been interpreted quasi-classically in terms of various scattering processes \cite{Crepieux2001}.
%%%%%%%%%%%%%%%%%%%%
%The topological contribution can be obtained by looking at the clean system.
%In this case the Hall conductivity is intrinsic, and can be determined by integrating the Berry curvature over the occupied bands.
%Next, the non-topological part is computed directly. 
%The topological term then follows as the difference $\hat{\sigma}^{II} = \hat{\sigma} - \hat{\sigma}^{I}$.
%%%%%%%%%%%%%%%%%%%%
%The Hall conductivity within the effective model \eqref{eq:2by2} is even in SOC and odd in altermagnetic splitting. 
%Indeed, the sign of the SOC without affecting the sign of $A$ as it appears in the effective Hamiltonian is achieved by the unitary transformation of rotation by $\pi$ radians, $U = \exp[i (\pi/2)\rho_z]$. 
%The Hall conductivity can also be shown to be odd in $A$, see App. 
%%%%%%%%%%%%%%%%%%%%
\section{Intrinsic AHC}
\label{sec:Intrinsic}
We start with reviewing the intrinsic anomalous Hall effect in the clean system \cite{Xiao2010},
%In this case, we have the only non-zero Hall conductivity \cite{Xiao2010},
\begin{align}\label{eq:sigma_Berry}
    \hat{\sigma}_{in} = -\sum_n \frac{e^2}{\hbar}\int \frac{d^3 k}{(2 \pi)^3} f_{n\mathbf{k}}\hat{\Omega}^{(n)}(\mathbf{k})\, ,
\end{align}
where $f_{n\mathbf{k}}$ is the Fermi-Dirac occupation number of the Bloch states at the band $n$ and momentum $\mathbf{k}$, and the anti-symmetric Berry curvature tensor per band, $n$ has components
\begin{align}\label{BC:define}
      \Omega^{(n)}_{\alpha \beta}(\mathbf{k})  =- 2 \Im  \langle \partial_{\alpha} n \mathbf{k} |\partial_{\beta}  n\mathbf{k}\rangle \, .
\end{align}
Here $|n \mathbf{k}\rangle$ denote the periodic parts of the Bloch wave-function, and $\partial_{\alpha} $ stands for $\partial/ \partial k_{\alpha}$.

\begin{figure}[ht]
    \includegraphics[width=0.38\textwidth]{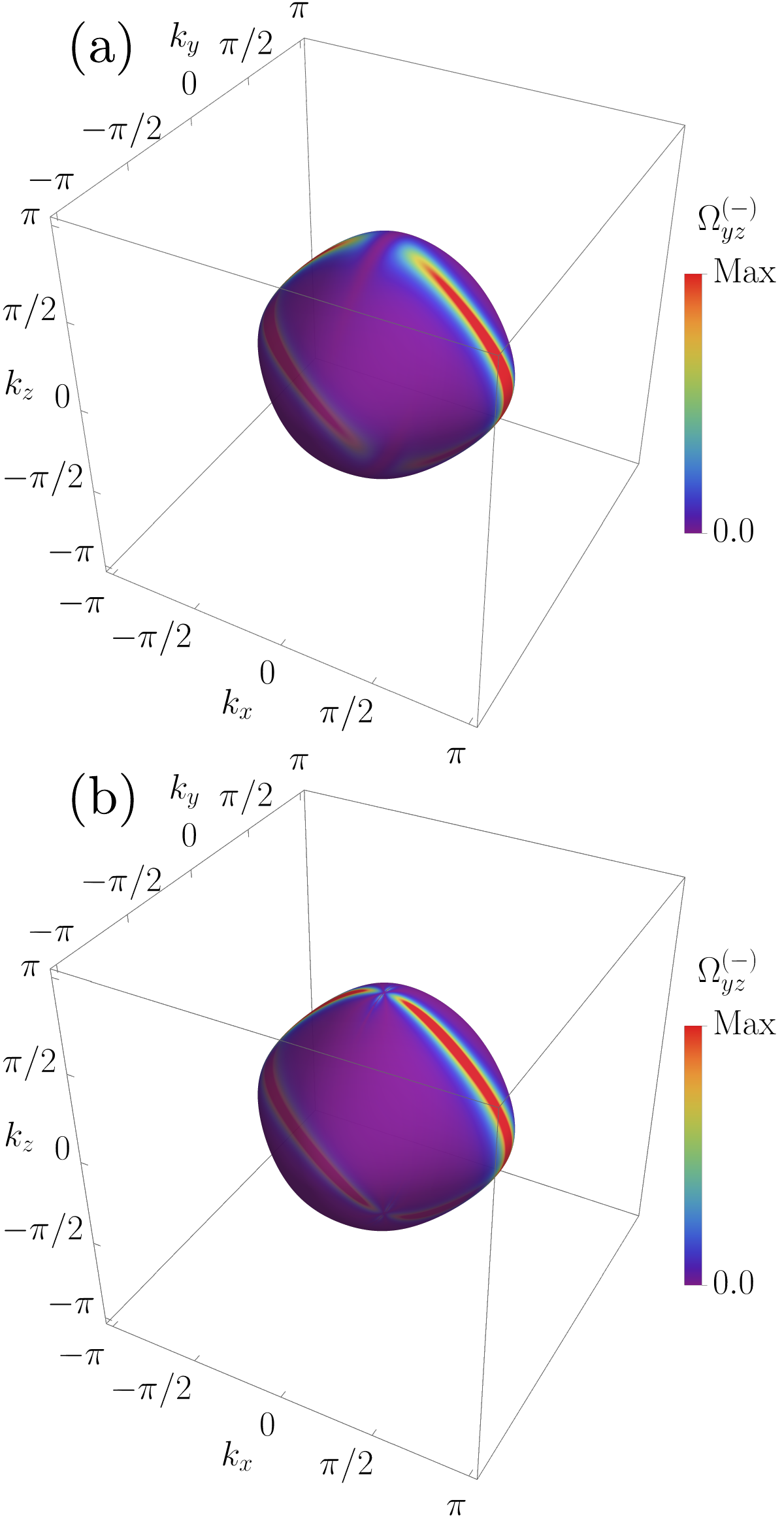}
    \caption{
    The Berry curvature $\Omega^{(-)}_{yz}(\mathbf{k})$ for the band $E_{\mathbf{k}}^-$ shown at the 
    outer Fermi surface, $E_{\mathbf{k}}^-=E_F$ in the two-band limit $1/|\mathbf{N}|=0$. The other two components of the $\hat{\Omega}$ tensor vanish for $\mathbf{N}\parallel \hat{y}$. The Berry curvature of the second inner band $E_{\mathbf{k}}^+$, $\hat{\Omega}^{(+)}(\mathbf{k}) = -\hat{\Omega}^{(-)}(\mathbf{k})$ is not shown.
    Panels (a) and (b) show $\Omega^{(-)}_{yz}(\mathbf{k})$ at the the larger of the two Fermi surfaces, $E_{\mathbf{k}}^-=E_F$ for the models representing class A and class B $d$-wave altermagnets, respectively as introduced in Sec.~\ref{sec:Model}. 
    The parameters, $E_0=E_F/3$, $t_A = 0.01 E_F$, $\lambda=\lambda_z = 0.24 t_A$ are the same for both panels, and are in the large altermagnetic splitting limit, $t_A\gg\mathrm{max}\{\lambda,\lambda_z\}$.
    Panels (a) and (b) are qualitatively similar.}
    \label{fig:Berry}
\end{figure}

The assumed inversion symmetry makes the second-rank Berry curvature tensor even in momentum, $\Omega^{(n)}_{\alpha \beta}(\mathbf{k})  = \Omega^{(n)}_{\alpha \beta}(-\mathbf{k})$.
For our choice of the N\'{e}el vector, $\mathbf{N}\parallel \hat{y}$ the mirror symmetries $m_x$, $\mathcal{T}m_y$ and $\mathcal{T}m_z$ enforce,
\begin{align}\label{Omega_symm}
\Omega^{(n)}_{yz}(k_x,k_y,k_z) & = \Omega^{(n)}_{yz}(-k_x,k_y,k_z)
\notag \\
\Omega^{(n)}_{yz}(k_x,k_y,k_z) & = \Omega^{(n)}_{yz}(-k_x,k_y,-k_z)
\notag \\
\Omega^{(n)}_{yz}(k_x,k_y,k_z) & = \Omega^{(n)}_{yz}(-k_x,-k_y,k_z)\, .
\end{align}
Combined with the $\mathcal{P}$ inversion symmetry, Eq.~\eqref{Omega_symm} implies that the $\Omega^{(n)}_{yz}(\mathbf{k})$ is symmetric under all three mirror operations $m_x$, $m_y$ and $m_z$, see Fig.~\ref{fig:Berry}.

%%%%%%%%%%%%%%%%%%%%%%%
%%%%%%%%%%%%%%%%%%%%%%%%%

We summarize the results in the limit of infinitely large exchange splitting in Sec.~\ref{sec:sigma_in_J=infty} followed by the discussion of the leading $1/|\mathbf{N}|$ corrections in Sec.~\ref{sec:sigma_in_J}.
%%%%%%%%%%%%%%%%%%%%%%%%%%%%%%%
\subsection{$\hat{\sigma}_{in}$ in the $1/|\mathbf{N}|=0$ limit}
\label{sec:sigma_in_J=infty}
In the limit $1/|\mathbf{N}| =0$ we can truncate the Hamiltonian, \eqref{eq:structure} down to the 2 by 2 Hamiltonian, $\check{H}_p$.
This is a two-band model with the band dispersions, 
\begin{align}\label{E+-}
    E^{\pm}_{\mathbf{k}} = E_0(\mathbf{k}) \pm h_{\mathbf{k}}\, .
\end{align}
Hereinafter, we omit the superscript of $\mathbf{h}^p_{\mathbf{k}}$ for brevity when discussion is focused on the positive energy bands.
The Berry curvature $\Omega^{(\pm)}(\mathbf{k})$ at the two bands $E_{\mathbf{k}}^\pm$, satisfies $\Omega^{(+)}(\mathbf{k}) = -\Omega^{(-)}(\mathbf{k})$.
Therefore, at low temperatures, Eq.~\eqref{eq:sigma_Berry} can be rewritten as the integral of $\Omega^{(-)}(\mathbf{k})$ over the volume $\Delta V$ enclosed between the smaller (larger) Fermi surface, FS$_{\pm}$ defined by $E^{\pm}_\mathbf{k} = E_F$, respectively. 
The well known expression for the Berry curvature allows us to write the intrinsic AHC, Eq.~\eqref{eq:sigma_Berry} in the form,
%\begin{align}\label{eq:BC2}
%    \Omega^{\pm}_{\alpha \beta} & = \mp \frac{\bm{h}}{2h^3}\cdot (\partial_{\alpha} \bm{h} \times \partial_{\beta} \bm{h})
%\end{align}
\begin{align}\label{eq:sigma_Berry_2}
\hat{\sigma}_{in;\alpha \beta} = - \frac{e^2}{\hbar}\int_{\Delta V} \!\frac{d^3 k}{(2 \pi)^3} 
\frac{\mathbf{h}_{\mathbf{k}}}{2 h^3_{\mathbf{k}}}\!\cdot\! (\partial_{\alpha} \mathbf{h}_{\mathbf{k}} \times \partial_{\beta} \mathbf{h}_{\mathbf{k}})\, .
\end{align}
%where the integration volume $V$ is enclosed between the smaller (larger) Fermi surface, FS$_{\pm}$ defined by $E^{\pm}_\mathbf{k} = E_F$, respectively. 
The Berry curvature at the $E_{\mathbf{k}}^-$ band, $\Omega^{(-)}(\mathbf{k})$, is shown for the two SLG in Fig.~\ref{fig:Berry}. 
%The Berry curvature on the $E_{\mathbf{k}}^+$ is $\Omega_{\mathbf{k}}^+=- \Omega_{\mathbf{k}}^-$.

As we consider the limit of the small spin splitting, $h \ll E_0$, we approximate the volume integration in Eq.~\eqref{eq:sigma_Berry_2} by an integral over the Fermi surface at zero spin splitting, $h=0$.  
\begin{align}\label{eq:sigma_Berry_3}
    \hat{\sigma}_{in;\alpha \beta} =  -\frac{e^2}{(2 \pi)^3} \oint \frac{d S}{v_F} \frac{\mathbf{h}_{\mathbf{k}}\cdot (\partial_{\alpha} \mathbf{h}_{\mathbf{k}} \times \partial_{\beta} \mathbf{h}_{\mathbf{k}})}{ h^2_{\mathbf{k}}}\, .
\end{align}

\begin{figure}[ht]
    \includegraphics[width=0.48\textwidth]{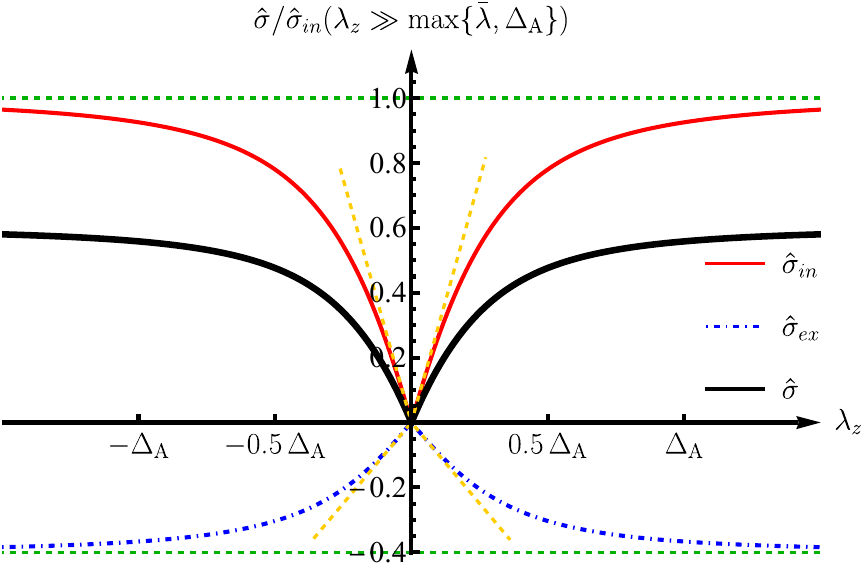}
    \caption{
     AHC $\hat{\sigma}$ of a class A representative as a function of the SOC, $\lambda_z$ in the two-band limit $1/|\mathbf{N}|=0$, $\bar{\lambda} = 0.09\lambda_z$, $\Delta_\mathrm{A}/E_F = 0.02$, and $k_F = 0.3$. 
     The magnetization is set to zero.
     The thin (red), dot-dashed (blue), and solid (black) curves show the intrinsic $\hat{\sigma}_{in}$, extrinsic $\hat{\sigma}_{ex}$, and total $\hat{\sigma}$ contributions, respectively. 
     The separate $\hat{\sigma}_{in}$ and $\hat{\sigma}_{ex}$ contributions are obtained numerically from Eqs.~\eqref{eq:sigma_Berry_3} and \eqref{sigma^I_b1}, respectively.
     Dashed straight (yellow) lines passing through the origin are plotted based on Eqs.~\eqref{h_expand6} and  ~\eqref{Extr7} valid at small $\lambda_z$.
     Dashed horizontal (green) lines validate Eqs.~\eqref{h_expand7} and \eqref{h_expand9} at large $\lambda_z$. 
     %\AO{Dashed cyan (green) straight lines denote the asymptotic behavior at small(large)} $\lambda_z$ given by Eqs.~\eqref{h_expand6},~\eqref{h_expand7},~\eqref{Extr7}, and~\eqref{h_expand9}. The conductivity is normalized by $\hat{\sigma}_{in}$ in the $\lambda_z \gg \max\{\bar{\lambda}, \Delta_\mathrm{A}\}$ limit, Eq.~\eqref{h_expand7}.
     }
     \label{fig:AHC_A}
\end{figure}

\begin{figure}   
    \includegraphics[width=0.48\textwidth]{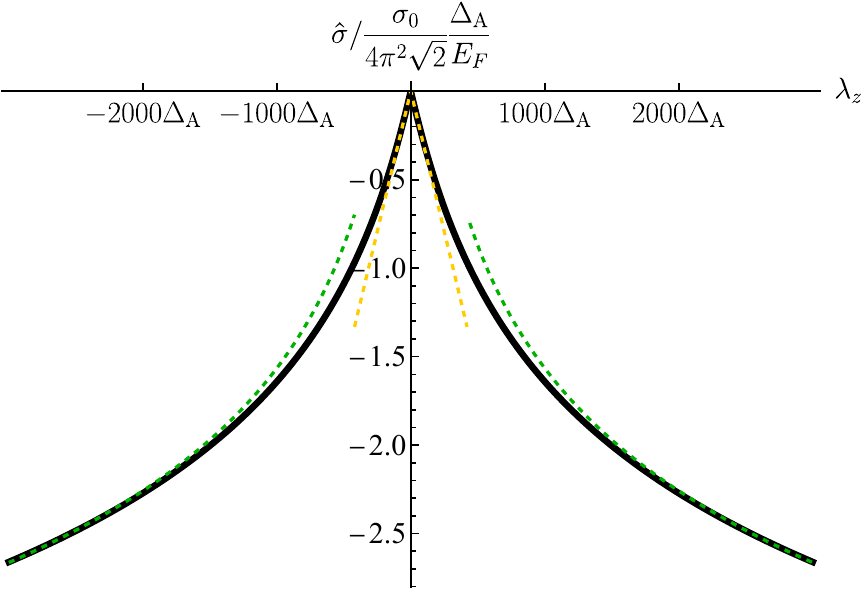}
    \caption{AHC $\hat{\sigma}$ of a class B representative as a function of the SOC, $\lambda_z$ in the two-band limit $1/|\mathbf{N}|=0$  for $\lambda = 0.1\lambda_z$, $\Delta_\mathrm{A}/E_F = 10^{-4}$.
     The magnetization is set to zero.
     The solid (black) curve is $\hat{\sigma}$ obtained numerically from Eq.~\eqref{eq:sigma_Berry_3}.
     Dashed straight (yellow) lines passing through the origin show the asymptotic scaling Eq.~\eqref{h_expand3a} as small $\lambda_z$.
     Dashed (green) line indicates the logarithmic scaling, Eq.~\eqref{h_expand8} at large $\lambda_z$ with properly adjusted additive constant. 
    %\AO{Dashed cyan (green) straight lines denote the asymptotic behavior at small(large)} $\lambda_z$ given by Eqs.~\eqref{h_expand3a} and~\eqref{h_expand8}. The unknown correction constant to the logarithmic asymptote was chosen in a wayo to smoothly connect the asymptote to the plot. The conductivity is normalized by asymptote prefactor $\hat\sigma_{in}$ of the Eq.~\eqref{h_expand8}.
}
    \label{fig:AHC_B}
\end{figure}

The result of the integration in Eq.~\eqref{eq:sigma_Berry_3} for class A and class B models is presented graphically on Figs.~\ref{fig:AHC_A} and \ref{fig:AHC_B}, respectively.  

The scaling of AHC at small and large SOC can be obtained analytically within the $\mathbf{k}\cdot\mathbf{p}$ approximation near $\Gamma$-point. 
As a result the integration in Eq.~\eqref{eq:sigma_Berry_3} is taken over the spherical Fermi sphere of a radius, $k_F$.
Here we state the main results relegating the derivation details to App.~\ref{app:Berry}.
%The results for the weak SOC assymptotics is obtained in App.~\ref{app:Berry}.
%%%%%%%%%%%
\subsubsection{Class A}
For our class A  ${}^2m{}^2m{}^1m$ representative we obtain in the limit of weak SOC $\Delta_{\mathrm{A}}\gg\lambda_z\gg\bar{\lambda}$ gives \cite{Attias2024}, 
\begin{align}\label{h_expand6}
    \hat{\sigma}_{in} =  \frac{\sigma_0}{16 \pi^2}\frac{\bar{\lambda}\mathrm{sgn}(\lambda_z\Delta_{\mathrm{A}})}{E_F} 
     \, ,
\end{align}
where $e^2 k_F = \sigma_0$.
Qualitatively, this result can be understood starting from Eq.~\eqref{eq:sigma_Berry_2}.
The region of the Brillouin zone with appreciable Berry curvature is a thin tube stretched along the $k_y=0$ meridian of a small cross-section, $\propto (\lambda_z /|\Delta_A|)k_F (\lambda_z/v_F)$ and the length, $\approx 2 \pi k_F$, see Fig.~\ref{fig:Berry}.
The numerator of Eq.~\eqref{eq:sigma_Berry_2} is estimated as $\lambda_z \bar{\lambda} \Delta_A/k_F^2$, while the denominator is typically $|\lambda_z|^3$.
Combining these estimates leads to Eq.~\eqref{h_expand6}.

In the opposite limit of the SOC dominated spin splitting, $\lambda_{z} \gg \{\bar{\lambda},\Delta_{\mathrm{A}}\}$ we obtain
\begin{align}\label{h_expand7}
    \hat{\sigma}_{in} =  \frac{\sigma_0 }{ 48\pi^2 E_F} 
    \frac{\Delta_{\mathrm{A}} \bar{\lambda} }{ \lambda_z }\, .
\end{align}

\subsubsection{Class B}
For the ${}^24/{}^1m{}^2m_y{}^1m_d$ candidate, in the weak SOC limit, 
\begin{align}\label{h_expand3a}
    \hat{\sigma}_{in}&  =-  \frac{\sigma_0}{16\pi^2} \frac{\bar{\lambda}\mathrm{sgn}(\lambda_z\Delta_{\mathrm{A}})}{E_F}
    \mathcal{F}(\lambda/\lambda_z)\, ,
\end{align}
where we have introduced the auxiliary function,
\begin{align}
    \mathcal{F}(x) =  \int_0^\pi d \theta 
      \frac{  \sin^2 \theta }
{     \sqrt{
   4  \sin^2 \theta + x^2
   \cos^2 \theta }}   \, .
\end{align}
In the opposite limit of large SOC we have
\begin{equation}\label{h_expand8}
    \hat{\sigma}_{in}\approx-\frac{\sigma_0}{4\pi^2\sqrt{2}}\frac{\Delta_\mathrm{A}\mathrm{sgn}(\lambda_z\bar{\lambda})}{E_F}
    \ln\!\left[\min\!\left(\left|\frac{\lambda_z}{\Delta_\mathrm{A}}\right|,\!\left|\frac{\lambda}{\Delta_\mathrm{A}}\right|\!\right)\!\right]
\end{equation}
valid up to an additive constant to within the logarithmic accuracy.

%%%%%%%%%

The main feature of AHC is its linear and non-analytic scaling at weak SOC, see Eqs.~\eqref{h_expand6} and \eqref{h_expand3a}.
%The intrinsic AHC is qualitatively similar for the two candidates. 
%Indeed, Fig.~\ref{fig:Berry}a and Fig.~\ref{fig:Berry}b are qualitatively similar in this limit.
This nonanalyticity results from the Berry curvature strongly peaked at the intersections of the nodal planes with the Fermi surfaces \cite{Smejkal2020}, see Fig.~\ref{fig:Berry}. 
As we see next, as the exchange energy is reduced this AHC coexists with an analytic contribution that is linear in SOC as well.
\subsection{Leading $1/|\mathbf{N}|$ corrections to  $\hat{\sigma}_{in}$}
\label{sec:sigma_in_J}
Here we investigate corrections to the intrinsic AHC up to the order $1/|\mathbf{N}|^2$.
We do it for two reasons. 
First, a finite magnetization, $\mathbf{M}$ is present  whenever AHC is allowed by symmetry and it is crucial to estimate it's contribution $\hat{\sigma}_{in}^M$ to AHC.

Second, we have to compare the magnetization contribution to the extrinsic AHC studied in this work.
In class B as long as the magnetization canting angle is small, the extrinsic contribution is by far more significant.
In class A both contributions are small in comparison with the intrinsic AHC.

The method we employ parallels the canonical Foldy–Wouthuysen transformation of the Dirac equation \cite{Bjorken1964}.
Within this analogy the $\check{H}_{p(n)}$ blocks of Eq.~\eqref{eq:structure} describe the particles and anti-particles, respectively.
According to Eq.~\eqref{Hud} these two sectors are separated by the large exchange splitting $2 |\mathbf{N}|$ playing the role of the Dirac gap.
Our expansion is, therefore very similar to finding the relativistic corrections to the Schr\"{o}dinger equation.
The effective Hamiltonian up to the order $1/|\mathbf{N}|^2$,
\begin{equation}\label{eq:H_effective_13}
 \check{H}^{\mathrm{eff}}_p\! =\! \check{H}_p\! +\! \frac{\check{V} \check{V}^\dagger}{2 |\mathbf{N}|} \!-\! \frac{1}{8|\mathbf{N}|^2} ( \check{V} \check{V}^\dagger \check{h}_p  \!-\! 2 \check{V} \check{h}_n \check{V}^\dagger  \!+ \! \check{h}_p \check{V} \check{V}^\dagger)
\end{equation}
is derived in App.~\ref{app:effective_Hamiltonian}.
The the first order $1/|\mathbf{N}|$ corrections to both the band dispersion and the Berry curvature are fully determined by the second term of Eq.~\eqref{eq:H_effective_13}.

To the next, second order in $1/|\mathbf{N}|$ corrections to the spectrum is captured by the last term of Eq.~\eqref{eq:H_effective_13}.
However, the Berry curvature is determined by the Bloch wave-functions.
To the order $(1/|\mathbf{N}|)^2$ apart from the corrections coming from Eq.~\eqref{eq:H_effective_13} a weak admixture of the negative energy wave-functions to the dominant positive energy bands contributes. 
These contributions read
\begin{align}\label{Omega_second_main}
    \tilde{\Delta} & \Omega^{(n)}_{\alpha\beta}  =  - 
    \frac{1}{2 |\mathbf{N}|^2}\Im \langle n \mathbf{k}| (\partial_\alpha \check{V}) (\partial_\beta \check{V}^\dagger) |n\mathbf{k}\rangle 
    \notag \\
    - & \frac{1}{4 |\mathbf{N}|^2}\Im \langle n\mathbf{k}|\left[ (\partial_\alpha \check{V})  \check{V}^\dagger -  \check{V}  (\partial_\alpha \check{V}^\dagger) \right] |\partial_\beta n\mathbf{k} \rangle 
    \notag \\
    - & \frac{1}{4 |\mathbf{N}|^2}\Im \langle \partial_\alpha  n \mathbf{k}| \left[ \check{V} (\partial_\beta \check{V}^\dagger)  - (\partial_\beta  \check{V} )V^\dagger)\right] |n \mathbf{k} \rangle ,
\end{align}
where $|n \mathbf{k}\rangle$ are the bands of positive energy Hamiltonian $\check{H}_p$ in the $1/|\mathbf{N}|=0$ limit.

We now specify the general results, Eqs.~\eqref{eq:H_effective_13} and \eqref{Omega_second_main} to our model of the off-diagonal block \eqref{check_V1} that couples the positive and negative energy solutions. 
For this model the correction up to the order $(1/|\mathbf{N}|)^2$ correction to the effective Hamiltonian reads,
\begin{align}\label{del_Hp}
 \check{H}^{\mathrm{eff}}_p\! -\! \check{H}_p\!  =
    \frac{t^2_{x,\mathbf{k}} +\lambda^2_{y,\mathbf{k}}}{2 |\mathbf{N}|}\left( 1 - \frac{\check{h}_p}{|\mathbf{N}|}\right)  + M \frac{t_{x,\mathbf{k}}}{|\mathbf{N}|}\rho_y \, .
\end{align}
At $M=0$, \eqref{del_Hp} a $\mathbf{k}$-dependent constant, an $\propto (t^2_{x,\mathbf{k}} +\lambda^2_{y,\mathbf{k}})$ and the term $\propto (t^2_{x,\mathbf{k}} +\lambda^2_{y,\mathbf{k}}) \check{h}_p$.
Neither one of these terms modifies the Berry curvature. 

The explicit Berry curvature correction, \eqref{Omega_second_main} reduces to 
\begin{align}\label{eq:Berry_correction_J}
\tilde{\Delta} \Omega^{(n)}_{\alpha\beta} & = \frac{1}{2 |\mathbf{N}|^2} \left(\partial_\alpha \lambda_{y,\mathbf{k}} \partial_\beta t_{x,\mathbf{k}} -
\partial_\beta \lambda_{y,\mathbf{k}} \partial_\alpha t_{x,\mathbf{k}} \right)\, 
\end{align}
in agreement with Ref.~\cite{Roig2025}.

\subsubsection{Corrections due to a finite $\mathbf{M}$}
\label{sec:sigma_in_J_B}
The magnetization enters the effective Hamiltonian \eqref{del_Hp} already at the order $1/N$.
Therefore, the effect of a weak magnetization follows from the last term of Eq.~\eqref{del_Hp}.
It implies that the effective Hamiltonian up to a trivial constant differs from the original Hamiltonian by the replacement of $\mathbf{h}^{p}_{\mathbf{k}}$ as given by Eq.~\eqref{hV_ud_A} by $\mathbf{h}'^{p}_{\mathbf{k}}$, such that 
$[\mathbf{h}^{p}_{\mathbf{k}}]_{x,z} = [\mathbf{h}'^{p}_{\mathbf{k}}]_{x,z}$, and
\begin{align}\label{replace2}
    [\mathbf{h}'^{p}_{\mathbf{k}}]_{y}= [\mathbf{h}^{p}_{\mathbf{k}}]_{y} -  \frac{M t_x}{N} \, .
\end{align}
Based on Eq.~\eqref{hV_ud_A}, Eq.~\eqref{replace2} implies that at finite $\mathbf{M}$ we can use the same model as for $\mathbf{M}=0$ with $\lambda_z$ replaced by $\lambda_z  + M t_x /N$.

Let us focus on the limit of large altermagnetic splitting, $\Delta_{\mathrm{A}} \gg \{ \lambda, \lambda_z \}$.
The assymptotic behaviour of AHC on SOC is controlled by the nodal lines where the degeneracy is lifted by SOC. 
Along these lines the altermagnetic splitting vanishes, and the method of effective Hamiltonian we employ here can only be valid if $M t_x /N \ll \max\{\lambda,\lambda_z\}$.

Let us first address class A. 
In the limit of weak SOC $M = c_M \lambda_z$ where $c_M$ some dimensionless constant that depends on the microscopic details.
Furthermore, we have $\lambda_z \gtrsim \lambda$ and the effective Hamiltonian approach holds in the regime $  c_M t_x /N \ll 1$ which is very reasonable. 
In this case a slight modification of the calculation for the $M=0$ limit yields,
\begin{align}\label{sigma_in_M_A}
    \hat{\sigma}_{in}^M \approx \frac{\sigma_0}{15 \pi } \sgn(t_A \lambda_z)\frac{M}{N} \frac{t_x}{E_F}\frac{\bar{\lambda}^3}{\lambda_z^3}\, .
\end{align}

For the class B similar procedure as applied to class A results in the correction,
\begin{align}\label{sigma_in_M_B}
\hat{\sigma}_{in}^M \approx - \frac{\sigma_0}{(4 \pi)^2 }\sgn(t_A \lambda_z)\frac{M}{N} \frac{t_x}{E_F} 
 \frac{\bar{\lambda}^3}{\lambda^3_z} \tilde{\mathcal{F}}(\bar{\lambda}/\lambda_z),
\end{align}
where $\tilde{\mathcal{F}}(x) = - \mathcal{F}'(x)/x$.

Remarkably, the correction due to magnetization in both classes, Eqs.~\eqref{sigma_in_M_A} and \eqref{sigma_in_M_B} depends on the atomic SOC only implicitly through the magnetization.
In classes A and B the magnetization scales with SOC linearly and quadratically, respectively \cite{Roig2025}.
We see that the same holds for the magnetization correction to AHC.
In other words, the magnetization plays a significantly more prominent role in AHC of materials belonging to the class A than in the materials of class B.

\subsubsection{Corrections to AHC the order $1/|\mathbf{N}|^2$ at $\mathbf{M}=0$}
\label{sec:sigma_in_J_A}
At $M=0$, \eqref{del_Hp} a $\mathbf{k}$-dependent constant, an $\propto (t^2_{x,\mathbf{k}} +\lambda^2_{y,\mathbf{k}})$ and the term $\propto (t^2_{x,\mathbf{k}} +\lambda^2_{y,\mathbf{k}}) \check{h}_p$.
Neither one of these terms modifies the Berry curvature. 
The second term decreases the spin splitting in the positive energy band, and in so doing modifies the AHC by a factor, $(1 - t_x^2/2N^2)$ where, agian in the $\mathbf{k} \cdot \mathbf{p}$ approximation $t^2_{x,\mathbf{k}} +\lambda^2_{y,\mathbf{k}} \approx t_x^2$.

The explicit correction to the Berry curvature, 
Eq.~\eqref{eq:Berry_correction_J} produces the linear in SOC contribution to AHC that is analytic.
It should be contrasted with Eqs.~\eqref{h_expand6} and \eqref{h_expand3a} that have a discontinuous derivative as a function of the SOC.

The Berry curvature correction $\tilde{\Delta} \Omega^{(n)}_{\alpha\beta}$ is the same for the two spin split subbands of the positive energy effective Hamiltonian. 
We therefore may estimate its contribution to the AHC as $\Delta \sigma_{in;yz} \propto \sigma_0 \bar{\lambda} t_x/ N^2$.
The ratio of the $1/N^2$ correction to the result in the limit of $1/N =0$ is $\Delta \sigma_{in;yz} /  \sigma_{in;yz} \propto  E_F t_x/ N^2$.
It is clear that even though $\Delta \sigma_{in;yz}$ sales with $1/N^2$ it grows with the Fermi energy. 
And despite being small in $1/N$ it can be much larger than $\sigma_{in}$ if $E_F \gg |\mathbf{N}|$. 
In this limit the analytic contribution to the AHC originating from the Berry curvature, \eqref{eq:Berry_correction_J} dominates. 
We believe this regime is realized for the set of parameters studied in Refs.~\cite{Roig2025v1}.

\section{Extrinsic AHC}
\label{sec:Extrinsic}
Here we compute Eq.~\eqref{eq:sigma^I} at finite disorder. 
%As we will see the extrinsic AHC originates from the intraband particle-hole excitations.
%It is therefore, justifiable to apply the two-band approximation.
The Kubo-St\v{r}eda formulation relies on knowledge of the Green functions.
The disorder free Green function takes the form,
\begin{align}\label{G8}
    G_{0\mathbf{k}}^{R(A)} = (\epsilon - E_{0,\mathbf{k}}- \mathbf{h}_{\mathbf{k}}\cdot \boldsymbol{\rho} +E_F \pm i 0)^{-1}\, ,
\end{align}
where the off shell energy $\epsilon$ is set to zero in the final expressions.
It is useful to rewrite Eq.~\eqref{G8} in the form, 
\begin{align}\label{G13}
     G_{0\mathbf{k}}^{R(A)} =  
    P_{\mathbf{k}+}G^{R(A)}_{\mathbf{k}+;0} + P_{\mathbf{k}-}^{R(A)}G^{R(A)}_{\mathbf{k}-;0}\, ,
\end{align}
where the band Green functions read
\begin{align}\label{G13a}
     G^{R(A)}_{\mathbf{k}\pm;0} =  
   \frac{1}{E_F - E^\pm_{\mathbf{k}} \pm i 0}\, .
\end{align}
The projection operators are expressed in terms of the eigenstates $|\pm \mathbf{k}\rangle$ of the Hamiltonian $H_u$,
\begin{align}\label{P+-}
    P_{\pm \mathbf{k}} = |\pm \mathbf{k}\rangle \langle \pm \mathbf{k} | = \frac{1}{2}\left( \rho_0 \pm \mathbf{n}_{\mathbf{k}}\cdot \boldsymbol{\rho} \right)\, ,
\end{align}
where $\mathbf{n}_{\mathbf{k}} = \mathbf{h}_{\mathbf{k}}/h_{\mathbf{k}}$.

We start with a discussion of the effect of the disorder on the Green function.
This will naturally leads one to the classification of Tab.~\ref{tab:AB}. 
%division into the two classes of altermagnets.
\subsubsection{Disorder averaged Green function}
Here we consider the disorder averaged Green function $G^{R(A)}_{\mathbf{k}}$.
It satisfies the Dyson equation, $[G^{R(A)}_{\mathbf{k}}]^{-1} = G^{-1}_{0\mathbf{k}} - \Sigma^{R(A)}_{\mathbf{k} }$.
In the limit of infinitesimally weak disorder the self energy, $\Sigma^{R(A)}_{\mathbf{k} }$ can be computed in the Born approximation. 
We parametrize it in the form,
\begin{align}\label{Sigma}
    \Sigma^{R(A)} & = - i \Gamma_0^{R(A)} \rho_0 - i \boldsymbol{\Gamma}^{R(A)}\boldsymbol{\rho} \, .
\end{align}

Within the Born approximation the scalar part of the self energy reads
\begin{align}\label{Sigma_scalar1}
    \Gamma_0^R =  \frac{n_{\mathrm{imp}} U^2}{(4 \pi)^2} \left( \oint_{\mathrm{FS}^+} \frac{d S }{v_+} +  \oint_{\mathrm{FS}^-} \frac{d S }{v_-} \right) \, ,
\end{align}
where the two Fermi surfaces, FS$^{\pm}$ are defined by $E_F = E^{\pm}_{\mathbf{k}}$, the velocities $\mathbf{v}_{\pm} = \nabla E^{\pm}_{\mathbf{k}}$ in the integration are evaluated at the corresponding Fermi surfaces.
It can also be written in the form, $ \Gamma_0 =   \pi n_{\mathrm{imp}} U^2 (\nu_+ + \nu_-)/2$, where $\nu_{\pm}$ are the densities of states at the two Fermi surfaces, FS$^{\pm}$ which turns into the standard Fermi Golden rule in the limit of zero spin splitting, $h = 0$.
In this case approximating the dispersion relation as $E_{0}(\mathbf{k}) = E_0 k^2$ we have 
\begin{align}\label{Gamma_0R}
    \Gamma_0^R \approx \frac{n_{\mathrm{imp}} U^2}{4 \pi} \frac{ k_F}{ E_0 } \, .
\end{align}

More crucial is the vector part of the self-energy, 
\begin{align}\label{Sigma_vector1}
    \boldsymbol{\Gamma}^R=  \frac{n_{\mathrm{imp}} U^2}{(4 \pi)^2}   
    \left( \oint_{\mathrm{FS}^+} \frac{d S }{v_+} \mathbf{n}_{\mathbf{k}} -  \oint_{\mathrm{FS}^-} \frac{d S }{v_-} \mathbf{n}_{\mathbf{k}}\right)\, .
\end{align}
The altermagnets of a class A and class B the vector component of the self energy, Eq.~\eqref{Sigma_vector1} is large (negligible), respectively.
This crucial distinction is simple to see within the $\mathbf{k}\cdot \mathbf{p}$ approximation.
In this limit to the leading order in the spin splitting $h_{\mathbf{k}}$, we can approximate  $\mathbf{n}_{\mathbf{k}} \approx \mathbf{n}_{\mathbf{k}=0} = \boldsymbol{\lambda}_0 / h_{\mathbf{k}}$ tabulated in Tab.~\ref{tab:AB}, such that Eq.~\eqref{Sigma_vector1} turns into
\begin{align}\label{Gamma_vec}
    \boldsymbol{\Gamma}^{R,A} = -\hat{y} \frac{\lambda_0}{2 E_F} \Gamma^{R,A}_0 
\end{align}
to the leading order in the spin splitting $h_{\mathbf{k}}$ (see App.~\ref{app:self} for details).

Having found the disorder averaged self energy, \eqref{Gamma_vec} we are in a position to present the Green function, $G^{R(A)}_{\mathbf{k}}(E)$.
It is similar in form to Eq.~\eqref{G13}, 
\begin{align}\label{G13_dis}
     G_{\mathbf{k}}^{R} =  
    \tilde{P}^{R}_{\mathbf{k}+}G^{R}_{\mathbf{k}+} + \tilde{P}_{\mathbf{k}-}^{R}G^{R}_{\mathbf{k}-}\, .
\end{align}
The band resolved Green functions acquire a finite width,
\begin{align}
    \Gamma^{\pm}_{\mathbf{k}} = \Gamma_0 \pm \mathbf{n}_{\mathbf{k}}\cdot \boldsymbol{\Gamma}
\end{align}
such that Eq.~\eqref{G13a} is replaced by
\begin{align}\label{G13a_dis}
     G^{R(A)}_{\mathbf{k} f} =  
   \frac{1}{E_F - E^f_{\mathbf{k}} \pm i \Gamma^f_{\mathbf{k}}}\, .
\end{align}
More importantly, the projection operators \eqref{P+-} acquire a finite imaginary component,
\begin{align}\label{G14_dis}
    \tilde{P}^{R,A}_{\mathbf{k} \pm } & = \frac{1}{2}\left( \rho_0 \pm \mathbf{n}^{R,A}_{\mathbf{k}}\cdot \boldsymbol{\rho} \right)\, ,
    \notag \\
    \mathbf{n}^{R,A}_{\mathbf{k}} & = \mathbf{n}_{\mathbf{k}} \pm i \delta \mathbf{n}_{\mathbf{k}}\, , \quad
    \delta \mathbf{n}_{\mathbf{k}} = 
    \frac{1}{h_\mathbf{k}}\left[ \mathbf{n}_{\mathbf{k}}(\boldsymbol{\Gamma} \cdot \mathbf{n}_{\mathbf{k}})-  \boldsymbol{\Gamma}\right]\, .
\end{align}
Equations \eqref{Gamma_vec} through \eqref{G14_dis} specify the Green function in the presence of the weak random short range disorder.  
The missing details of the derivation of this result are summarized in App.~\ref{app:Green}.
It will be convenient to present the Green function in the following form, 
\begin{subequations}\label{Green_dis_1}
\begin{align}\label{Green_dis_1R}
         G^{R}_{\mathbf{k}} 
    = \sum_{f= \pm} P_{f\mathbf{k}} G_{f\mathbf{k}}^{R}  + \frac{1}{2} i \delta \mathbf{n}_{\mathbf{k}}\boldsymbol{\rho}(G_{\mathbf{k}+}^{R} -G_{\mathbf{k}-}^{R})\, ,
\end{align}
\begin{align}\label{Green_dis_1A}
       G^{A}_{\mathbf{k}}
    = \sum_{f= \pm} P_{f\mathbf{k}} G_{f\mathbf{k}}^{A}   - \frac{1}{2} i \delta \mathbf{n}_{\mathbf{k}}\boldsymbol{\rho}(G_{\mathbf{k}+}^{A} -G_{\mathbf{k}-}^{A})\, .
\end{align}
\end{subequations}

Here we focus on the significance of the vector part of the self energy, $\boldsymbol{\Gamma}$ in Eq.~\eqref{Sigma}.
As is clear from Eq.~\eqref{Sigma_vector1}, it originates from the $\mathbf{k}$-independent SOC, $\boldsymbol{\lambda}_0$ at least close to $\mathbf{k}=0$.
Such SOC acts as the constant magnetic field that causes spin flip and spin conserving transitions into the continuum of states with $\mathbf{k}' \neq \mathbf{k}$. 

Let us fix $\mathbf{k}$ and make a decomposition, $\boldsymbol{\Gamma} = \boldsymbol{\Gamma}_{\parallel}+\boldsymbol{\Gamma}_{\perp}$ such that $\boldsymbol{\Gamma}_{\parallel} \parallel \mathbf{h}_{\mathbf{k}} $ and $\boldsymbol{\Gamma}_{\perp} \perp \mathbf{h}_{\mathbf{k}} $ describes the spin conserving and spin flip transitions, respectively. 
The latter processes are similar to the spin flip transitions of the two-level atom caused by the dipole coupling to the electro-magnetic radiation \cite{Carmichael1993}.
In our problem similar processes are elastic and have rate $\boldsymbol{\Gamma}_{\perp}$. 
Precisely these processes make $\delta \mathbf{n}_{\mathbf{k}}$ in Eq.~\eqref{G14_dis}, and consequently extrinsic AHC finite.

In the case of two-level atom, the real part of the self energy describes the Lamb shift of spectral lines.
In our discussion we omit the real parts of self energy even though it causes the band renormalization.
The reason is that the effect of the real part of the self energy is analytic in the impurity density.
In result, it does not contribute to the extrinsic AHC per our definition \eqref{sigma_ex}.

%%%%%%%%%
\begin{figure}[ht]
    \includegraphics[width=0.48\textwidth]{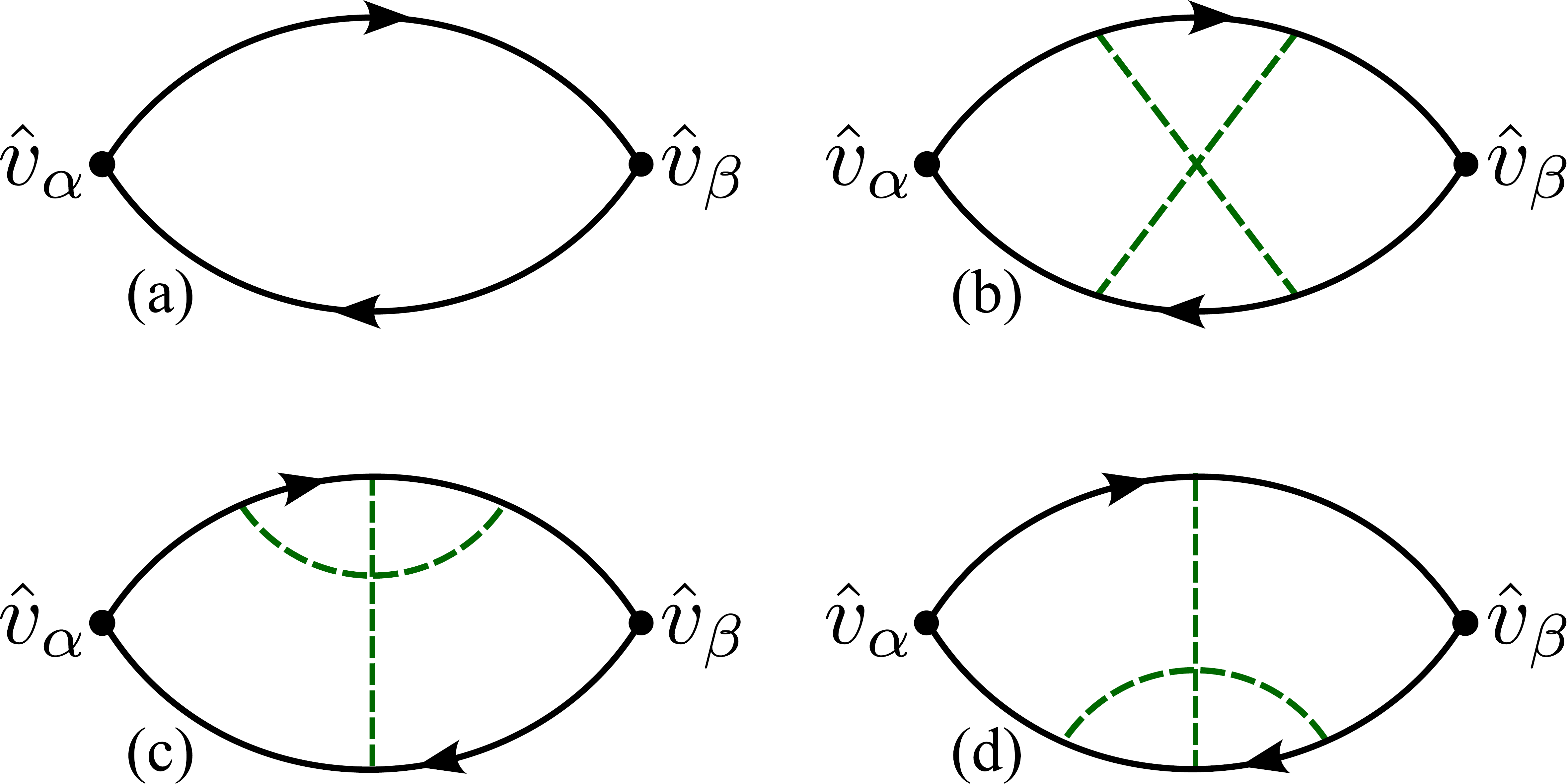}
    \caption{The diagrammatic contributions to Eq.~\eqref{eq:sigma^I} that gives rise to the extrinsic Hall conductivity via Eq.~\eqref{sigma_ex}. Panel (a) represents the classical contribution. 
    Panel (b) is the $\sigma^X_{ex}$ contribution from the $X$-diagram due to diffraction scattering off the two impurities close by. Panels (c) and (d) show the two $\Psi$-diagrams which produce the contribution that is similar to the $X$-diagram.}
    \label{fig:diagrams}
\end{figure}

%%%%%%%%%
%\begin{figure}[ht]
%    \includegraphics[width=0.48\textwidth]{Crystall2.pdf}
%    \caption{The schematic picture of crystalline structures of type A (a) and B (b). The direction of the Neel vector $\mathbf{N}$ is along the $y$-axis and the magnetization vector $\mathbf{M}$ is along the $x$-axis.\AO{Maybe the $\mathbf{N}$ and $\mathbf{M}$ should be a little bit repositioned. If the transparency is too low -- tell me}}
%    \label{fig:2grid}
%\end{figure}
\subsection{The extrinsic Hall conductivity}
Here we compute the extrinsic AHC, \eqref{eq:sigma^I} presented diagrammatically in Fig.~\ref{fig:diagrams}.
It contains the standard non-crossing (nc) diagram Fig.~\ref{fig:diagrams}a as well as the diffractive contributions Fig.~\ref{fig:diagrams}b-d describing the effect of the diffraction scattering off the two impurities separated by a distance of the order of the de Broglie wavelength.
The importance of the diffractive contributions on par with the more standard non-crossing ones is a peculiar feature of AHC \cite{Ado2016,Konig2016,Konig2021}.

Keeping this in mind, we start with the non-crossing contributions, which graphically is represented by the bubble diagram with the impurity lines that do not cross each other (see Fig.~\ref{fig:diagrams}a).  
The dispersion relation is even in $\mathbf{k}$.
Correspondingly, the current operator is odd.
Therefore, the short-range disorder produces no current vertex corrections. 
In addition, the skew scattering is excluded for the Gaussian disorder considered here. 

%Summarizing, we are left with only one diagrammatic contribution shown in Fig.~\ref{fig:diagrams}.
In result, $\sigma_{ex}$  contains four parts, $\sigma_{ex} = \sigma^{nc}_{ex} + \sigma^{X}_{ex} + \sigma^{\Psi_1}_{ex} + \sigma^{\Psi_2}_{ex}$ corresponding to panels (a), (b), (c) and (d) in Fig.~\ref{fig:diagrams}, respectively.
The non-crossing contribution splits in two parts, $\sigma^{nc}_{ex}= \sigma^{nc_1}_{ex} + \sigma^{nc_2}_{ex}$, 
\begin{subequations}\label{sigma^I}
    \begin{align}\label{sigma^I_a}
    \sigma^{nc_1}_{ex;\alpha \beta} 
& =   \frac{ e^2}{2 (2\pi)^3}\left[  \oint_{\mathrm{FS}^+} \frac{d S}{v_+ (1 + \mathbf{n}_{\mathbf{k}}\boldsymbol{\gamma})} - \oint_{\mathrm{FS}^-} \frac{d S}{v_- (1 - \mathbf{n}_{\mathbf{k}}\boldsymbol{\gamma})} \right]
\notag \\
\times & \left( \frac{\partial \mathbf{h}_{\mathbf{k}}}{\partial k_{\alpha}} \times \frac{\partial \mathbf{h}_{\mathbf{k}}}{\partial k_{\beta}}\right) \cdot 
\frac{1}{h_\mathbf{k}}
\left[ \mathbf{n}_{\mathbf{k}}(\boldsymbol{\gamma} \mathbf{n}_{\mathbf{k}})  -\boldsymbol{\gamma}  \right]
\, ,
\end{align}
\begin{align}\label{sigma^I_b}
   & \sigma^{nc_2}_{ex;\alpha \beta} 
 =   \frac{ e^2}{2 (2\pi)^3}\left[  \oint_{\mathrm{FS}^+} \frac{d S}{v_+ (1 + \mathbf{n}_{\mathbf{k}}\boldsymbol{\gamma})} +\oint_{\mathrm{FS}^-} \frac{d S}{v_- (1 - \mathbf{n}_{\mathbf{k}}\boldsymbol{\gamma})} \right]
\notag \\
& \times  \left[
    \left(
     \frac{\partial E_0(\mathbf{k})}{\partial k_\alpha} \frac{\partial \mathbf{h}_{\mathbf{k}}}{\partial k_\beta }-
    \frac{\partial E_0(\mathbf{k})}{\partial k_\beta} \frac{\partial \mathbf{h}_{\mathbf{k}}}{\partial k_\alpha } 
     \right) \times \mathbf{n}_{\mathbf{k}} \right] \cdot   \frac{\boldsymbol{\gamma}}{h_\mathbf{k}} \, ,
\end{align}
\end{subequations}
where $\boldsymbol{\gamma} = \boldsymbol{\Gamma}/\Gamma_0$ is independent of disorder concentration.
The quantum contribution, $\sigma^{X}_{ex} $ is calculated in the App.~\ref{app:Cooperon}.

To estimate the relative importance of the extrinsic contribution let us assume for clarity that all the spin splitting energy scales $\lambda_z,\bar{\lambda},\Delta_{\mathrm{A}}$ are of the same order of magnetude and estimated as a typical spin splitting energy scale, $h$.
In terms of this energy scale, 
$\sigma^{nc_1}_{ex} \propto \sigma_0 (h/E_F)^3$. %and $\sigma^{nc_2}_{ex} \sim \sigma_0 (h/E_F)$.
The quantum diffraction contributions shown in Fig.~\ref{fig:diagrams} b,c,d $\sigma^{X,\Psi_1,\Psi_2}_{ex} \propto \sigma_0 (\lambda_z\bar{\lambda}\Delta_{\mathrm{A}})/E_F^3$, see App.~\ref{app:Cooperon} for a proof.
Based on the results of Sec.~\ref{sec:Intrinsic}, the intrinsic AHC , $\sigma_{in} \propto \sigma_0 (h/E_F)$ is comparable to the extrinsic contribution, $\sigma^{nc_2}_{ex}$ given by Eq.~\eqref{sigma^I_b}. 
%As is demonstrated in the App.~\ref{app:Cooperon}, $ \sigma^{X}_{ex} \sim \sigma_0 (\lambda_z\bar{\lambda}\Delta_{\mathrm{A}})/E_F^3$.
In summary, in the limit, $h \ll E_F$,  $\sigma_{ex} \approx \sigma^{nc}_{ex} \approx \sigma^{nc_2}_{ex}$.
Therefore, to the leading order in $h/E_F$,
\begin{align}\label{sigma^I_b1}
   & \sigma_{ex;\alpha \beta} 
 =   \frac{ e^2}{ (2\pi)^3}\oint_{\mathrm{FS}} \frac{d S}{v}  
\notag \\
& \times  \left[
    \left(
     \frac{\partial E_0(\mathbf{k})}{\partial k_\alpha} \frac{\partial \mathbf{h}_{\mathbf{k}}}{\partial k_\beta }-
    \frac{\partial E_0(\mathbf{k})}{\partial k_\beta} \frac{\partial \mathbf{h}_{\mathbf{k}}}{\partial k_\alpha } 
     \right) \times \mathbf{n}_{\mathbf{k}} \right] \cdot   \frac{\boldsymbol{\gamma}}{h_\mathbf{k}} \, .
\end{align}

For the ${}^2m{}^2m{}^1m$ class A representative we obtain in the limit of weak SOC $\Delta_{\mathrm{A}}\gg\lambda_z\gg\bar{\lambda}$
\begin{align}\label{Extr7}
\hat{\sigma}_{ex}
%=-\frac{\sigma_{0}}{48\pi^{2}}\frac{\bar{\lambda}\mathrm{sgn}(\lambda_{z}\Delta_{\mathrm{A}})}{E_{F}}
=-\frac{1}{3}\hat{\sigma}_{in}\, .
\end{align}
In the complimentary limit of the SOC dominated spin splitting, $\lambda_{z} \gg \max\{\bar{\lambda},\Delta_{\mathrm{A}}\}$,
\begin{align}\label{h_expand9}
    \hat{\sigma}_{ex} 
    %= -\frac{\sigma_{0}}{120\pi^{2}}\frac{\bar{\lambda}\Delta_{\mathrm{A}}}{E_{F}\lambda_{z}}
    = -\frac{2}{5}\hat{\sigma}_{in}.
\end{align}

The ratio of the extrinsic and intrinsic contributions is model dependent and can vary for more realistic band structures and/or when the spin splitting is comparable to the Fermi energy.
Our main conclusion is that generally the two contributions are of the same order of magnitude even at infinitesimally weak disorder. 
This result applies universally to class A altermangnets listed in the Tab.~\ref{tab:AB}. 
In class B altermagnets extrinsic AHC is generally negligible.  
%So the relatively large induced magnetization is concomitant with the 
%Even infinitesimally weak disorder has stronger effects on AHC than a finite magnetization.

\section{Discussion}
\label{sec:Conclusions}
Our main conclusion is that the extrinsic AHC is as important as the intrinsic one in roughly the half of the altermagnetic candidates, see Tab.~\ref{tab:AB} in the limit of large exchange splitting.
In materials with the non-zero Dzyaloshinskii–Moriya type interaction linear in SOC, the extrinsic contribution is essential in the clean limit, and is largely irrelevant otherwise.
The extrinsic AHC arises from the intra-band disorder scattering and is prominent in the two-band limit resulting from the large exchange splitting. 

The strong short range disorder results into the in-gap states localized at the impurity sites \cite{Gondolf2025}.
The long range strain disorder leads to the magnetic field induced reentrance transition into the altemagnetic state \cite{Chakraborty2025}.
Here we argue that the extrinsic Hall conductivity is finite even for the infinitesimal disorder concentration as long the thermodynamic limit applies. 
This non-analiticity and the non-analytic dependence of the intrinsic AHC on SOC are interrelated.
The origin of the two non-analiticities is in lifting of the spin degeneracy along the nodal lines that is specific to altermagnets by a weak SOC. 
We find that the same reason underlines the linear scaling of the intrinsic AHC with magnetization. 

It is important to point out that when the exchange splitting is comparable to the band width the AHC mostly originates from the inter-band processes.
In this limit the intrinsic AHC is linear and analytic in SOC.
Moreover, it only depends on a part of the SOC that polarizes spins along N\'{e}el vector.
This has been shown in Ref.~\cite{Roig2025} based on the expression for the Berry curvature for the full four band model.
In this four-band regime the extrinsic contribution is expected to be analytic as well, and can be ignored if the system is not too dirty. 

From this perspective, the non-analytic AHC arises from the terms of the Berry curvature that are nominally quadratic in SOC. 
These terms are finite in the limit of large exchange splitting and have the denominators vanishing along the nodal lines hosting the spin degeneracy protected by the spin group symmetry. 
For the same reason, the extrinsic AHC is substantial in this two-band limit.

The question might arise how our results compare to other well known systems exhibiting the AHC.
In the case of Weyl semimetals the considerations similar to those presented here lead to very different conclusions \cite{Burkov2014}. 
In Weyl semimetals unless they are strongly doped to make the Fermi surfaces comparable to the separation between the Weyl nodes the extrinsic contribution is negligible.
This is a consequence of the two facts. 
First, the AHC in the limit where the FS contains just Weyl nodes, is purely intrinsic $\hat{\sigma}=\hat{\sigma}^{II}$, topological and insensitive to the disorder. 
Second, the dispersion relation close to the nodes is linear, and this suppresses the Fermi surface contribution because of the effective time reversal symmetry at the nodes. 
%The extrinsic contribution has been obtained as the difference of the ac diffusive and ballistic responses which is the same as our definition, Eq.~\eqref{sigma_ex}.
The extrinsic contribution is finite, yet small unless the Fermi surfaces become relatively large.
In contrast, in our system $\sigma^{II}$ is small in the limit of small spin splitting, and as a result extrinsic contribution turned out to be comparable to the intrinsic contribution as both are basically Fermi surface effects.

It is often emphasized that the finite magnetization makes a contribution to the AHC on par with the intrinsic AHC.
We point out that in cases where the magnetization is relatively large the extrinsic contribution is comparable to the intrinsic one, and has to be included if the quantitative comparison to the experimental data is done.
The other potentially important sources of the AHC such as phonons and chiral magnons \cite{Steward2023,Smejkal2023} are beyond the scope of this work.
We relegate these effects to separate studies.

\section*{Acknowledgements}

A. O. and M. K. acknowledge the financial support from the Israel Science Foundation, Grant No.~2665/20. A. L. acknowledges the financial support by the National Science Foundation Grant No. DMR-2452658 and H. I. Romnes Faculty Fellowship provided by the University of Wisconsin-Madison Office of the Vice Chancellor for Research and Graduate Education with funding from the Wisconsin Alumni Research Foundation.

\begin{appendix}
\section{Intrinsic AHC for two models representing classes A and B}
\label{app:Berry}
%In this appendix we present the details of the calculation of the intrinsic Hall conductivity in the simplified model \eqref{eq:2by2}.
%The goal is to find small $t$ and small $A$ assymptotics of the intrinsic Hall conductivity Eq.~\eqref{eq:intr3}.
%We do it for the two systems of interest.
Here we compute the intrinsic Hall conductivity in the limit of small spin splitting and large exchange interaction, $1/|\mathbf{N}|=0$.
%We do it for the two representative models of the systems in class A and class B.
In this case, the intrinsic Hall conductivity is given by Eq.~\eqref{eq:sigma_Berry_3}.
For completeness, we estimate AHC for all possible relations between the three energy scales, $\Delta_{\mathrm{A}}$, $\lambda_z$, and $\bar{\lambda}$.

%the details of this calculation performed in the $\mathbf{k} \cdot \mathbf{p}$ approximation assuming the N\'eel vector is along $y$-axis.
%%%%%%%%%%%%%%%%%%%%%%%%%%%%%%%%
\subsection{Intrinsic Hall conductivity for ${}^2m_x{}^2m_y{}^1m_z $}\label{IntrinsicFeSb2}
We start by expansion of Eq.~\eqref{hV_ud_A} around $\mathbf{k}=0$ based on Eq.~\eqref{eq:SOC_A},
\begin{align}\label{h_expand1}
    \mathbf{h}_{\mathbf{k}} \approx (\frac{\lambda}{4}k_x k_z, - \lambda_z, t_A k_x k_y)\, .
\end{align}
With this expansion  Eq.~\eqref{eq:sigma_Berry_3}  yields
\begin{align}\label{h_expand2}
    \hat{\sigma}_{in;y z} =  \frac{e^2}{(2 \pi)^3} \oint \frac{d S}{v_F}
    \frac{\lambda_z \lambda t_A k_x^2/4}{\lambda_z^2 + t_A^2 k_x^2 k_y^2+(\lambda k_x k_z/4)^2 }\, .
\end{align}
%To the leading order in the spin splitting the integration is over the spherical Fermi surface. 
In spherical coordinates,  Eq.~\eqref{h_expand2}  takes the form 
\begin{align}\label{h_expand3}
    & \hat{\sigma}_{in;y z} =  \frac{e^2}{2 \pi^2} \frac{k_F^2}{v_F} \oint \frac{d \Omega}{4 \pi}
    \\
    \times & \frac{\lambda_z \lambda t_A k_F^2 \sin^2 \theta \cos^2\phi/4}{\lambda_z^2 + (t_A k_F^2 \sin^2 \theta \cos \phi \sin \phi)^2 +(\lambda k_F^2 \cos\theta \sin \theta \cos \phi /4)^2}\, .
    \notag 
\end{align}
\subsubsection{The limit \( \Delta_{\mathrm{A}} \gg \max\{\lambda_z,\bar{\lambda}\} \) }
\label{sec:limit1}
At weak SOC, $\Delta_{\mathrm{A}} \gg \{ \lambda_z, \bar{\lambda}  \}$ the integrand in Eq.~\eqref{h_expand3} is strongly peaked in the two angular intervals, $|\phi|, |\phi - \pi| \lesssim \max\{ \lambda_z, \bar{\lambda}  \}/\Delta_{\mathrm{A}} \ll \pi$, and therefore,
%Therefore, we obtain, 
\begin{align}\label{h_expand4}
    \hat{\sigma}_{in;y z} =  \frac{\sigma_0}{32\pi^2} \frac{\mathrm{sgn}(\Delta_{\mathrm{A}}) \lambda_z\bar{\lambda} }{E_F}  \!\!  \intop_{0}^{\pi} \!\!
    \frac{d \theta\sin\theta }
    {\sqrt{\lambda_z^2+(\bar{\lambda} \cos\theta \sin \theta  /4)^2 } }.
\end{align}
The salient qualitative feature of Eq.~\eqref{h_expand4} is the non-analiticity in a small SOC.

In the regime $\bar{\lambda} \ll \lambda_z$, Eq.~\eqref{h_expand4} yields Eq.~\eqref{h_expand6} of the main text.
%\begin{align}\label{h_expand5}
%    \hat{\sigma}_{in;y z} =  \frac{\sigma_0}{16 \pi^2 } \frac{\bar{\lambda}\mathrm{sgn}(\lambda_z\Delta_{\mathrm{A}})}{E_F}
%     \, .
%\end{align}
In the case $\lambda_z \ll \bar{\lambda}$ the integral over $\theta$ in Eq.~\eqref{h_expand4} 
can be estimated by introducing the cutoff $C$ satisfying $1\gg C\gg\lambda_z/\bar{\lambda}$. 
In the domain $\theta\in[\pi/2-C;\pi/2+C]$ one can expand $\cos\theta\approx(\pi/2-\theta)$, while for $\theta$ outside this domain one can neglect $\lambda_z$.
In summary, 
\begin{multline}\label{alpha3}
    \hat{\sigma}_{in;yz}\!=\!\frac{\sigma_{0}}{16\pi^{2}}\!
    \begin{cases}
    \frac{\bar{\lambda}\mathrm{sgn}(\lambda_{z}\Delta_{\mathrm{A}})}{E_{F}}, & \!\!\!\Delta_{\mathrm{A}}\gg\lambda_{z}\gg\bar{\lambda}\\
    \frac{4\lambda_{z}\mathrm{sgn}(\bar{\lambda}\Delta_{\mathrm{A}})}{E_{F}}\ln\frac{\bar{\lambda}}{\lambda_{z}}, & \!\!\!\Delta_{\mathrm{A}}\gg\bar{\lambda}\gg\lambda_{z}.
\end{cases}
\end{multline}

\subsubsection{The limit \( \lambda_z \gg \max\{\Delta_{\mathrm{A}},\bar{\lambda} \}\)  }
\label{sec:limit2}
In this limit Eq.~\eqref{h_expand4} results in Eq.~\eqref{h_expand7} of the main text,
\begin{equation}\label{alpha2}
    \hat{\sigma}_{in;yz} = \frac{\sigma_{0}}{48\pi^{2}}\frac{\bar{\lambda}\Delta_{\mathrm{A}}}{E_{F}\lambda_{z}}\, . %,\  \lambda_{z}\gg\bar{\lambda},\Delta_{\mathrm{A}}
\end{equation}

\subsubsection{The limit \( \bar{\lambda} \gg \max\{\Delta_{\mathrm{A}}, \lambda_z\}\) }
\label{sec:limit3}
%For completeness we investigate the intrinsic AHC in other limits.
To explore this regime, we simplify Eq.~\eqref{h_expand2} using the different parametrization of the unit sphere, $k_x=k_F\cos\alpha$, $k_y= k_F\sin\alpha\cos\beta$, $k_z= k_F\sin\alpha\sin\beta$, $\alpha\in[0;\pi]$ and $\beta\in[0;2\pi)$. 
The integration over $\beta$ can be done analytically using the relation,
\begin{equation}\label{alpha}
    \intop_{0}^{2\pi}\frac{d\beta}{a+b\sin^2\beta}=\frac{2\pi}{\sqrt{a(a+b)}},\ a,b>0.
\end{equation}
In result,
\begin{multline}\label{alpha1}
    \hat{\sigma}_{in;yz} = \frac{\sigma_{0}}{32\pi^{2}}\frac{\lambda_{z}\bar{\lambda}\Delta_{\mathrm{A}}}{E_{F}}\times\\
    \intop_{0}^{\pi} \!\! \frac{d\alpha\sin\alpha\cos^{2}\alpha }{
    \sqrt{\lambda_{z}^{2}+\left(\Delta_{\mathrm{A}}\sin\alpha\cos\alpha\right)^{2}}
    \sqrt{\lambda_{z}^{2}+\left(\bar{\lambda}\sin\alpha\cos\alpha/4\right)^{2}}},
\end{multline}
%For this integral we obtain the following asymptotic behavior in different limiting cases. 
%In the limit $\Delta_{\mathrm{A}}\gg\lambda_{z} \gg \bar{\lambda}$ considered in Sec.~\ref{sec:limit1},
%Eq.~\eqref{alpha1} again reproduces Eq.~\eqref{h_expand6}.
%we reproduce result Eq.~\eqref{h_expand4} and Eq.~\eqref{h_expand5}. 
%In the limit of the dominant $\lambda_z$ considered in Sec.~\ref{sec:limit2} we similarly recover Eq.~\eqref{h_expand7}. 
%\begin{equation}\label{alpha2}
%    \hat{\sigma}_{in;yz} = \frac{\sigma_{0}}{48\pi^{2}}\frac{\bar{\lambda}\Delta_{\mathrm{A}}}{E_{F}\lambda_{z}},\  \lambda_{z}\gg\bar{\lambda},\Delta_{\mathrm{A}}
%\end{equation}
%
%Due to the symmetry of the integral Eq.~\eqref{alpha1} with respect to $\Delta_{\mathrm{A}}\leftrightarrow\bar{\lambda}/4$ in the limit $\bar{\lambda}\gg\lambda_{z},\Delta_{\mathrm{A}}$we obtain:
which in the considered limit reduces to
\begin{equation}\label{alpha4}
    \hat{\sigma}_{in;yz} = \frac{\sigma_{0}}{8\pi^{2}}\frac{\lambda_{z}\Delta_{\mathrm{A}}\mathrm{sgn}(\bar{\lambda})}{E_{F}}
    \intop_{0}^{\pi}  \frac{d\alpha\sin\alpha}{
    \sqrt{\lambda_{z}^{2}+\left(\Delta_{\mathrm{A}}\sin\alpha\cos\alpha\right)^{2}}}\, .
\end{equation}
Finally we estimate, 
\begin{equation}
    \hat{\sigma}_{in;yz}\!=\!\frac{\sigma_{0}}{4\pi^{2}}\begin{cases}
    \frac{\Delta_{\mathrm{A}}\mathrm{sgn}(\lambda_{z}\bar{\lambda})}{E_{F}}, & \!\!\! \bar{\lambda}\gg\lambda_{z}\gg\Delta_{\mathrm{A}}\\
    \frac{\lambda_{z}\mathrm{sgn}(\bar{\lambda}\Delta_{\mathrm{A}})}{E_{F}}\ln\!\frac{4\Delta_{\mathrm{A}}}{\lambda_{z}}, & \!\! \! \bar{\lambda}\gg\Delta_{\mathrm{A}}\gg\lambda_{z}.
\end{cases}
\end{equation}

\subsection{Intrinsic Hall conductivity for \( {}^24/{}^1m{}^2m_y{}^1m_d \).}
In this case, instead of Eq.~\eqref{h_expand1} we have
\begin{align}\label{h_expand1a}
    \mathbf{h}_{\mathbf{k}} \approx \left(\frac{\lambda}{4}k_x k_z, \frac{\lambda_z}{2}(k_x^2 -k_y^2), t_A k_x k_y \right)\, .
\end{align}
Substitution of Eq.~\eqref{h_expand1a} into Eq.~\eqref{eq:sigma_Berry_3} results in the expression 
\begin{align}\label{h_expand2a}
    \hat{\sigma}_{in;y z}& =  -\frac{2 e^2}{(2 \pi)^3} \oint \frac{d S}{v_F}
   \notag \\
\times &    \frac{ \lambda_z \lambda t_A \left(k_x^4 +  k_x^2 k_y^2\right) }{16 t_A^2 k_x^2 k_y^2+
   4 \lambda_z^2 \left(k_x^2-k_y^2\right)^2+\lambda ^2
   k_x^2 k_z^2 }\, 
\end{align}
replacing Eq.~\eqref{h_expand2}.
Similarly to the considerations of a class A representative we have in the limit of weak SOC Eq.~\eqref{h_expand2a} results in Eq.~\eqref{h_expand3a}.

In the opposite limit of strong SOC the main contribution comes from the nodal lines $k_x = \pm k_y$ confined to $k_z=0$ planes, where the SOC vanishes, and the denominator of Eq.~\eqref{h_expand2a} reaches minimum.
This consideration allows us to estimate the integration in Eq.~\eqref{h_expand2a} with the logarithmic accuracy.
This leads to the Eq.~\eqref{h_expand8} of the main text. 
%\begin{equation}\label{h_expand4a}
%    \hat{\sigma}_{in;yz}\approx\frac{\sigma_0}{4\pi^2\sqrt{2}}\frac{\Delta_\mathrm{A}\mathrm{sgn}(\lambda_z\bar{\lambda})}{E_F}\ln{\left(\min{\left[\frac{\lambda_z}{A},\frac{\lambda}{A}\right]}\right)}.
%\end{equation}
%This result is valid in the logarithmic approximation i.e., it is valid up to an additional constant to the logarithm which depends weakly with respect to the ratio $\lambda/\lambda_z$. 

\subsection{The effect of $|\mathbf{M}|\neq 0$ on intrinsic AHC}
Here we provide the details of the derivation of Eqs.~\eqref{sigma_in_M_A} and \eqref{sigma_in_M_B}
in the limit of large altermagnetic splitting.
As detailed in Sec.~\ref{sec:sigma_in_J_B} the effect of magnetization can be accounted for if one makes a replacement $\lambda_z \rightarrow \lambda_z + M t_x /N$ to construct the effective Hamiltonian to the linear order in $\mathbf{M}$.
Having made the above replacement one can expand Eq.~\eqref{h_expand4} to the linear order in $M$ which gives  
\begin{align}\label{h_expand4_M}
    \hat{\sigma}^M_{in;y z} \approx \frac{e^2}{ \pi} \frac{M\mathrm{sgn}(t_A)}{N v_F} \oint \frac{d \Omega}{4 \pi}
\frac{t_x \bar{\lambda}^3 \sin ^2\theta \cos
   ^2\theta}{\left(\bar{\lambda}^2 \sin ^2\theta \cos ^2\theta+16
   \lambda_z^2\right)^{3/2}}.
\end{align}
In the regime $\lambda_z \gtrsim \bar{\lambda}$ we can neglect $\bar{\lambda}$ in Eq.~\eqref{h_expand4_M}.
Finally, the angular integration yields the result \eqref{sigma_in_M_A}.
The derivation of Eq.~\eqref{sigma_in_M_B} is similar.
%%%%%%%%%%%%%%%%

\section{Effective Hamiltonian by $1/|\mathbf{N}|$ expansion}
\label{app:effective_Hamiltonian}
The idea of the method is to apply a unitary transformation $\hat{U}$ to the original 4 by 4 Hamiltonian, Eq.~\eqref{eq:structure} 
\begin{align*}
    \hat{H}_4 = \begin{pmatrix}
        \check{H}_p & \check{V} \\ \check{V}^\dagger & \check{H}_n 
    \end{pmatrix}\, \tag{\ref{eq:structure}}
\end{align*}
such that the transformed Hamiltonian, $\hat{H}'_4 =\hat{U} \hat{H}_4 \hat{U}^\dagger$ is simpler than $\hat{H}_4$.
In the present context the simplification is achieved by making the off-diagonal blocks of Eq.~\eqref{eq:structure} as small as possible.

The construction is based on the energy scale separation.
The N\'eel vector is assumed to be the large energy scale, $N$ that enters the diagonal blocks of Eq.~\eqref{eq:structure} as in the Eq.~\eqref{eq:check_H}
\begin{align}%\label{eq:check_H}
    \check{H}_{p(n)} = \pm N + \check{h}_{p(n)} \tag{\ref{eq:check_H}}\, .
\end{align}
We also assume that $\check{h}_{u(d)}$ and $\check{V}$ are of the same order of magnitude, such that the both are estimated as the typical band splitting, $h$.

The considered approach is perturbative in $h/N \ll 1$.
We stress that it differs from a similar construction of the effective Hamiltonian used in quantum optics to describe the coupling of atomic degrees of freedom and radiation \cite{Cohen1992}.
The light-matter interaction is weak thanks to the smallness of the fine structure constant which in the case of optics is a well motivated expansion parameter.
In our problem instead we rely on large splitting between the quasi-degenerate manifolds.

We construct the unitary transformation, $\hat{U} =e^{i \hat{S}}$ with $\hat{S}^{\dagger} = \hat{S}$ and look for $\hat{S}$ as the expansion, 
\begin{align}\label{eq:S_expand}
    \hat{S} = \sum_{l=1}^{\infty} \frac{\hat{S}_l}{N^l}.
\end{align}
Each consecutive term $\hat{S}_l \propto h^l$ of the expansion is set to reduce 
the off-diagonal blocks of the Hamiltonian, Eq.~\eqref{eq:structure} $V$ and $V^\dagger$ to the operators scaling as $h(h/N)^l$. 

To fix the operators $\hat{S}_l$ uniquely we impose on them the block-off-diagonal structure,
\begin{align}\label{S_l}
    \hat{S}_l = \begin{pmatrix}
        \check{0} & \check{S}_l \\ \check{S}_l^\dagger & \check{0}
    \end{pmatrix}\, .
\end{align}
The two effective Hamiltonians, $\check{H}^{\mathrm{eff}}_{u(d)}$ are the up (down) diagonal blocks  of the transformed Hamiltonian $\hat{H}'$.
To the expansion \eqref{eq:S_expand} corresponds the expansion of the effective Hamiltonian,
\begin{align}
    \check{H}^{\mathrm{eff}}_{p,n} = \check{H}_{p,n} + \sum_{l=1}^\infty \frac{\check{H}_{p,n}^{(l)}}{N^l}\, .
\end{align}

For our purposes it is sufficient to compute $\check{H}^{\mathrm{eff}}_{p,n}$ up to $(h/N)^2$.
Therefore, it is enough to keep the following terms of the transformed Hamiltonian,
\begin{align}\label{eq:H'_expand}
    \hat{H}'_4 & \approx \hat{H}_4 +\frac{i}{N}[\hat{S}_1,\hat{H}_4]+ \frac{(-1)}{2 N^2} [\hat{S}_1,[\hat{S}_1,\hat{H}_4]]
    + \frac{i}{N^2}[\hat{S}_2,\hat{H}_4]
     \notag \\
    + & \frac{(-1)}{2 N^3} [\hat{S}_1,[\hat{S}_2,\hat{H}_4]] +  \frac{(-1)}{2 N^3} [\hat{S}_2,[\hat{S}_1,\hat{H}_4]]\, .
\end{align}

We start with finding $\hat{S}_1$.
To this end the following commutation relation is useful, 
\begin{align}\label{eq:commute1}
   \hat{C}^l= [\hat{S}_l, \hat{H}_4] = \begin{pmatrix}
        \check{S}_l \check{V}^\dagger - \check{V} \check{S}_1^\dagger & \check{S}_l \check{H}_n - \check{H}_p \check{S}_l \\ \check{S}_l^\dagger \check{H}_p - \check{H}_n \check{S}_l^\dagger & \check{S}_l^\dagger \check{V} - \check{V}^\dagger \check{S}_l
    \end{pmatrix}\, .
\end{align}
It follows that to eliminate the off-diagonal blocks to the order $(1/N)^0$ one has to impose condition 
%\begin{align}\label{eq:S1}
%     \langle i u|  S_1 |k d \rangle  = - \frac{i}{2 }  \langle i u |  V | k d \rangle \, .
%\end{align}
\begin{align}\label{eq:S1}
  \check{S}_1   = - \frac{i}{2 }   \check{V}  \, .
\end{align}
Such an elimination is possible because the off-diagonal commutators 
\begin{align}
    \check{C}^{l}_{12} & = -2 \check{S}_l N + \check{S}_l \check{h}_n - \check{h}_p \check{S}_l
    \notag \\
    \check{C}^{l}_{21} & = 2 \check{S}_l^\dagger N + \check{S}_l^\dagger \check{h}_p - \check{h}_n \check{S}_l^\dagger
\end{align}
contain term linear in $N$.
Hereinafter we have employed a natural parametrization of the commutator, Eq.~\eqref{eq:commute1},
\begin{align}\label{eq:commute2}
   \hat{C}^l= \begin{pmatrix}
        \check{C}^l_{11} & \check{C}^l_{12}\\ \check{C}^l_{21} & \check{C}^l_{22}
    \end{pmatrix}\, .
\end{align}
Knowledge of $\hat{S}_1$ allows us to fix the effective Hamiltonian to the order $1/N$ using Eq.~\eqref{eq:H'_expand}.
Note, that in addition to the second term of Eq.~\eqref{eq:H'_expand}, also the third term needs to be included. 
To see this we quote the expression for the nested commutator, 
\begin{equation}\label{eq:commute3}
[\hat{S}_{l'},\! [\hat{S}_l, \hat{H}_4] ]\!=\! \begin{pmatrix}\!
        \check{S}_{l'}\check{C}^l_{21} -\check{C}^l_{12} \check{S}_{l'}^\dagger & \check{S}_{l'} \check{C}_{22} - \check{C}_{11} \check{S}_{l'}
       \\ \check{S}_{l'}^\dagger \check{C}_{11} - \check{C}_{22} \check{S}_{l'}^\dagger & \check{S}_{l'}^\dagger \check{C}_{12} -\check{C}_{21} \check{S}_l
       \!
    \end{pmatrix}\!.
\end{equation}
Since $\check{C}^{l=1}_{21} \propto N$ it follows from Eq.~\eqref{eq:commute3} that the nested commutator \eqref{eq:commute3} with $l = l' = 1$ contributes to the effective Hamiltonian.
We have for the order $ (1/N)^1$ correction to the effective Hamiltonian, 
\begin{align}\label{eq:Heff_1}
    \check{H}_{p}^{(1)} = \frac{i}{N}(\check{S}_1 \check{V}^\dagger - \check{V} \check{S}_1^\dagger) - \frac{2}{ N} \check{S}_1  \check{S}_{1}^\dagger\, .
\end{align}
Applying \eqref{eq:S1} we obtain the expression
\begin{align}\label{eq:Heff_2}
    \check{H}^{\mathrm{eff}}_{p} = \check{H}_p +  \frac{\check{V} \check{V}^\dagger}{2 N  }    + \mathcal{O}((1/N)^2)
\end{align}
that turns into the standard result of the non-degenerate perturbation theory in the case of a one-dimensional manifold. 

To fix the $\hat{S}_2 $ we have to eliminate the off-diagonal elements to the order $(1/N)^1$.
This gives the matrix solution,
\begin{align}\label{eq:S2}
\check{S}_2= \frac{1}{2}( \check{S}_1 \check{h}_n -\check{h}_p \check{S}_1)\, .
\end{align}
%To solve this matrix equation we specify to the matrix elements, 
%\begin{align}
%   \frac{i}{N}( S_1 h_d -h_u S_1) +  \frac{i}{N^2}(-2N) S_2  =0
%\end{align}
Equation \eqref{eq:S2} allows us to compute the correction to the effective Hamiltonian to the order $1/N^2$,
\begin{align}
 \check{H}_{p}^{(2)} =   -\frac{1}{8} ( \check{V} \check{V}^\dagger \check{h}_p  - 2 \check{V} \check{h}_n \check{V}^\dagger    + \check{h}_p \check{V} \check{V}^\dagger  ).
\end{align}
For convenience, we summarize the effective Hamiltonian to the order $1/N^2$, 
\begin{align}\label{eq:Heff_summary}
    \check{H}^{\mathrm{eff}}_{p} & =   \check{H}_p +  \frac{\check{V} \check{V}^\dagger}{2 |\mathbf{N}|  }  
     \\
   - & \frac{1}{8 |\mathbf{N}|^2} ( \check{V} \check{V}^\dagger \check{h}_p  - 2 \check{V} \check{h}_n \check{V}^\dagger    + \check{h}_p \check{V} \check{V}^\dagger  ) + \mathcal{O}((1/N)^3). \notag
\end{align}

\subsection{Berry curvature to the order $1/N^2$}
Here we show how to determine the Berry curvature to the accuracy $(1/N)^2$.
As before, we focus on the two positive energy bands related adiabatically to the solutions $|\mathbf{k} \pm \rangle$ of $\check{H}_{p}$. 
The unitary transformation, $\hat{U}$ has been designed to bring the Hamiltonian into the form,
\begin{align*}
    \hat{H}'_4 = \begin{pmatrix}
        \check{H}^{\mathrm{eff}}_{p} & \check{V}'\\ \check{V}'^\dagger & \check{H}^{\mathrm{eff}}_{n}
    \end{pmatrix}\, ,
\end{align*}
where $\check{V}' = \mathcal{O}(h(h/N)^2)$ and $\check{H}^{\mathrm{eff}}_{p} $ is computed up to the same order. 
%The unitary transformation to the same order reads,
%\begin{align}
%    U \approx \mathbb{1}+ i \frac{\hat{S}_1}{N} +i \frac{\hat{S}_2}{N^2} - \frac{\hat{S}_1^2}{2 N^2}
%\end{align}
%We assume that the eigenfunctions of the effective Hamiltonian given up to any specified order, $\ell$ can be explicitly computed.
%We denote them as $|n_\ell\rangle$.
%For instance,  $|n_{\ell=0}\rangle$ are the wavefunctions of the orignal Hamiltonian $H_{u}$, and  $|n\rangle_{\ell=1}$ are the eigenfunction of the Hamiltonian $H_u + H_u^{(1) }/N$, etc.
%%
%are a series in $1/N$, 
%\begin{align}
%    |n\rangle = |n\rangle_0 + \frac{1}{N}|n\rangle_1 + \frac{1}{N^2}|n\rangle_2+\ldots \, .
%\end{align}
%which just reflects the fact that the effective Hamiltonian is such series as well. 
%In other words, the term of the 

%To the order $1/N$ the Berry curvature is just that computed with the wave-functions obtained from the Hamiltonian $H_u + H_u^{(1) }/N$. 
%The unitary transformation does not contribute to the order $1/N$.
%The reason for this is the structure of the unitary transformation \eqref{S_l} which implies that $S_1$ produces the tails of the wave-function.
%These tails proportional to $1/N$ are also refereed to as contamination of the wave function in quantum optics context. 

The Berry curvature, $\Omega_{\alpha \beta}$ depends solely on the eigenfunctions of the original Hamiltonian $\hat{H}_{4}$ which we denote by $|n\rangle$.
Again we are only interested in the two eigenfunctions that transform smoothly into the wave-functions of $\check{H}_p$ as $\check{V}$ tends to zero.
These solutions are obtained from the eigenfunctions $|n'\rangle$ of the effective Hamiltonian $\check{H}^{\mathrm{eff}}_p$ by applying the inverse unitary transformation, $|n\rangle = U^{\dagger} |n'\rangle$. 
%With the eigenfunctions known to the second order in $(h/N)$, the Berry curvature follows, 
By definition, Eq.~\eqref{BC:define},
\begin{align}\label{Omega_eff1}
    \Omega_{\alpha \beta} = -2\Im \langle \partial_\alpha ( \hat{U}^\dagger n')| \partial_\beta (\hat{U}^\dagger n') \rangle\, .
\end{align}
In this equation the eigenstates $|n'\rangle$ are obtained by solving the effective Hamiltonian that is correct to the second order included, and the unitary transformation that to this order takes the form, 
\begin{align}\label{eq:U}
    \hat{U} \approx \begin{pmatrix}
        \check{\mathbb{1}}-\check{S}_1 \check{S}^\dagger_1/2N^2 & i \check{S}_1/N  + i \check{S}_2/N^2 \\
        i \check{S}^\dagger_1/N  + i \check{S}^\dagger_2/N^2 & \check{\mathbb{1}} - \check{S}^\dagger_1 \check{S}_1/2N^2 
    \end{pmatrix} \, .
\end{align}

Equation \eqref{Omega_eff1} breaks into four parts,
\begin{align}
    \Omega_{\alpha \beta} = \Omega^{\mathrm{eff}}_{\alpha \beta} + \Omega^a_{\alpha \beta} + \Omega^b_{\alpha \beta} + \Omega^c_{\alpha \beta}\, ,
\end{align}
where the first part is just the Berry curvature computed on the bands of the effective Hamiltonian,
\begin{align}\label{Omega_eff_17}
   \Omega^{\mathrm{eff}}_{\alpha \beta}  =-2 \Im \langle \partial_\alpha n'| \partial_\beta  n' \rangle\, .
\end{align}
Eq.~\eqref{Omega_eff_17} ignores the unitary transformation.
The consistency of the approach requires that the accuracy of the expression \eqref{Omega_eff_17} does not exceed the accuracy of the effective Hamiltonian.
The remaining parts originate from the unitary transformation we used to affect the transformation to the effective Hamiltonian.
They are 
\begin{align}
     \Omega^a_{\alpha \beta}  &= -2
    \Im \langle n'| (\partial_\alpha \hat{U}) (\partial_\beta \hat{U}^\dagger) |n'\rangle 
 \notag \\
 \Omega^b_{\alpha \beta}  &= -2
    \Im \langle n'| (\partial_\alpha \hat{U}^\dagger)  \hat{U} |\partial_\beta n' \rangle
\notag \\ 
     \Omega^c_{\alpha \beta}  &= -2 \Im \langle \partial_\alpha  n'| \hat{U} (\partial_\beta \hat{U}^\dagger) |n' \rangle. 
\end{align}

With the unitary transformation \eqref{eq:U} we have 
\begin{align}
   \Omega^a_{\alpha \beta} & = -\frac{2}{N^2}\Im \langle n | (\partial_\alpha S_1) (\partial_\beta S_1^\dagger) |n\rangle 
    \\
   \Omega^b_{\alpha \beta} & = -\frac{2}{N^2}\Im \langle n| (\partial_\alpha S_1)  S_1^\dagger |\partial_\beta n \rangle +\frac{1}{N^2}\Im \langle n|\partial_\alpha (S_{1}S_1^\dagger ) |\partial_\beta n \rangle
  \notag  \\
   \Omega^c_{\alpha \beta} & = - \frac{2}{N^2}\Im \langle \partial_\alpha  n| S_1 (\partial_\beta S_1^\dagger) |n \rangle +\frac{1}{N^2}\Im \langle \partial_\alpha n| \partial_\beta(S_{1}S_1^\dagger ) | n \rangle 
   \notag 
\end{align}
We can simplify the last two contributions as
\begin{align}
    \Omega^b_{\alpha \beta} & = \frac{1}{N^2}\Im \langle n|\left[  S_1  (\partial_\alpha S_1^\dagger) - (\partial_\alpha S_1)  S_1^\dagger \right] |\partial_\beta n \rangle 
  \notag  \\
   \Omega^c_{\alpha \beta} & = \frac{1}{ N^2}\Im \langle \partial_\alpha  n| \left[   (\partial_\beta  S_1 )S_1^\dagger) - S_1 (\partial_\beta S_1^\dagger)\right] |n \rangle. 
\end{align}

Finally, we summarize the Berry phase up to the order $1/N^2$ as follows, 
\begin{align}\label{Omega_second}
    \Omega_{\alpha\beta} & = \Omega^{\mathrm{eff}}_{\alpha \beta} - 
    \frac{1}{2 N^2}\Im \langle n | (\partial_\alpha \check{V}) (\partial_\beta \check{V}^\dagger) |n\rangle 
    \notag \\
    - & \frac{1}{4 N^2}\Im \langle n|\left[ (\partial_\alpha \check{V})  \check{V}^\dagger -  \check{V}  (\partial_\alpha \check{V}^\dagger) \right] |\partial_\beta n \rangle 
    \notag \\
    - & \frac{1}{4 N^2}\Im \langle \partial_\alpha  n| \left[ \check{V} (\partial_\beta \check{V}^\dagger)  - (\partial_\beta  \check{V} )V^\dagger)\right] |n \rangle,
\end{align}
where $\Omega^{\mathrm{eff}}_{\alpha \beta}$ is computed with the effective Hamiltonian correct to the order $(h/N)^2$, and in the remaining terms $|n\rangle$ are the band wavefunctions of the original Hamiltonian, $\check{H}_{p}$.
Again, although the first term contains in principle corrections to all orders in $(h/N)$ only the terms up to order $(h/N)^2$ should be kept in this expressions as required by the consistency with our expansion.

\section{Disorder averaged Green function}
\subsection{Self energy in the $\mathbf{k}\cdot \mathbf{p}$ approximation}
\label{app:self}
In this appendix we derive the expression Eq.~\eqref{Gamma_vec} based on our effective two-band model with the dispersion, Eq.~\eqref{E+-}.
The implicit equations fixing the two Fermi surfaces, $E^{\pm}_\mathbf{k} = E_F$ can be in principle solved for the two Fermi momenta, $\mathbf{k}_{\pm}(\Omega)$.
As the polar angles $\Omega = (\phi,\theta)$ range over the intervals $\phi \in [0,2\pi)$, $\theta \in [0,\pi]$ respectively,
$\mathbf{k}_{\pm}(\Omega)$ span the two Fermi surface. 
In the limit of zero spin splitting the two bands are degenerate with the dispersion relation $E_0(\mathbf{k}) = E_{0,k}(\Omega)$.
Correspondingly, Fermi surfaces coincide  $k_{\pm}(\Omega)=k_F(\Omega)$ in the same limit.

As here we assume that typical spin splitting, $h_{\mathbf{k}} $ is much smaller than the Fermi energy, $E_F$, we can approximate,
\begin{align}\label{k+-}
    k_{\pm}(\Omega) \approx k_F(\Omega)  \mp \frac{h_{{k}}(\Omega)}{\partial E_{0,k}(\Omega)/\partial k  }
\end{align}
with the derivative evaluated at $k = k_F(\Omega)$.

Only the $y$-component of the vector $\Gamma_y$ survives the angular integration to the leading order in $h/E_F$.
We chose to write this component in the spherical coordinates, 
\begin{align}\label{eq:Gammay_y}
  \Gamma_{y}\!=\!\frac{n_{\text{imp}}U^{2}}{4\pi}\!\! \int \!\!\frac{d\Omega}{4\pi} \! \left[n_{\mathbf{k}_+,y}g_{+}(\Omega)-n_{\mathbf{k}_-,y}g_{-}(\Omega)\right],
\end{align}
where we have introduced the notation,
\begin{align}\label{eq:g+-}
    g_{\pm}(\Omega) =\frac{k_\pm^2}{|\partial E_{\pm}(k_{\pm},\Omega)/\partial k_{\pm} |}\, .
\end{align}
Near $\Gamma$-point only the constant part of the SOC survives the angular momentum integration, and we write 
\begin{align}\label{n+-1}
    n_{\mathbf{k}_{\pm},y} \approx \lambda_0/ h_{\mathbf{k}_\pm}\, .
\end{align}

Our goal is to expand Eq.~\eqref{eq:Gammay_y} around $h=0$.
Within $\mathbf{k}\cdot\mathbf{p}$ approximation the degenerate Fermi surfaces have the common 
spin-independent dispersion relation, $E_{0}(\mathbf{k}) = E_0 k^2$.
In result, Eq.~\eqref{k+-} simplifies to 
\begin{align}\label{k+-1}
    k_{\pm}(\Omega) \approx k_F   \mp \frac{h_{k_F}(\Omega)}{2 E_0 k_F  }\, .
\end{align}
The expansion of Eq.~\eqref{eq:g+-} to the first order in $h/E_F$ using Eq.~\eqref{k+-1} reads,
\begin{align}\label{eq:g+-_1}
    g_{\pm}(\Omega) & =\frac{k_\pm^2}{|\partial E_{\pm}(k_{\pm},\Omega)/\partial k_{\pm} |} 
    \notag \\
    \approx & \frac{k}{2E_{0}}\left[
    1 \mp \frac{d h_{k}(\Omega)/dk }{2E_{0}k}\mp\frac{ h_{k}(\Omega)}{2E_{F}}\right]_{k =k_F}\, .
\end{align}
Similarly, 
\begin{align}\label{1/h+-}
\frac{1}{h_{k_{\pm}}(\Omega)} & \approx \frac{1}{h_{k_{F}}(\Omega)}-
\frac{dh_{k_{F}}(\Omega)/dk_F}
{h^{2}_{k_{F}}(\Omega)}\left(k_{\pm}-k_{F}\right)
\notag \\
\approx &\frac{1}{h_{k_{F}}(\Omega)}\left[ 1\pm\frac{dh_{k_{F}}(\Omega)/dk_F}{2E_{0}k_{F}}\right].
\end{align}
Combining Eqs.~\eqref{n+-1}, \eqref{eq:g+-_1} and \eqref{1/h+-} we can approximate Eq.~\eqref{eq:Gammay_y} as
\begin{align}\label{eq:Gammay_y1}
  \Gamma_{y}\!= -\frac{\lambda_0}{E_F} \frac{n_{\text{imp}}U^{2}}{4\pi}\!\! \int \!\!\frac{d\Omega}{4\pi} \! 
  \frac{k_F}{2 E_0}.
\end{align}
Comparison of Eq.~\eqref{eq:Gammay_y1} with Eq.~\eqref{Gamma_0R} yields the expression \eqref{Gamma_vec}.
%%%%%%%%%%%%%%%%%%%%%%%%
%%%%%%%%%%%%%%%%%%%%%%%%
\subsection{Green function in a weak disorder limit}
\label{app:Green}
With the self-energy in the form of Eq.~\eqref{Sigma} we can write the exact expression for the Green function as
\begin{align}\label{G35}
    G_{\mathbf{k}}^R(E) = P'_{+\mathbf{k}}\frac{1}{E - E'^+_{\mathbf{k}}} + P'_{-\mathbf{k}}\frac{1}{E - E'^-_{\mathbf{k}}}\, ,
\end{align}
where the two poles are positioned at 
\begin{align}\label{G32}
    E^{'\pm}_{\mathbf{k}} = E_{0,\mathbf{k}} - i \Gamma_0 \pm \sqrt{(\mathbf{h}_{\mathbf{k}} - i \boldsymbol{\Gamma})^2}
\end{align}
and  
\begin{align}\label{G36}
    P'_{\pm \mathbf{k}} = \frac{1}{2}\left[ 1 \pm \frac{(\mathbf{h}_{\mathbf{k}}- i \boldsymbol{\Gamma})\cdot \boldsymbol{\rho}}{\sqrt{(\mathbf{h}_{\mathbf{k}}- i \boldsymbol{\Gamma})^2} } \right].
\end{align}
are projection operators only in the clean case, Eq.~\eqref{P+-}.
As eventually we are interested in the clean limit, we make an expansion to the leading order in $n_{\mathrm{imp}}$ which results in Eqs.~\eqref{G13_dis} through \eqref{G14_dis}.

\section{Extrinsic Hall conductivity in non-crossing approximation}
\label{app:derivation}
\subsection{Analytic expression for $\sigma^{nc}_{ex}$}
To derive the classical contributions Eq.~\eqref{sigma^I} one uses Eq.~\eqref{Green_dis_1} to write Eq.~\eqref{eq:sigma^I} as the sum of the two terms, 
\( \sigma^{I} = \sigma^{I,0} + \sigma_{ex} \). 
The first term,
\begin{align}\label{eq:sigma^I0}
    \sigma^{I,0}_{\alpha \beta} = & 
    \frac{e^2}{4 \pi} \Tr \left[ \hat{v}_{\alpha}\left( \sum_{f= \pm} P_{f\mathbf{k}} G_{f\mathbf{k}}^{R} \right)\hat{v}_{\beta} \left(\sum_{f= \pm} P_{f\mathbf{k}} G_{f\mathbf{k}}^{A} \right)\right] 
    \notag \\
    & - (\alpha \leftrightarrow \beta)\, ,
\end{align}
in the clean limit cancel the clean system result in Eq.~\eqref{sigma_ex}.
This implies that the remaining contribution is the extrinsic AHC,
\begin{align}\label{sigma_ex_1}
     \sigma_{ex;\alpha \beta} & =
     \frac{i e^2}{4 \pi} \Tr \left[v_\alpha   \delta \mathbf{n}\boldsymbol{\rho}\delta G_{\mathbf{k}}^{A} v_\beta \left( \sum_{f= \pm} P_{f\mathbf{k}} G_{f\mathbf{k}}^{R} \right) 
     \right.
     \notag \\
     -  v_\alpha & \left.  \left( \sum_{f= \pm} P_{f\mathbf{k}} G_{f\mathbf{k}}^{A} \right) v_\beta \delta \mathbf{n}\boldsymbol{\rho}\delta G_{\mathbf{k}}^{R} \right]
      - (\alpha \leftrightarrow \beta)\, ,
\end{align}
where we denote 
%$\delta G^{R,A}_{\mathbf{k}}  = ( G_{\mathbf{k}+}^{R,A} - G_{\mathbf{k}-}^{R,A} )/2$.
%The derivation of Eq.~\eqref{sigma^I} starting from Eq.~\eqref{sigma_ex_1} is outlined in App.~\ref{app:derivation}. 
\begin{align}\label{sum_diff}
    \bar{G}_{ \mathbf{k}}^{R,A} & = \frac{1}{2}\left(G_{\mathbf{k}+}^{R,A} + G_{\mathbf{k}-}^{R,A}\right)\, , 
    \notag \\
    \delta G^{R,A}_{\mathbf{k}} & = \frac{1}{2} \left( G_{\mathbf{k}+}^{R,A} - G_{\mathbf{k}-}^{R,A} \right)\, .
\end{align}
we have the representation, 
\begin{align}\label{decomp}
    \sum_{f= \pm} P_{\mathbf{k}f} G_{\mathbf{k}f}^{R,A} =  \bar{G}_{\mathbf{k}}^{R,A} + \delta G^{R,A}_{\mathbf{k}} \mathbf{n}_{\mathbf{k}}\boldsymbol{\rho}\, .
\end{align}
Corresponding to the decomposition \eqref{decomp}, Eq.~\eqref{sigma_ex_1} splits into two contributions, $\sigma_{ex} = \sigma_{ex}^{(a)} + \sigma_{ex}^{(b)}$, 
\begin{subequations}
\begin{align}\label{sigma1_19a}
    \sigma^{nc_1}_{ex;\alpha \beta} = & \frac{i e^2}{4 \pi} \Tr \left[v_\alpha   \delta \mathbf{n}\boldsymbol{\rho}\delta G_{\mathbf{k}}^{A} v_\beta \bar{G}_{\mathbf{k}}^{R} 
     %\right.
     %\notag \\
     -  v_\alpha 
     %& \left.  
     \bar{G}_{\mathbf{k}}^{A} v_\beta \delta \mathbf{n}\boldsymbol{\rho}\delta G_{\mathbf{k}}^{R} \right]
     \notag \\
      & - (\alpha \leftrightarrow \beta) \, ,
\end{align}
%\begin{align}\label{sigma1_19a}
%     \sigma^{(b)}_{ex;\alpha \beta} 
%   = &
%     \frac{  i e^2}{4 \pi} \Tr \left[v_\alpha   \delta \mathbf{n}\boldsymbol{\rho}\delta G^{A}  v_\beta \delta G^{R}\mathbf{n}\boldsymbol{\rho}  - v_\alpha  \delta G^{A} \mathbf{n}\boldsymbol{\rho}  v_\beta   \delta \mathbf{n}\boldsymbol{\rho}\delta G^{R} \right]
% \notag \\
%  &    - (\alpha \leftrightarrow \beta)\, .
%\end{align}
\begin{align}\label{sigma1_19b}
     \sigma^{nc_2}_{ex;\alpha \beta} 
   = &
     \frac{  i e^2}{4 \pi} \Tr \left[ 
     \delta G^{R} \delta G^{A} \left(
     v_\alpha   \delta \mathbf{n}\boldsymbol{\rho}  v_\beta \mathbf{n}\boldsymbol{\rho}  - v_\alpha   \mathbf{n}\boldsymbol{\rho}  v_\beta   \delta \mathbf{n}\boldsymbol{\rho} \right) \right]
 \notag \\
  &    - (\alpha \leftrightarrow \beta)\, .
\end{align}
\end{subequations}
We show that Eq.~\eqref{sigma1_19a} is identical to Eq.~\eqref{sigma^I_a} and Eq.~\eqref{sigma1_19b} is identical to Eq.~\eqref{sigma^I_b}, respectively.

First, we derive Eq.~\eqref{sigma^I_a} starting from Eq.~\eqref{sigma1_19a}.
As the disorder is taken to zero at the end of the calculation, 
we replace $\bar{G}^A \delta G^{R} $ and $\bar{G}^R \delta G^{A} $ by $(G_+^A G_+^{R} - G_-^A G_-^{R})/4$. 
This simplifies \eqref{sigma1_19a},
\begin{align}\label{sigma1_21}
    \sigma^{nc_1}_{ex;\alpha \beta} &=    
     \frac{i e^2}{8 \pi} \Tr\left[  (G_+^R G_+^A - G_-^R G_-^A)  [v_\beta , v_\alpha]_-   \delta \mathbf{n}_{\mathbf{k}}\boldsymbol{\rho}   \right]\, .
\end{align}
Since the velocity reads 
\begin{align}\label{sigma1_22}
    v_{\alpha} = \frac{\partial E_{0,\mathbf{k}}}{\partial k_{\alpha}} + 
    \frac{\partial \mathbf{h}_{\mathbf{k}}}{\partial k_{\alpha}}\cdot \boldsymbol{\rho}
\end{align}
the commutation relation read,
\begin{align}\label{sigma1_23}
    [v_\alpha,v_\beta]_- = 2 i  \left(\frac{\partial \mathbf{h}_{\mathbf{k}}}{\partial k_{\alpha}} \times \frac{\partial \mathbf{h}_{\mathbf{k}}}{\partial k_{\beta}} \right) \cdot \boldsymbol{\rho}
\end{align}
As a result, the trace over the band indices gives, 
\begin{align}\label{sigma1_24}
    \Tr  \left[ [v_\alpha , v_\beta]_- \left\{  \delta \mathbf{n}_{\mathbf{k}}\boldsymbol{\rho} \right\}  \right] = 4 i \left(\frac{\partial \mathbf{h}_{\mathbf{k}}}{\partial k_{\alpha}} \times \frac{\partial \mathbf{h}_{\mathbf{k}}}{\partial k_{\beta}} \right) \cdot \delta \mathbf{n}_{\mathbf{k}}.
\end{align}
Substitution of Eq.~\eqref{sigma1_24} in Eq.~\eqref{sigma1_21} gives
\begin{align}\label{sigma1_25}
    \sigma^{nc_1}_{ex;\alpha \beta} &=   
     \frac{e^2}{ 2\pi}\int \frac{d^3 k}{(2 \pi)^3}  (G_+^R G_+^A - G_-^R G_-^A) 
     \notag \\
& \times     \left(\frac{\partial \mathbf{h}_{\mathbf{k}}}{\partial k_{\alpha}} \times \frac{\partial \mathbf{h}_{\mathbf{k}}}{\partial k_{\beta}} \right) \cdot \delta \mathbf{n}_{\mathbf{k}}\, .
\end{align}
To take the limit of vanishing disorder we rely on the approximation,
\begin{align}\label{sigma1_26}
    G_{\pm}^R G_{\pm}^A = \frac{\pi}{\Gamma_\pm} \delta(E - E^{\pm}_\mathbf{k})\, 
\end{align}
in order to evaluate \eqref{sigma1_25} as follows, 
\begin{align}\label{sigma1_27}
    \sigma^{nc_1}_{ex;\alpha \beta} = &   
\frac{ e^2}{2 (2\pi)^3}\left(  \oint \frac{d S_+}{v_+ \Gamma_+} - \oint \frac{d S_-}{v_- \Gamma_-} \right)
\notag \\
& \times \left(\frac{\partial \mathbf{h}_{\mathbf{k}}}{\partial k_{\alpha}} \times \frac{\partial \mathbf{h}_{\mathbf{k}}}{\partial k_{\beta}} \right) \cdot \delta \mathbf{n}_{\mathbf{k}}
\, .
\end{align}
Now with Eq.~\eqref{G14_dis}, Eq.~\eqref{sigma1_27} reproduces Eq.~\eqref{sigma^I_a}. 

%%%%%%%%%%%%%%%%%%

We now turn to the derivation of Eq.~\eqref{sigma^I_b} starting from Eq.~\eqref{sigma1_19b}.
The only contribution surviving the trace over the pseudo-spin indices contains the two terms of the velocity operator,  Eq.~\eqref{sigma1_22},
\begin{align}\label{sigma1_19a_1}
     & \sigma^{nc_2}_{ex;\alpha \beta} 
      =  
     \frac{e^2}{2 \pi i}  \Tr 
     \left\{ \left(\delta G^{A}_{\mathbf{k}} \delta G^{R}_{\mathbf{k}} \right)
     \left[
     \frac{\partial E_0(\mathbf{k})}{\partial k_\alpha} 
     (  \mathbf{n}_{\mathbf{k}}\boldsymbol{\rho} )
      \right. \right.
     \notag \\
    & \left. \left. \times \left(\frac{\partial \mathbf{h}_{\mathbf{k}}}{\partial k_\beta} \boldsymbol{\rho} \right)( \delta \mathbf{n}_{\mathbf{k}}\boldsymbol{\rho} ) 
     + \left(\frac{\partial \mathbf{h}_{\mathbf{k}}}{\partial k_\alpha}  \boldsymbol{\rho} \right) \left(  \mathbf{n}_{\mathbf{k}} \boldsymbol{\rho}  \right) \frac{\partial E_0(\mathbf{k})}{\partial k_\beta} \left(  \delta \mathbf{n}_{\mathbf{k}}\boldsymbol{\rho} \right) 
     \right] \right\} 
     \notag \\
    & - (\beta \leftrightarrow \alpha) \, .
\end{align}
Use the expression for the trace
\begin{align}
   \Tr [(\mathbf{A}\boldsymbol{\rho} )(\mathbf{B}\boldsymbol{\rho})(\mathbf{C}\boldsymbol{\rho})] =2 i(\mathbf{A}\times \mathbf{B} )\cdot \mathbf{C}  
\end{align}
\begin{align}\label{sigma1_19a_2}
     \sigma^{nc_2}_{ex;\alpha\beta} 
      = & 
     \frac{e^2}{ \pi} \Tr \left\{ \delta G^{A}_{\mathbf{k}} \delta G^{R}_{\mathbf{k}}
     \left[\left(\frac{\partial E_0(\mathbf{k})}{\partial k_\beta} \frac{\partial \mathbf{h}_{\mathbf{k}}}{\partial k_\alpha } 
     \right. \right. \right.
     \notag \\
     -  & \left. \left. \left. \frac{\partial E_0(\mathbf{k})}{\partial k_\alpha} \frac{\partial \mathbf{h}_{\mathbf{k}}}{\partial k_\beta }
     \right) \times \mathbf{n} \right]\right\} \cdot \delta \mathbf{n}_{\mathbf{k}} - (\beta \leftrightarrow \alpha)\, .
\end{align}
Applying \eqref{sigma1_26} again we arrive at the result Eq.~\eqref{sigma^I_b}.
\subsection{Asymptotic behavior in the small spin splitting limit}
In the small spin splitting regime $h\ll E_F$ as we have stressed out in the main text $\sigma^{nc_1}_{ex;yz}\ll\sigma^{nc_2}_{ex;yz}$ and we will consider only $\sigma^{nc_2}_{ex;yz}$ of the classical extrinsic conductivity throughout this appendix. We neglect the differences between Fermi spheres, Fermi velocities $v_\pm$ and the corrections due to $\boldsymbol{\gamma}$. For the particular model Eq.\eqref{h_expand1} and with the use of Eq.\eqref{eq:Gammay_y1}, Eq.~\eqref{Gamma_0R} we get the following result for classical extrinsic Hall conductivity:
\begin{equation}\label{Extr}
\hat{\sigma}_{ex;yz}=-\frac{e^{2}}{(2\pi)^{3}}\oint\frac{dS}{v_{F}}\frac{\lambda_{z}\lambda t_A}{4k_{F}^{2}}\frac{k_{x}^{2}\left(k_{y}^{2}+k_{z}^{2}\right)}{\lambda_z^2 + t_A^2 k_x^2 k_y^2+(\lambda k_x k_z/4)^2 }
\end{equation}
In the spherical coordinates this integral has the form:
\begin{multline}\label{Extr1}
    \hat{\sigma}_{ex;yz}=-\frac{\sigma_{0}}{16\pi^{2}}\frac{\lambda_{z}\bar{\lambda}\Delta_{\mathrm{A}}}{E_{F}}\oint\frac{d\Omega}{4\pi}\\\frac{\sin^{2}\theta\cos^{2}\phi\left(1-\sin^{2}\theta\cos^{2}\phi\right)}{\lambda_{z}^{2}+(t_Ak_{F}^{2}\sin^{2}\theta\cos\phi\sin\phi)^{2}+(\lambda k_{F}^{2}\cos\theta\sin\theta\cos\phi/4)^{2}}
\end{multline}
Similarly to the App.\ref{IntrinsicFeSb2} with a proper rotation of the unit sphere we obtain:
\begin{multline}\label{Extr2}
    \hat{\sigma}_{ex;yz} = -\frac{\sigma_{0}}{32\pi^{2}}\frac{\lambda_{z}\bar{\lambda}\Delta_{\mathrm{A}}}{E_{F}}\times\\
    \intop_{0}^{\pi} d\alpha \frac{\sin^{3}\alpha\cos^{2}\alpha}{
    \sqrt{\lambda_{z}^{2}+\left(\Delta_{\mathrm{A}}\sin\alpha\cos\alpha\right)^{2}}
    \sqrt{\lambda_{z}^{2}+\left(\bar{\lambda}\sin\alpha\cos\alpha/4\right)^{2}}}
\end{multline}
Here we present the asymptotic behavior of the Eq.\eqref{Extr2} in the same limits as it was done for $\hat{\sigma}_{in;yz}$ in App.\ref{IntrinsicFeSb2}: 
\begin{equation}\label{Extr3}
    \hat{\sigma}_{ex;yz} = -\frac{\sigma_{0}}{120\pi^{2}}\frac{\bar{\lambda}\Delta_{\mathrm{A}}}{E_{F}\lambda_{z}}
    = -\frac{2}{5}\hat{\sigma}_{in;yz},\ \lambda_{z}\gg\{\bar{\lambda},\Delta_{\mathrm{A}}\}.
\end{equation}
In the limit $\Delta_{\mathrm{A}}\gg\{\lambda_{z},\bar{\lambda}\}$ we get:
\begin{multline}
    \hat{\sigma}_{ex;yz} = -\frac{\sigma_{0}}{32\pi^{2}}\frac{\lambda_{z}\bar{\lambda}\mathrm{sgn}(\Delta_{\mathrm{A}})}{E_{F}}\times\\
    \intop_{0}^{\pi} d\alpha \frac{\sin\alpha\cos^{2}\alpha}{
    \sqrt{\lambda_{z}^{2}+\left(\bar{\lambda}\sin\alpha\cos\alpha/4\right)^{2}}}
\end{multline}
For the limiting cases we obtain:
\begin{equation}\label{Extr4}
    \hat{\sigma}_{ex;yz}=\begin{cases}
    -\frac{\sigma_{0}}{48\pi^{2}}\frac{\bar{\lambda}\mathrm{sgn}(\lambda_{z}\Delta_{\mathrm{A}})}{E_{F}}=-\frac{1}{3}\hat{\sigma}_{in;yz}, & \!\!\Delta_{\mathrm{A}}\gg\lambda_{z}\gg\bar{\lambda}\\
    -\frac{\sigma_{0}}{4\pi^{2}}\frac{\lambda_{z}\mathrm{sgn}(\bar{\lambda}\Delta_{\mathrm{A}})}{E_{F}}\ll\hat{\sigma}_{in,yz}, & \!\!\Delta_{\mathrm{A}}\gg\bar{\lambda}\gg\lambda_{z}
\end{cases}
\end{equation}\label{Extr5}
Finally, for $\bar{\lambda}\gg\{\lambda_{z},\Delta_{\mathrm{A}}\}$ the result has the form:
\begin{multline}
    \hat{\sigma}_{ex;yz} = -\frac{\sigma_{0}}{8\pi^{2}}\frac{\lambda_{z}\Delta_{\mathrm{A}}\mathrm{sgn}(\bar{\lambda})}{E_{F}}\times\\
    \intop_{0}^{\pi} d\alpha \frac{\sin\alpha\cos^{2}\alpha}{
    \sqrt{\lambda_{z}^{2}+\left(\Delta_{\mathrm{A}}\sin\alpha\cos\alpha\right)^{2}}}
\end{multline}
The limiting cases are given by:
\begin{equation}\label{Extr6}
    \hat{\sigma}_{ex;yz}=\begin{cases}
    -\frac{\sigma_{0}}{12\pi^{2}}\frac{\Delta_{\mathrm{A}}\mathrm{sgn}(\lambda_{z}\bar{\lambda})}{E_{F}}=-\frac{1}{3}\hat{\sigma}_{in;yz}, & \!\!\bar{\lambda}\gg\lambda_{z}\gg\Delta_{\mathrm{A}}\\
    -\frac{\sigma_{0}}{4\pi^{2}}\frac{\lambda_{z}\mathrm{sgn}(\bar{\lambda}\Delta_{\mathrm{A}})}{E_{F}}\ll\hat{\sigma}_{in,yz}, & \!\!\bar{\lambda}\gg\Delta_{\mathrm{A}}\gg\lambda_{z}
\end{cases}
\end{equation}

%%%%%%%%%%%%%%%%%%%%%
%%%%%%%%%%%%%%%%%%%%%
%%%%%%%%%%%%%%%%%%%%%
%%%%%%%%%%%%%%%%%%%%%
%\begin{widetext}
\section{Quantum contribution to extrinsic AHC}
\label{app:Cooperon}
\subsubsection{Derivation of the X-diagram}
In this appendix, we provide a detailed derivation of the estimate for the X-diagram shown in Fig.~\ref{fig:diagrams}b.
For the regime of small spin splitting, we assert that $\sigma_{\alpha\beta}^{X} \ll \sigma_{ex;\alpha\beta}$.
The proof is based on analyzing the scaling of this diagram with respect to the spin-splitting parameter $h/E_{F}$.
Within the Kubo–St\v{r}eda formalism, we focus solely on the $\sigma_{\alpha\beta}^{I}$ part contribution to the conductivity in the weak-disorder limit.
Using the expression for $\sigma_{\alpha\beta}^{I}$ given in Eq.~\eqref{eq:sigma^I}, we obtain:
\begin{multline}
\sigma_{\alpha\beta}^{X}=\frac{e^{2}}{4\pi}\left(n_{\mathrm{imp}}U^{2}\right)^{2}\Tr\left[\hat{v}_{\alpha\mathbf{k}}\hat{G}_{\mathbf{k}}^{R}\hat{G}_{\mathbf{q}}^{R}\hat{G}_{\mathbf{p}}^{R}\times\right.\\\left.\hat{v}_{\beta\mathbf{p}}\hat{G}_{\mathbf{p}}^{A}\hat{G}_{\mathbf{Q-q}}^{A}\hat{G}_{\mathbf{k}}^{A}\right]-(\alpha\leftrightarrow\beta),
\end{multline}
here $\mathbf{Q=k+p}$.

We rewrite this expression so that each velocity vertex $\hat{v}_{\alpha,\beta}$ is sandwiched between Green’s functions with the same momentum on both sides:
\begin{multline}
\sigma_{\alpha\beta}^{X}=\frac{e^{2}}{4\pi}\left(n_{\mathrm{imp}}U^{2}\right)^{2}\Tr\left[\left(\hat{G}_{\mathbf{k}}^{A}\hat{v}_{\alpha\mathbf{k}}\hat{G}_{\mathbf{k}}^{R}\right)\hat{G}_{\mathbf{q}}^{R}\times\right.\\\left.\left(\hat{G}_{\mathbf{p}}^{R}\hat{v}_{\beta\mathbf{p}}\hat{G}_{\mathbf{p}}^{A}\right)\hat{G}_{\mathbf{Q-q}}^{A}\right]-(\alpha\leftrightarrow\beta).
\end{multline}

In the weak-disorder limit $U \to 0$, and since $\Gamma_{0} \propto n_{\mathrm{imp}} U^{2}$,
a nonzero contribution to $\sigma_{\alpha\beta}^{X}$ arises only if the expression inside the $\Tr$ scales at least as $1/\Gamma_{0}^{2}$.
From Eq.~\eqref{sigma1_26}, the vertices provide exactly this power of $\Gamma_{0}$, while, as we will show below, the leading-order contributions from the integrals over $\mathbf{q}$ are independent of disorder.
This allows us to use the disordered Green’s function in the form of Eq.~\eqref{G13_dis}, retaining the pole structure given in Eq.~\eqref{G13a_dis}, and to neglect the distinction between retarded and advanced projectors in Eq.~\eqref{G14_dis}.
Corrections arising from the difference between retarded and advanced projectors are proportional to $\Gamma_{0}$ and therefore vanish in the weak-disorder limit.
Thus, employing Eq.~\eqref{P+-}, the combinations of Green’s functions around the vertices that yield a finite contribution to $\sigma_{\alpha\beta}^{X}$ take the form:

\begin{multline}
\hat{G}_{\mathbf{k}}^{A}\hat{v}_{\alpha\mathbf{k}}\hat{G}_{\mathbf{k}}^{R}=\sum_{f_{1,2}=\pm}G_{f_{1}\mathbf{k}}^{A}\left|f_{1}\mathbf{k}\right\rangle \left\langle f_{1}\mathbf{k}\right|\hat{v}_{\alpha\mathbf{k}}\times\\ G_{f_{2}\mathbf{k}}^{R}\left|f_{2}\mathbf{k}\right\rangle \left\langle f_{2}\mathbf{k}\right|.
\end{multline}

According to Eq.~\eqref{sigma1_26}, after integrating over $\mathbf{k}$, only the combinations of Green’s functions with $f_{1} = f_{2}$ produce a $1/\Gamma_{0}$ factor.
Therefore, for our calculations we retain only the relevant terms that give this leading contribution, discarding all others:
\begin{equation}
\hat{G}_{\mathbf{k}}^{A}\hat{v}_{\alpha\mathbf{k}}\hat{G}_{\mathbf{k}}^{R}=\sum_{f=\pm}\left\langle f\mathbf{k}\right|\hat{v}_{\alpha\mathbf{k}}\left|f\mathbf{k}\right\rangle G_{f\mathbf{k}}^{A}G_{f\mathbf{k}}^{R}\left|f\mathbf{k}\right\rangle \left\langle f\mathbf{k}\right|,
\end{equation}

Since $\hat{v}_{\alpha\mathbf{k}} = \partial \hat{H}_{\mathbf{k}} / \partial k_{\alpha}$,
the Feynman–Hellmann theorem then gives:

\begin{gather}
\hat{G}_{\mathbf{k}}^{A}\hat{v}_{\alpha\mathbf{k}}\hat{G}_{\mathbf{k}}^{R}=\sum_{f=\pm}T_{f\mathbf{k}}^{(\alpha)}\left|f\mathbf{k}\right\rangle \left\langle f\mathbf{k}\right|,\\
T_{f\mathbf{k}}^{(\alpha)}=\frac{\partial E_{f\mathbf{k}}}{\partial k_{\alpha}}G_{f\mathbf{k}}^{A}G_{f\mathbf{k}}^{R},
\end{gather}

Therefore, in the weak-disorder limit, the X diagram is expressed as:
\begin{multline}
\sigma_{\alpha\beta}^{X}=\frac{e^{2}}{4\pi}\left(n_{\mathrm{imp}}U^{2}\right)^{2}\sum_{f_{1,2}=\pm}T_{f_{1}\mathbf{k}}^{(\alpha)}T_{f_{2}\mathbf{p}}^{(\beta)}\times\\\Tr\left[\left|f_{1}\mathbf{k}\right\rangle \left\langle f_{1}\mathbf{k}\right|\hat{G}_{\mathbf{q}}^{R}\left|f_{2}\mathbf{p}\right\rangle \left\langle f_{2}\mathbf{p}\right|\hat{G}_{\mathbf{Q-q}}^{A}\right]-(\alpha\leftrightarrow\beta).\label{eq:SigmaXGqQ}
\end{multline}

Unlike the case of the vertices, it is convenient to use the representation of the Green function in the $\mathbf{q}$ sector of the X diagram given by Eq.~\eqref{decomp}, together with the identity:

\begin{equation}\label{eq:projector}
\left|f\mathbf{k}\right\rangle \left\langle f\mathbf{k}\right|=\frac{1}{2}\left(\rho_{0}+f\mathbf{n}_{\mathbf{k}}\boldsymbol{\rho}\right).
\end{equation}

We analyze separately the different contributions to $\sigma_{\alpha\beta}^{X}$ arising from the number of $\delta G_{\mathbf{q}}$ or $\delta G_{\mathbf{Q-q}}$ terms in Eq.~\eqref{eq:SigmaXGqQ}, beginning with the zeroth order:

\begin{multline}
\sigma_{\alpha\beta}^{X,0}=\frac{e^{2}}{4\pi}\left(n_{\mathrm{imp}}U^{2}\right)^{2}\sum_{f_{1,2}=\pm}T_{f_{1}\mathbf{k}}^{(\alpha)}T_{f_{2}\mathbf{p}}^{(\beta)}\left\{\right.\\\left.\bar{G}_{\mathbf{q}}^{R}\bar{G}_{\mathbf{Q-q}}^{A}\Tr\left[\left|f_{1}\mathbf{k}\right\rangle \left\langle f_{1}\mathbf{k}\left|f_{2}\mathbf{p}\right.\right\rangle \left\langle f_{2}\mathbf{p}\right|\right]\right\}-(\alpha\leftrightarrow\beta).
\end{multline}

Under the change of variables $\mathbf{k} \leftrightarrow \mathbf{p}$ and $f_{1} \leftrightarrow f_{2}$, the first term maps onto itself with $\alpha \leftrightarrow \beta$,
and hence $\sigma_{\alpha\beta}^{X,0} = 0$.

The next-order contribution takes the form:
\begin{multline}
\sigma_{\alpha\beta}^{X,1}=\frac{e^{2}}{4\pi}\left(n_{\mathrm{imp}}U^{2}\right)^{2}\sum_{f_{1,2}=\pm}T_{f_{1}\mathbf{k}}^{(\alpha)}T_{f_{2}\mathbf{p}}^{(\beta)}\left\{\right.\\  \bar{G}_{\mathbf{Q-q}}^{A}\delta G_{\mathbf{q}}^{R}\Tr\left[\left|f_{1}\mathbf{k}\right\rangle \left\langle f_{1}\mathbf{k}\right|\mathbf{n}_{\mathbf{q}}\boldsymbol{\rho}\left|f_{2}\mathbf{p}\right\rangle \left\langle f_{2}\mathbf{p}\right|\right]+\\
+\left.\bar{G}_{\mathbf{q}}^{R}\delta G_{\mathbf{Q-q}}^{A}\Tr\left[\left|f_{1}\mathbf{k}\right\rangle \left\langle f_{1}\mathbf{k}\left|f_{2}\mathbf{p}\right.\right\rangle \left\langle f_{2}\mathbf{p}\right|\mathbf{n}_{\mathbf{Q-q}}\boldsymbol{\rho}\right]\right\}\\-(\alpha\leftrightarrow\beta).
\end{multline}

Subtracting the term with $\alpha \leftrightarrow \beta$ is equivalent to performing the subtraction inside the trace of the terms with $\mathbf{k} \leftrightarrow \mathbf{p}$ and $f_{1} \leftrightarrow f_{2}$, while keeping the other factors unchanged.
This clearly corresponds to subtracting a Hermitian-conjugated matrix product.
By changing variables $\mathbf{q} \to \mathbf{Q-q}$, one sees that the relative sign between $\bar{G}_{\mathbf{Q-q}}^{A}\delta G_{\mathbf{q}}^{R}$ and $\bar{G}_{\mathbf{Q-q}}^{R}\delta G_{\mathbf{q}}^{A}$ is negative, yielding:

\begin{multline}
\sigma_{\alpha\beta}^{X,1}=\frac{e^{2}}{4\pi}\left(n_{\mathrm{imp}}U^{2}\right)^{2}\sum_{f_{1,2}=\pm}T_{f_{1}\mathbf{k}}^{(\alpha)}T_{f_{2}\mathbf{p}}^{(\beta)}\times\\\left(\bar{G}_{\mathbf{Q-q}}^{A}\delta G_{\mathbf{q}}^{R}-\bar{G}_{\mathbf{Q-q}}^{R}\delta G_{\mathbf{q}}^{A}\right)\times\\
\Tr\left[\left|f_{1}\mathbf{k}\right\rangle \left\langle f_{1}\mathbf{k}\right|\mathbf{n}_{\mathbf{q}}\boldsymbol{\rho}\left|f_{2}\mathbf{p}\right\rangle \left\langle f_{2}\mathbf{p}\right|-h.c.\right].\label{eq:SigmaX1Tr}
\end{multline}

For the subsequent calculation, we make use of the following identities:
\begin{gather}
\Tr\left[\rho_{i}\rho_{j}\right]=2\delta_{ij},\label{eq:2rho}\\
\Tr\left[\rho_{i}\rho_{j}\rho_{l}\right]=2i\varepsilon_{ijl},\label{eq:3rho}\\
\Tr\left[\rho_{i}\rho_{j}\rho_{l}\rho_{k}\right]=2\left(\delta_{ij}\delta_{lk}-\delta_{il}\delta_{jk}+\delta_{ik}\delta_{jl}\right),\label{eq:4rho}
\end{gather}

The trace in Eq.~\eqref{eq:SigmaX1Tr} can be evaluated using Eqs.~(\ref{eq:projector}, \ref{eq:2rho}--\ref{eq:4rho}):
\begin{multline}
\Tr\left[\left|f_{1}\mathbf{k}\right\rangle \left\langle f_{1}\mathbf{k}\left|f_{2}\mathbf{p}\right.\right\rangle \left\langle f_{2}\mathbf{p}\right|\mathbf{n}_{\mathbf{q}}\boldsymbol{\rho}-h.c.\right]=\\if_{1}f_{2}\left(\mathbf{n}_{\mathbf{k}},\mathbf{n}_{\mathbf{p}},\mathbf{n}_{\mathbf{q}}\right).
\end{multline}

This leads to:
\begin{multline}
\sigma_{\alpha\beta}^{X,1} =\frac{e^{2}}{4\pi}\left(n_{\mathrm{imp}}U^{2}\right)^{2}\sum_{f_{1,2}=\pm}if_{1}f_{2}T_{f_{1}\mathbf{k}}^{(\alpha)}T_{f_{2}\mathbf{p}}^{(\beta)}\times\\\left(\bar{G}_{\mathbf{Q-q}}^{A}\delta G_{\mathbf{q}}^{R}-\bar{G}_{\mathbf{Q-q}}^{R}\delta G_{\mathbf{q}}^{A}\right)\left(\mathbf{n}_{\mathbf{p}},\mathbf{n}_{\mathbf{k}},\mathbf{n}_{\mathbf{q}}\right).
\end{multline}

The final contribution, $\sigma_{\alpha\beta}^{X,2}$, is given by:
\begin{multline}
\sigma_{\alpha\beta}^{X,2}=\frac{e^{2}}{4\pi}\left(n_{\mathrm{imp}}U^{2}\right)^{2}\sum_{f_{1,2}=\pm}T_{f_{1}\mathbf{k}}^{(\alpha)}T_{f_{2}\mathbf{p}}^{(\beta)}\left\{\delta G_{\mathbf{q}}^{R}\delta G_{\mathbf{Q-q}}^{A}\times\right.\\\ \left.\Tr\left[\left|f_{1}\mathbf{k}\right\rangle \left\langle f_{1}\mathbf{k}\right|\mathbf{n}_{\mathbf{q}}\boldsymbol{\rho}\left|f_{2}\mathbf{p}\right\rangle \left\langle f_{2}\mathbf{p}\right|\mathbf{n}_{\mathbf{Q-q}}\boldsymbol{\rho}\right]\right\}-\\(\alpha\leftrightarrow\beta).\label{eq:SigmaXGqQ-1}
\end{multline}

Subtracting the term with $\alpha \leftrightarrow \beta$ is equivalent to performing the subtraction inside the trace of the terms with $\mathbf{k} \leftrightarrow \mathbf{p}$ and $f_{1} \leftrightarrow f_{2}$ while keeping the other factors unchanged.
This, in turn, corresponds to the change of variables $\mathbf{q} \to \mathbf{Q-q}$ within the trace:
\begin{multline}
\sigma_{\alpha\beta}^{X,2} =\frac{e^{2}}{4\pi}\left(n_{\mathrm{imp}}U^{2}\right)^{2}\sum_{f_{1,2}=\pm}T_{f_{1}\mathbf{k}}^{(\alpha)}T_{f_{2}\mathbf{p}}^{(\beta)}\delta G_{\mathbf{q}}^{R}\delta G_{\mathbf{Q-q}}^{A}\times\\\Tr\left[\left|f_{1}\mathbf{k}\right\rangle \left\langle f_{1}\mathbf{k}\right|\mathbf{n}_{\mathbf{q}}\boldsymbol{\rho}\left|f_{2}\mathbf{p}\right\rangle \left\langle f_{2}\mathbf{p}\right|\mathbf{n}_{\mathbf{Q-q}}\boldsymbol{\rho}-(\mathbf{q}\to\mathbf{Q-q})\right].
\end{multline}

Due to Eqs.~(\ref{eq:projector}, \ref{eq:2rho}--\ref{eq:4rho}), combinations containing an even number of $\hat{\rho}$ matrices form symmetric tensors and thus cancel each other.
Consequently, only terms with three $\hat{\rho}$ matrices contribute, and using Eq.~\eqref{eq:3rho}, we obtain:
\begin{multline}
\Tr\left[\left|f_{1}\mathbf{k}\right\rangle \left\langle f_{1}\mathbf{k}\right|\mathbf{n}_{\mathbf{q}}\boldsymbol{\rho}\left|f_{2}\mathbf{p}\right\rangle \left\langle f_{2}\mathbf{p}\right|\mathbf{n}_{\mathbf{Q-q}}\boldsymbol{\rho}-(\mathbf{q}\to\mathbf{Q-q})\right]=\\i\left(f_{1}\mathbf{n}_{\mathbf{k}}-f_{2}\mathbf{n}_{\mathbf{p}},\mathbf{n}_{\mathbf{q}},\mathbf{n}_{\mathbf{Q-q}}\right).
\end{multline}

Thus, for $\sigma_{\alpha\beta}^{X,2}$, we get:
\begin{multline}
\sigma_{\alpha\beta}^{X,2}=\frac{e^{2}}{4\pi}\left(n_{\mathrm{imp}}U^{2}\right)^{2}\sum_{f_{1,2}=\pm}iT_{f_{1}\mathbf{k}}^{(\alpha)}T_{f_{2}\mathbf{p}}^{(\beta)}\times\\\delta G_{\mathbf{q}}^{R}\delta G_{\mathbf{Q-q}}^{A}\left(f_{1}\mathbf{n}_{\mathbf{k}}-f_{2}\mathbf{n}_{\mathbf{p}},\mathbf{n}_{\mathbf{q}},\mathbf{n}_{\mathbf{Q-q}}\right).\label{eq:SigmaX2Im}
\end{multline}

To verify that the result in Eq.~\eqref{eq:SigmaX2Im} is real, one can add to it the same expression with the substitution $\mathbf{q} \to \mathbf{Q-q}$ and divide by two.
Hence:
\begin{multline}
\sigma_{\alpha\beta}^{X,2}=\frac{e^{2}}{4\pi}\left(n_{\mathrm{imp}}U^{2}\right)^{2}\sum_{f_{1,2}=\pm}\frac{i}{2}T_{f_{1}\mathbf{k}}^{(\alpha)}T_{f_{2}\mathbf{p}}^{(\beta)}\times\\\left(\delta G_{\mathbf{q}}^{R}\delta G_{\mathbf{Q-q}}^{A}-\delta G_{\mathbf{Q-q}}^{R}\delta G_{\mathbf{q}}^{A}\right)\times\\\left(f_{1}\mathbf{n}_{\mathbf{k}}-f_{2}\mathbf{n}_{\mathbf{p}},\mathbf{n}_{\mathbf{q}},\mathbf{n}_{\mathbf{Q-q}}\right).
\end{multline}

The complete expression for $\sigma_{\alpha\beta}^{X}$ is given by:
\begin{multline}
\sigma_{\alpha\beta}^{X}=\frac{e^{2}}{4\pi}\left(n_{\mathrm{imp}}U^{2}\right)^{2}\int\frac{d^{3}k}{(2\pi)^{3}}\frac{d^{3}p}{(2\pi)^{3}}\frac{d^{3}q}{(2\pi)^{3}}\sum_{f_{1,2}=\pm}i\times\\T_{f_{1}\mathbf{k}}^{(\alpha)}T_{f_{2}\mathbf{p}}^{(\beta)}\{f_{1}f_{2}\left(\bar{G}_{\mathbf{Q-q}}^{A}\delta G_{\mathbf{q}}^{R}-\bar{G}_{\mathbf{Q-q}}^{R}\delta G_{\mathbf{q}}^{A}\right)\times\\\left(\mathbf{n}_{\mathbf{p}},\mathbf{n}_{\mathbf{k}},\mathbf{n}_{\mathbf{q}}\right)
+\frac{1}{2}\left(\delta G_{\mathbf{q}}^{R}\delta G_{\mathbf{Q-q}}^{A}-\delta G_{\mathbf{Q-q}}^{R}\delta G_{\mathbf{q}}^{A}\right)\times\\\left(f_{1}\mathbf{n}_{\mathbf{k}}-f_{2}\mathbf{n}_{\mathbf{p}},\mathbf{n}_{\mathbf{q}},\mathbf{n}_{\mathbf{Q-q}}\right)\}.\label{eq:SigmaXAnalytical}
\end{multline}

Up to this point, the calculation of the X-diagram has been exact.
We now introduce the notation near the Fermi surface, $h_{i}(\mathbf{k}) = H_{i} s_{i}(\hat{\mathbf{k}})$,
where $H_{i}$ sets the energy scale of the $i$-th component of $\mathbf{h}$, and $s_{i}(\hat{\mathbf{k}})$ describes its dependence on the unit vector $\hat{\mathbf{k}}$.
In particular, for a class A material, see Eqs.~(\ref{eq:tx_tz}--\ref{eq:SOC_A}), we have $H_{x} = \bar{\lambda}$, $H_{y} = \lambda_{z}$, and $H_{z} = \Delta_{A}$.
We note that Eq.~\eqref{eq:SigmaXAnalytical} contains three powers of the spin-splitting parameter $h/E_{F} \ll 1$, leading to the following estimate:

\begin{equation}
\sigma_{\alpha\beta}^{X}\sim\sigma_{0}\frac{H_{x}H_{y}H_{z}}{E_{F}^{3}}.
\end{equation}
\subsubsection{Vertex integral estimate}
We start with the analysis of the integrals over $\mathbf{k}$ and $\mathbf{p}$.
Without loss of generality, consider the integral over $\mathbf{k}$ in the form:
\begin{equation}
\mathcal{V}_{1}=\sum_{f=\pm}f\int\frac{d^{3}k}{(2\pi)^{3}}T_{f\mathbf{k}}^{(\alpha)}n_{i\mathbf{k}}F(\mathbf{k}),
\end{equation}
where $F(\mathbf{k})$ represents the part of the integral in Eq.~\eqref{eq:SigmaXAnalytical} that depends on the vector $\mathbf{Q} = \mathbf{p} + \mathbf{k}$, and $i$ denotes one of the components of the vector $\mathbf{n_{k}}$ appearing in either $\left(\mathbf{n}_{\mathbf{p}},\mathbf{n}_{\mathbf{k}},\mathbf{n}_{\mathbf{q}}\right)$ or $\left(\mathbf{n}_{\mathbf{k}},\mathbf{n}_{\mathbf{q}},\mathbf{n}_{\mathbf{Q-q}}\right)$.

Using Eq.~\eqref{sigma1_26}, we express $\mathcal{V}$ as a surface integral:
\begin{equation}
\mathcal{V}_{1}=\sum_{f=\pm}\frac{1}{8\pi^{2}}f\oint_{\text{FS}^{f}}\frac{dS_{f}}{v_{f}\Gamma_{f}}\frac{\partial E_{f}}{\partial k_{\alpha}}n_{i\mathbf{k}_{f}}F\left(\mathbf{k}_{f}\right).\label{eq:VertexIntegralSurface}
\end{equation}

We express both integrals in spherical coordinates, $\Omega = (\phi, \theta)$ (see Appendix~\ref{app:self} for definitions) which gives:

\begin{equation}
\mathcal{V}_{1}=\sum_{f=\pm}f\oint\frac{d\Omega}{8\pi^2}\frac{g_{f}}{\Gamma_{f}}\frac{\partial E_{f}}{\partial k_{\alpha}}n_{i\mathbf{k}_{f}}F\left(\mathbf{k}_{f}\right).\label{eq:VertexIntegralSphere}
\end{equation}

For small spin splitting, the Fermi surfaces of the $+$ and $-$ bands are nearly identical.
Hence, to leading order in $h/E_{F} \ll 1$, $\mathcal{V}_{1}$ vanishes.
To obtain a nonzero result, all quantities in Eq.~\eqref{eq:VertexIntegralSurface} must be expanded to first order in the spin-splitting parameter $h/E_{F}$.
This gives:
\begin{equation}
k_{f}(\Omega)=k_{F}(\Omega)-f\frac{h_{k_{F}}(\Omega)}{\partial E_{0,k_{F}}(\Omega)/\partial k},\label{eq:Expansions1}
\end{equation}
\begin{multline}
\frac{\partial E_{f}(k_{f})}{\partial k_{\alpha}}=\frac{\partial E_{0,k_{F}}}{\partial k_{\alpha}}+f\frac{\partial h_{k_{F}}}{\partial k_{\alpha}}-\\f\left.\frac{\partial}{\partial k}\left(\frac{\partial E_{0,k}}{\partial k_{\alpha}}\right)\right|_{k_{F}}\frac{h_{k_{F}}(\Omega)}{\partial E_{0,k_{F}}(\Omega)/\partial k},\label{eq:Expansions2}
\end{multline}
\begin{multline}
\frac{g_{f}(\Omega)}{h_{k_{f}}(\Omega)}=\frac{k_{F}^{2}(\Omega)}{h_{k_{F}}(\Omega)\partial E_{0,k_{F}}(\Omega)/\partial k}\\\left(1-fh_{k_{F}}(\Omega)\frac{k_{F}(\Omega)\frac{\partial^{2}E_{0,k_{F}}(\Omega)}{\partial k^{2}}-2\frac{\partial E_{0,k_{F}}(\Omega)}{\partial k}}{k_{F}(\Omega)\left[\partial E_{0,k_{F}}(\Omega)/\partial k\right]^{2}}\right),\label{eq:Expansions3}
\end{multline}
\begin{equation}
\frac{1}{\Gamma_{f}}=\frac{1}{\Gamma_{0}}\left(1-f\frac{\mathbf{n}_{\mathbf{k}}\mathbf{\Gamma}}{\Gamma_{0}}\right).\label{eq:Expansions4}
\end{equation}

Let us now illustrate the order of magnitude of the result in the simplest case of a spherical Fermi surface with $F(\mathbf{k}) \approx \text{const}$.
For a nearly spherical Fermi surface, the following estimates can be made: each term containing $E_{0,k}$ can be approximated by $E_{F}$, derivatives scale as $\partial/\partial k \sim 1/k_{F}$, and the density of states is $g_{F} \sim k_{F}^{3}/E_{F}$.
Since the nonzero contribution arises at first order in the spin-splitting parameter $h_{k_{F}}/E_{F}$, we obtain:
\begin{equation}
\mathcal{V}_{1}\sim\frac{k_{F}^{3}/E_{F}}{\Gamma_{0}}\frac{E_{F}}{k_{F}}\frac{H_{i}}{h_{k_{F}}}\frac{h_{k_{F}}}{E_{F}}\sim\frac{H_{i}k_{F}^{2}}{\Gamma_{0}E_{F}}.\label{eq:Vestimate}
\end{equation}

We now discuss in more detail the validity of this estimate.
As mentioned earlier (see Sec.~\ref{sec:sigma_in_J=infty}), the main contribution to the conductivity comes from regions where $h \approx\underset{\text{FS}}{\min}\ h$.
In Eqs.~(\ref{eq:Expansions1}--\ref{eq:Expansions4}), nearly all correction terms are proportional to $h_{k_{F}}(\Omega)$.
However, the terms $f\partial h/\partial k_{\alpha}$ in Eq.~\eqref{eq:Expansions2} and $f\mathbf{n_{k}}\mathbf{\Gamma}/\Gamma_{0}$ are of a different type and therefore require separate consideration.

We begin with the term $\partial h / \partial k_{\alpha}$.
To obtain a contribution larger than the estimate in Eq.~\eqref{eq:Vestimate}, one must consider a situation where one of the energy scales—without loss of generality, $H_{y}$ --- is much larger than the other two, and the corresponding component $h_{y}$ vanishes (see, for example, Eq.~\eqref{h_expand1a} with $\lambda_{z} \gg \bar\lambda, 
\Delta_\mathrm{A}$).

For each component of $\mathbf{h}$ in the $\mathbf{k} \cdot \mathbf{p}$ approximation, the region where it vanishes is a plane. In its vicinity, $h_{y}(\mathbf{k}) \propto (\mathbf{a} \cdot \hat{\mathbf{k}} + b) , \Phi(\mathbf{k})$,
where $\Phi(\mathbf{k}) \neq 0$ on the plane, and $\mathbf{a}$ is a vector perpendicular to the plane with $|\mathbf{a}| < |b|$.

The most singular behavior in Eq.~\eqref{eq:VertexIntegralSphere} arises when all other factors in the integrand do not vanish on the plane $\mathbf{a} \cdot \hat{\mathbf{k}} + b = 0$.
Thus, the leading singular contribution comes from integrals of the form:
\begin{equation}
\oint d\Omega\frac{\partial h/\partial k_{\alpha}}{h}\sim\text{v.p.}\oint_{\mathcal{R}}d\Omega\frac{a_{\alpha}}{\mathbf{\mathbf{a}}\cdot\hat{\mathbf{k}}+b}+O(1),\label{eq:SingularInt}
\end{equation}
where $\mathcal{R}$ denotes a ring on the sphere of small thickness $\epsilon$, formed by the intersection of the plane $\mathbf{a} \cdot \hat{\mathbf{k}} + b = 0$ with the sphere.
The principal value of the integral comes from neglecting the $h_{x,z}$ components in $h = \sqrt{h_{x}^{2} + h_{y}^{2} + h_{z}^{2}}$.
In our approximation, $H_{x,z} / H_{y} \ll \epsilon \ll 1$.
By an appropriate rotation of the sphere, the first integral in Eq.~\eqref{eq:SingularInt} becomes:
\begin{multline}
\oint_{\mathcal{R}}d\Omega\frac{a_{\alpha}}{\mathbf{\mathbf{a}}\cdot\hat{\mathbf{k}}+b}=\frac{a_{\alpha}}{|\mathbf{a}|}\text{v.p.}\int_{\pi-\arccos(b/|\mathbf{a}|)-\epsilon}^{\pi-\arccos(b/|\mathbf{a}|)+\epsilon}d\theta\times\\\intop_{0}^{2\pi}d\varphi\frac{\sin\theta}{\cos\theta+b/|\mathbf{a}|}<\infty.\label{eq:dh/hestim}
\end{multline}
One can see that, although this integral exhibits a logarithmic divergence near $\theta = \pi - \arccos(b / |\mathbf{a}|)$, the singularities cancel because the denominator $\cos\theta + b / |\mathbf{a}|$ changes sign.
This argument rules out any logarithmic factors of the form $\ln(H_{x,z} / H_{y})$ in Eq.~\eqref{eq:Vestimate} arising from the nodal planes of the spin–orbit interaction and altermagnetism.

The same analysis applies to the term $f\mathbf{n}_{\mathbf{k}}\mathbf{\Gamma}/\Gamma_{0}$ in Eq.~\eqref{eq:Expansions4}.
Our estimate in Eq.~\eqref{eq:Vestimate} relies on the approximation $\mathbf{n}_{\mathbf{k}}\mathbf{\Gamma}/h\Gamma_{0}\sim1/E_{F}$, which we show remains valid even when $h$ varies significantly across the Fermi sphere, as discussed above.

The dominant contribution to the integral in Eq.~\eqref{eq:VertexIntegralSphere} comes from the same scenario considered for the term $f\partial h/\partial k_{\alpha}$.
Thus, again assuming $H_{y} \gg H_{x,z}$ and $h_{y} \propto (\mathbf{a} \cdot \hat{\mathbf{k}} + b) \Phi(\mathbf{k})$ near the nodal plane, and applying the same $h/E_{F} \ll 1$ approximation as in the derivation of Eq.~\eqref{eq:Gammay_y1}, we obtain:

\begin{equation}
\oint\frac{d\Omega}{4\pi}\frac{\mathbf{n}_{\mathbf{k}}\mathbf{\Gamma}}{h\Gamma_{0}}\sim\frac{1}{E_{F}}\text{v.p.}\oint_{\mathcal{R}}d\Omega\cdot\frac{1}{\mathbf{\mathbf{a}}\cdot\hat{\mathbf{k}}+b}\sim\frac{1}{E_{F}}.\label{eq:nkGammaestim}
\end{equation}

This confirms that the estimate in Eq.~\eqref{eq:Vestimate} remains valid even in the regions where $h$ varies significantly across the Fermi surface.

We also provide here the estimate for a vertex of the form:
\begin{equation}
\mathcal{V}_{2}=\sum_{f=\pm}\int\frac{d^{3}k}{(2\pi)^{3}}T_{f\mathbf{k}}^{(\alpha)}.\label{eq:V2}
\end{equation}

Due to Eq.~\eqref{sigma1_26}, we get:
\begin{multline}
\mathcal{V}_{2}=\sum_{f=\pm}\oint\frac{d\Omega}{8\pi^2}\frac{g_{f}}{\Gamma_{f}}\frac{\partial E_{f}}{\partial k_{\alpha}}\sim\frac{k_{F}^{3}}{\Gamma_{0}E_{F}}\frac{E_{F}}{k_{F}}=\frac{k_{F}^{2}}{\Gamma_{0}}.\label{eq:V2estim}
\end{multline}
\subsubsection{The estimate of the integral over $\mathbf{q}$}
We now turn to the estimate of the integral over $\mathbf{q}$.
To perform this estimate in the small spin-splitting limit, we use the following approximation:
\begin{gather}
\delta G_{\mathbf{q}}^{R(A)}\approx h_{\mathbf{q}}\left(\mathcal{G}_{\mathbf{q}}^{R(A)}\right)^{2},\\ \mathcal{G}_{\mathbf{q}}^{R(A)}=\frac{1}{E_{F}-E_{0}\left(\mathbf{q}\right)\pm i0}.\label{eq:deltaGApprox}
\end{gather}

The main contribution to the integral over $\mathbf{q}$ comes from the points where the denominator of the Green function is small.
We provide the following arguments to justify the expansion in Eq.~\eqref{eq:deltaGApprox}.

First, our theory is defined near the $\Gamma$-point, where $E_{0}(\mathbf{q})$ can be expanded in even powers of $\mathbf{q}$ due to inversion symmetry. We restrict ourselves to the quadratic term. This argument also applies to anharmonicities in $E_{0}(\mathbf{q})$, which lead to corrections proportional to $h_{\mathbf{q}}$, since the $h_{\mathbf{q}}$-independent corrections cancel in $\delta G_{\mathbf{q}}$.

Second, in the quadratic spectrum approximation, $E_{0}(\mathbf{q})\approx(m^{-1})_{ij}q_{i}q_{j}$ is a bilinear form in $\mathbf{q}$. We can perform a linear transformation of the variables $\mathbf{p}, \mathbf{q}, \mathbf{k}$ to diagonalize $E_{0}(\mathbf{q}) = \mathbf{q}^{2} / 2m$. The Jacobian of this transformation is constant and therefore acts as a multiplicative factor in the integral in Eq.~\eqref{eq:SigmaXAnalytical}.

Finally, it is sufficient to treat $h_{\mathbf{q}}$ as constant. Indeed, if the integral becomes singular for some spherical angle $\Omega$ due to $h_{\mathbf{q}}$, the spherical symmetry only amplifies this singularity. Since the integral receives the dominant contribution around $q \approx k_{F}$, we can approximate $h_{\mathbf{q}} = h(|\mathbf{q}|) \approx h(k_{F})$.

Consider the integral of the form:
\begin{multline}
\mathcal{J}\left(\mathbf{Q},E_{F_{1}},E_{F_{2}}\right)=\\\int\frac{d^{3}q}{\left(2\pi\right)^{3}}\frac{1}{E_{F_{1}}-\frac{\mathbf{q}^{2}}{2m}+i0}\frac{1}{E_{F_{2}}-\frac{(\mathbf{Q}-\mathbf{q})^{2}}{2m}-i0}.\label{eq:JIntegral}
\end{multline}

This integral can be evaluated exactly and is given by:
\begin{equation}
\mathcal{J}=\frac{m^{2}}{2\pi Q}\left[\pi\Theta(Q-|\delta k|)+i\ln\left|\frac{Q+\delta k}{Q-\delta k}\right|\right],\label{eq:JIntegralAns}
\end{equation}
here $\delta k=k_{F_{1}}-k_{F_{2}}$ and $\Theta$ is the Heaviside function.

Since, for the typical values of $\mathbf{p}$ and $\mathbf{q}$ in the integral, $Q \sim k_{F}$, the corresponding values of $\mathcal{J}$ can be approximated as:
\begin{equation}
\mathcal{J}\left(\mathbf{Q},E_{F},E_{F}\right)\sim k_{F}^{3}/E_{F}^{2}.\label{eq:Jestim}
\end{equation}

We can relate an integral containing $\delta G_{\mathbf{q}}$ in Eq.~\eqref{eq:SigmaXAnalytical} to $\mathcal{J}$ in the following way:
\begin{multline}
\int\frac{d^{3}q}{\left(2\pi\right)^{3}}\delta G_{\mathbf{q}}^{R}\mathcal{G}_{\mathbf{Q-q}}^{R/A}\approx\\\int\frac{d^{3}q}{\left(2\pi\right)^{3}}\frac{h}{\left(E_{F}-\frac{\mathbf{q}^{2}}{2m}+i0\right)^{2}}\frac{1}{E_{F}-\frac{(\mathbf{Q-q})^{2}}{2m}+i0}=\\-h\left.\frac{\partial\mathcal{J}}{\partial E_{F_{1}}}\right|_{E_{F_{1}}=E_{F_{2}}}.\label{eq:Jder}
\end{multline}

Since $\mathcal{J}$ is convergent, its derivative is well-defined.
To justify the approximation in Eq.~\eqref{eq:deltaGApprox}, we need to show that the integral over $\mathbf{p}$ and $\mathbf{k}$ in Eq.~\eqref{eq:SigmaXAnalytical} is convergent. Let us change variables to $\mathbf{Q} = \mathbf{p} + \mathbf{k}$.

The integral over $\mathbf{Q}$ has potential singularities at $Q = 0$ and $Q = \infty$. The point $Q = \infty$ does not contribute due to Eq.~\eqref{sigma1_26}, which restricts $Q < 2 k_{F}$. At $Q = 0$, the first and second derivatives of $\mathcal{J}$ behave as $1/Q^{2}$. However, the integration measure $Q^{2} dQ$ and the fact that, at $Q = 0$ and due to inversion symmetry, $\mathbf{n}_{\mathbf{k}}=\mathbf{n_{\mathbf{p}}}$ and $\mathbf{n}_{\mathbf{q}}=\mathbf{n}_{\mathbf{Q-q}}$ in Eq.~\eqref{eq:SigmaXAnalytical} ensure that the contribution from this region to the full answer for $\sigma_{\alpha\beta}^{X}$ is small and can be neglected. Therefore, the expansion in Eq.~\eqref{eq:deltaGApprox} holds.

Due to Eq.~\eqref{eq:Jestim} and Eq.~\eqref{eq:Jder}, we can estimate the integrals containing $\delta G_{\mathbf{q}}$ as:
\begin{gather}
d^{3}q\delta G_{\mathbf{q}}^{R}\mathcal{G}_{\mathbf{Q-q}}^{R/A}\sim\frac{h_{\mathbf{q}}k_{F}^{3}}{E_{F}^{3}},\\
d^{3}q\delta G_{\mathbf{q}}^{R}\delta G_{\mathbf{Q-q}}^{R/A}\sim\frac{h_{\mathbf{q}}h_{\mathbf{Q-q}}k_{F}^{3}}{E_{F}^{4}}.\label{eq:deltaGestim}
\end{gather}

We are now able to perform the full estimate of the X-diagram expression in Eq.~\eqref{eq:SigmaXAnalytical}.
Since $\left(\mathbf{n}_{\mathbf{k_{1}}},\mathbf{n}_{\mathbf{k_{2}}},\mathbf{n}_{\mathbf{k_{3}}}\right)\sim H_{x}H_{y}H_{z}/h_{\mathbf{k_{1}}}h_{\mathbf{k}_{2}}h_{\mathbf{k_{3}}}$, using Eqs.~\eqref{eq:deltaGestim}, \eqref{eq:Vestimate}, and \eqref{eq:V2estim}, we obtain:
\begin{multline}
\sigma_{\alpha\beta}^{X}\sim e^{2}\left(n_{\mathrm{imp}}U^{2}\right)^{2}\frac{H_{x}k_{F}^{2}}{\Gamma_{0}E_{F}}\frac{H_{y}k_{F}^{2}}{\Gamma_{0}E_{F}}\frac{H_{z}k_{F}^{3}}{E_{F}^{3}}+\\e^{2}\left(n_{\mathrm{imp}}U^{2}\right)^{2}\frac{H_{x}k_{F}^{2}}{\Gamma_{0}E_{F}}\frac{k_{F}^{2}}{\Gamma_{0}}\frac{H_{y}H_{z}k_{F}^{3}}{E_{F}^{4}}.\label{eq:sigmaXestim}
\end{multline}

Both estimates arise from the first and second terms in Eq.~\eqref{eq:SigmaXAnalytical}.
Without the loss of generality, we have assumed $\mathbf{h}_{\mathbf{k}}\to h_{x}$, $\mathbf{h}_{\mathbf{p}}\to h_{y}$, $\mathbf{h}_{\mathbf{q}}\to h_{z}$, 
$\mathbf{h}_{\mathbf{Q-q}}\to h_{y}$. Using Eq.~\eqref{Gamma_0R}, we finally get:
\begin{equation}
\sigma_{\alpha\beta}^{X}\sim e^{2}k_{F}\cdot\frac{H_{x}H_{y}H_{z}}{E_{F}^{3}}=\sigma_{0}\frac{H_{x}H_{y}H_{z}}{E_{F}^{3}}.\label{eq:sigmaXFinalEstim}
\end{equation}
Applying the same procedure to the $\Psi$-diagram yields the same estimate, $\sigma_{\alpha\beta}^{\Psi} \sim \sigma_{\alpha\beta}^{X}$.
%\end{widetext}

\end{appendix}

\end{document}